\documentclass[aip,graphicx]{revtex4-1}

\usepackage{empheq}
\usepackage{graphicx}
\usepackage{subfigure}
\newcommand{\beq}{\begin{equation}}
\newcommand{\eeq}{\end{equation}}

\newcommand{\eps}{\varepsilon}

\newcommand{\epsc}{\underline{\varepsilon}}
\newcommand{\muc}{\underline{\mu}}

\renewcommand{\rho}{\varrho}
\renewcommand{\theta}{\vartheta}
\renewcommand{\phi}{\varphi}

\newcommand{\Ev}{\ensuremath{ \vec{E} }}

\newcommand{\Hv}{\ensuremath{ \vec{H} }}
\newcommand{\Bv}{\ensuremath{ \vec{B} }}

\newcommand{\Zc}{\ensuremath{\protect \underline{Z} }}

\newcommand{\vv}{\ensuremath{ \vec{v} }}

\newcommand{\dsv}{\ensuremath{ \mathrm{d}\vec{s}} }

\newcommand{\dV}{\ensuremath{ \mathrm{d}V}}

\newcommand{\dz}{\ensuremath{ \mathrm{d}z}}

\newcommand{\wegdamit}[1]{} 

\newlength{\lwveryfine}   
\newlength{\lwfine}   
\newlength{\lwnormal} 
\newlength{\lwthick}  
\newlength{\lwverythick}  

\begin{document}


\title{Analytic Modeling, Simulation and Interpretation of Broadband Beam Coupling Impedance Bench Measurements} 



\author{Uwe Niedermayer}
\email[]{niedermayer@temf.tu-darmstadt.de}
\affiliation{Institut f\"ur Theorie elektromagnetischer Felder, Technical University Darmstadt, Schlossgartenstr. 8 D-64289 Darmstadt, Germany}

\author{Lewin Eidam}
\affiliation{GSI Helmholtzzentrum f\"ur Schwerionenforschung, Planckstr. 1, D-64291 Darmstadt, Germany}

\author{Oliver Boine-Frankenheim}
\affiliation{GSI Helmholtzzentrum f\"ur Schwerionenforschung, Planckstr. 1, D-64291 Darmstadt, Germany}
\altaffiliation{Institut f\"ur Theorie elektromagnetischer Felder, Technical University Darmstadt, Schlossgartenstr. 8 64289 Darmstadt, Germany}


\date{\today}

\begin{abstract}
In the first part of the paper a generalized theoretical approach towards beam coupling impedances and stretched-wire measurements is introduced.
Applied to a circular symmetric setup, this approach allows to estimate the systematic measurement error due to the presence of the wire. 
Further, the interaction of the beam or the TEM wave, respectively, with dispersive material such as ferrite is discussed.
The dependence of the obtained impedances on the relativistic velocity $\beta$ is investigated and found as material property dependent. 
The conversion formulas for the TEM scattering parameters from measurements to impedances are compared with each other and the analytical impedance solution. 
In the second part of the paper the measurements are compared to numerical simulations of wakefields and scattering parameters. 
In practice, the measurements have been performed for the circularly symmetric example setup. 
The optimization of the measurement process is discussed. 
The paper concludes with a summary of systematic and statistic error sources for
impedance bench measurements and their diminishment strategy. 
\end{abstract}


\maketitle 


\section{Introduction}
The field distribution of a single particle in free space approaches the one of a lossless coaxial TEM transmission line in the ultrarelativistic limit. This motivates measuring the longitudinal or transverse beam coupling impedance of accelerator components by replacing the beam with one or two wires, respectively. The transmission line measurement technique has been introduced by Sands and Rees \cite{Sands1974} for the determination of beam energy loss factors in Time Domain (TD) by pulse excitation. When using modern Vector Network Analyzers (VNA) the beam coupling impedance can be determined in Frequency Domain (FD) by sweeping a narrow-band signal. Especially when looking at particular sidebands that are susceptible to beam instabilities rather than on the total energy loss the FD method is to be preferred. 

In both TD and FD one has to make sure not to measure effects of the setup.
The de-embedding process to measure only the accelerator device under test (DUT) is investigated especially for lumped impedances by Hahn and Pedersen \cite{Hahn1978}. In order to enable de-embedding with a reference (REF) measurement of an empty box or beam pipe, the impedance mismatch from the cables to the measurement box has to be minimized. 
At high frequency one can also use Time Domain Gating to disregard the mismatch reflections \cite{Caspers1992}, but this requires a very high bandwidth of the VNA to properly represent the spectrum of the window-function. Another option is to damp multiple reflections with RF attenuation foam.

Walling et al.\cite{Walling1989} first introduced an approximative formula for measuring distributed impedances which was later replaced by the exact one by Vaccaro \cite{Vaccaro1994} and Jensen\cite{Jensen2000}. 

This paper covers analytical and numerical models for longitudinal and transverse impedance measurement of strongly lossy and broadband structures. 
The models will be applied to the example case of a dispersive Ferrite ring. 
Starting from a 2D analytical model, its limitations are illustrated by a 3D numerical model for finite length. 

The analytical models imply also that there cannot be a general formula to scale the impedance with the beam velocity. Also the bench measurements cannot be scaled for $\beta<1$, but the measurements can be used to validate numerical simulations \cite{CST}\cite{Doliwa2007a}\cite{Niedermayer2012b}, that allow velocity scaling. Numerical simulations for $\beta=1$ are also important to avoid wrong a priori assumptions in the measurements. The analytical model for the dispersive material presented here motivates also a simplified low frequency (LF) approach ("radial model") that plays an important role for the interpretation of LF impedance in general and in particular of coil measurements for transverse impedance\cite{Roncarolo2009}.

The paper is structured as follows: Section \ref{sect_ana} starts with the analytical model for the beam impedance and for the measurement, i.e. a model with excitation and an Eigenvalue problem, respectively. Both are solved for circularly symmetric 2D geometry. In Sect. \ref{sect_meas_tech} the way to determine the impedance from scattering parameters is discussed (see also \cite{Hahn2004a}). Section \ref{sect_ana_res} then draws an intermediate conclusion, comparing the analytical results only. These are the beam models for different velocity and the measurement model with different S-parameter conversion formulas and wire thicknesses.

The real Ferrite ring, as it was measured, was simulated with a particle beam (TD) and a wire (TD/FD), as described in Sect. \ref{sect_num}. This is followed by the discussion of measurement results in Sect. \ref{sect_meas}. Section \ref{sect_trans} points out the commonalities and differences for the longitudinal and transverse measurements.

The paper concludes with summarizing measurement error sources and discussion of the interplay between measurements and simulations, also for $\beta<1$ in Sect. \ref{sect_conclusion}.

\section{Analytical model}
\label{sect_ana}
In a first analytical approach, the beam and the wire setup are considered as purely two dimensional. 
It will be seen in section \ref{sect_meas_tech}, that this is justified for large longitudinal electrical length. 
From Maxwell's equations we find the 2D Helmholtz equation 
\beq
		( \Delta_\perp +k_\perp^2 ) E_z=rhs, \label{equ_helmholz}
\eeq
and the dispersion relation
\beq
 k_\perp^2+k_z^2 = \omega ^2 \underline{\mu} \underline{\epsilon}
\label{dispersionrelation}
\eeq
which will be solved for three different assumptions:
\begin{enumerate}
	\item Beam model
	\beq
		k_z=\frac{\omega}{\beta c}
	\eeq
	\beq
			rhs=-\frac{i\omega}{\beta^2\gamma^2}\mu_0\frac{q}{\pi a^2}H(a-r)
	\eeq
	 with beam radius $a$ and $H$ being the Heaviside step function
	\item Radial model
	obtained from beam model with $\beta\rightarrow\infty$, i.e.
	\beq
			k_z=0, \;\; \gamma=0,\;\; \beta\gamma=i, \;\; \Ev_\perp=0
	\eeq
	\item Coaxial line model
	\beq	
		E_z(r\leq r_0)=0\;,\;\;\rm{Quasi-TEM-Eigenmode} \;(rhs=0),
	\eeq
	where the Eigenvalue $k_z$ is obtained from the equation
	\beq
		(\Delta_\perp +\omega^2 \muc\epsc ) E_z= k_z^2 E_z. \label{Eigenvalue_eq}
	\eeq
\end{enumerate}
The range of validity of the radial model is also discussed in \cite{Niedermayer2012} and \cite{Hahn2010}. 

Before solving Eq. \ref{equ_helmholz} we take a closer look on the dispersion relation \ref{dispersionrelation}, rewritten for the beam model as 
\beq
 k_\perp^2=\frac{\omega^2}{c_0^2}(\muc_r\epsc_r-\frac{1}{\beta^2}).
   \label{separationEq}
\eeq
The material properties are presented as 
\beq
\muc=\mu'-i\mu'' \;\; \rm{and} \;\;
\epsc=\eps'-i\eps''+\frac{\kappa}{i\omega}
\eeq
with $\kappa$ being the conductivity and $\mu''$ and $\eps''$ being magnetization and polarization losses. Note that all these material properties are considered as functions of the frequency. Furthermore we define the lossless refraction index and the loss tangents as 
\beq
n=\sqrt{\mu_r' \eps_r'} \;, \;\;\; \tan \delta_\mu=\frac{\mu_r''}{\mu_r'} \;\;\; \rm{and} \;\;\; \tan \delta_\eps=\frac{\eps_r''+\kappa/\omega\eps_0}{\eps_r'}.
\eeq 
This allows to rewrite Eq. \ref{separationEq} as
\beq
 k_\perp^2=\frac{\omega^2}{c_0^2}\left[n^2(1-\tan \delta_\mu\tan \delta_\eps)-\frac{1}{\beta^2}-i n^2(\tan \delta_\mu+\tan \delta_\eps)\right]
\label{dispersion_tandelta}
\eeq
which shows that in the lossless case one has transversely propagating waves exactly when the the Cerenkov-condition $\beta n > 1$ is fulfilled.
This still holds in the case of dielectric losses and nonconducting ferrites, but the product of the tangents cannot be dropped in the case of electrically conducting magnetic material such as Magnetic Alloys.
For lossy material it makes sense to plot $k_\perp^2$ in the complex plane parametrically, as a function of $\omega$ and $\beta$.
Figure \ref{fig:kPerpSquare} shows the properties of the different quadrants in the complex $k_\perp^2$-plane. 
\begin{figure}[htbp]
	\centering
		\includegraphics[width=0.4\textwidth]{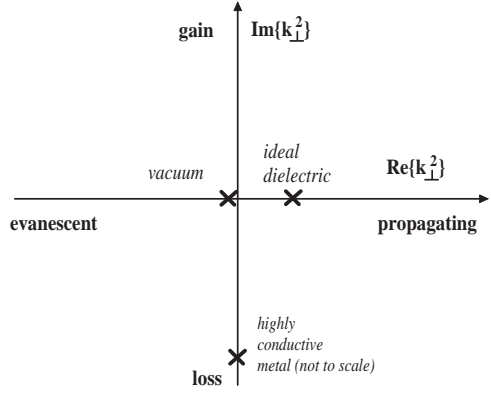}
		\caption{Complex $k_\perp^2$ plane (transverse propagation plot). The vertical axis represents the Cerenkov-condition.}
	\label{fig:kPerpSquare}
\end{figure}
\begin{figure}[ht]
		\includegraphics[angle=-90, width=0.47\textwidth]{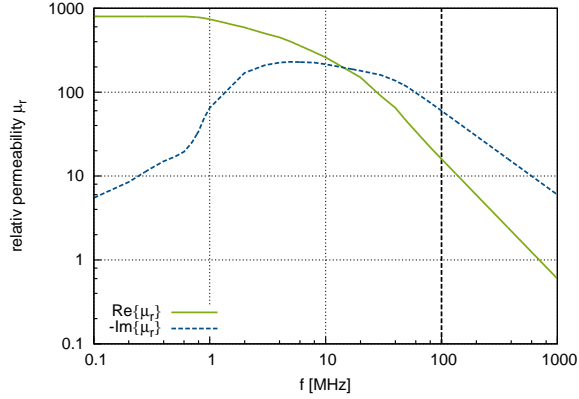}
	\caption{Material properties of the example ferrite material \cite{Mischung43}. The permittivity is roughly constant, $\epsilon_r=10 $. Above 100MHz a power law extrapolation has been applied.}
	\label{mischung43}
\end{figure}
For further considerations we will focus on some material with properties shown in Fig. \ref{mischung43}.
\begin{figure}[htbp]
	\centering
		\includegraphics[angle=-90,width=0.47\textwidth]{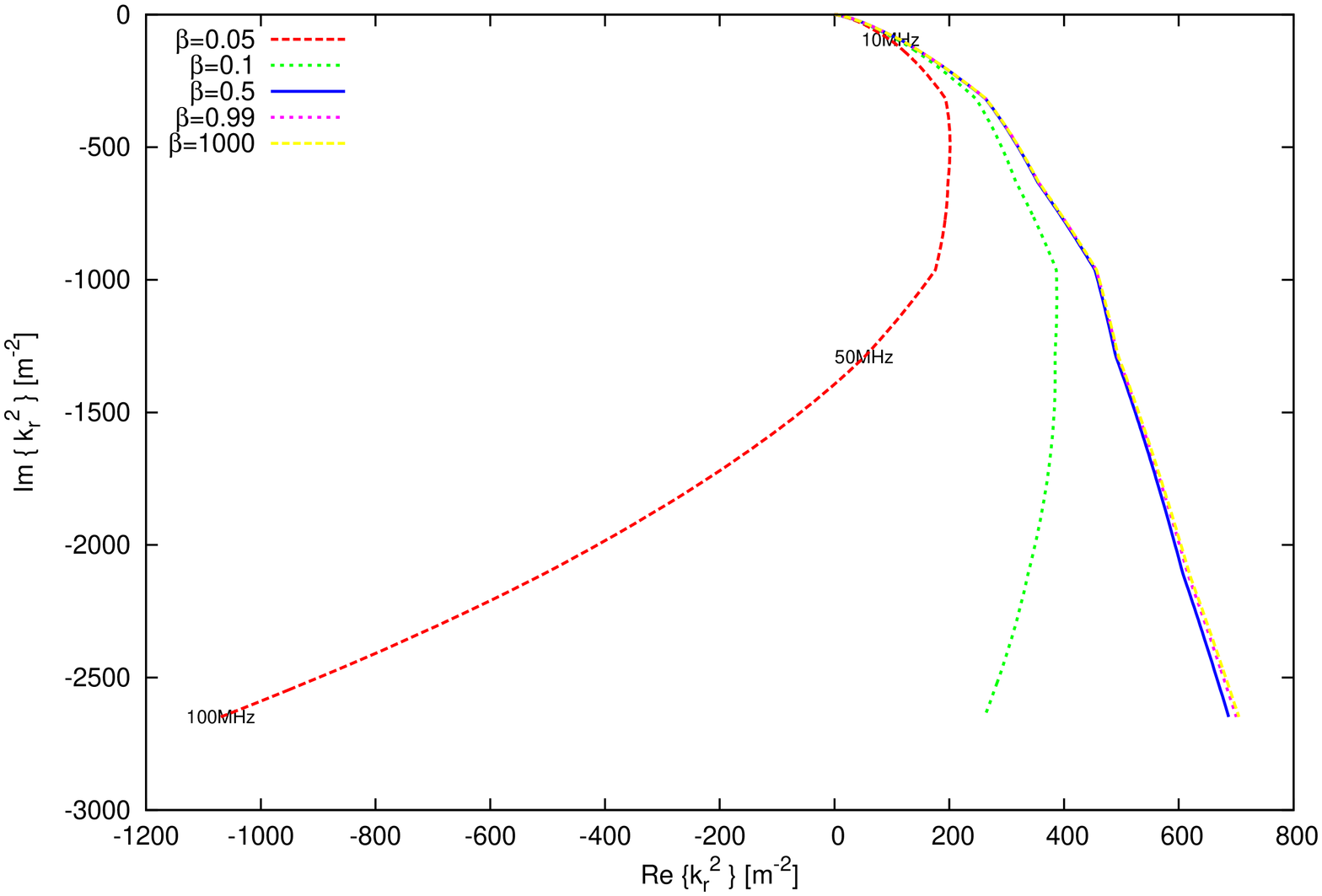}
		\includegraphics[angle=-90,width=0.47\textwidth]{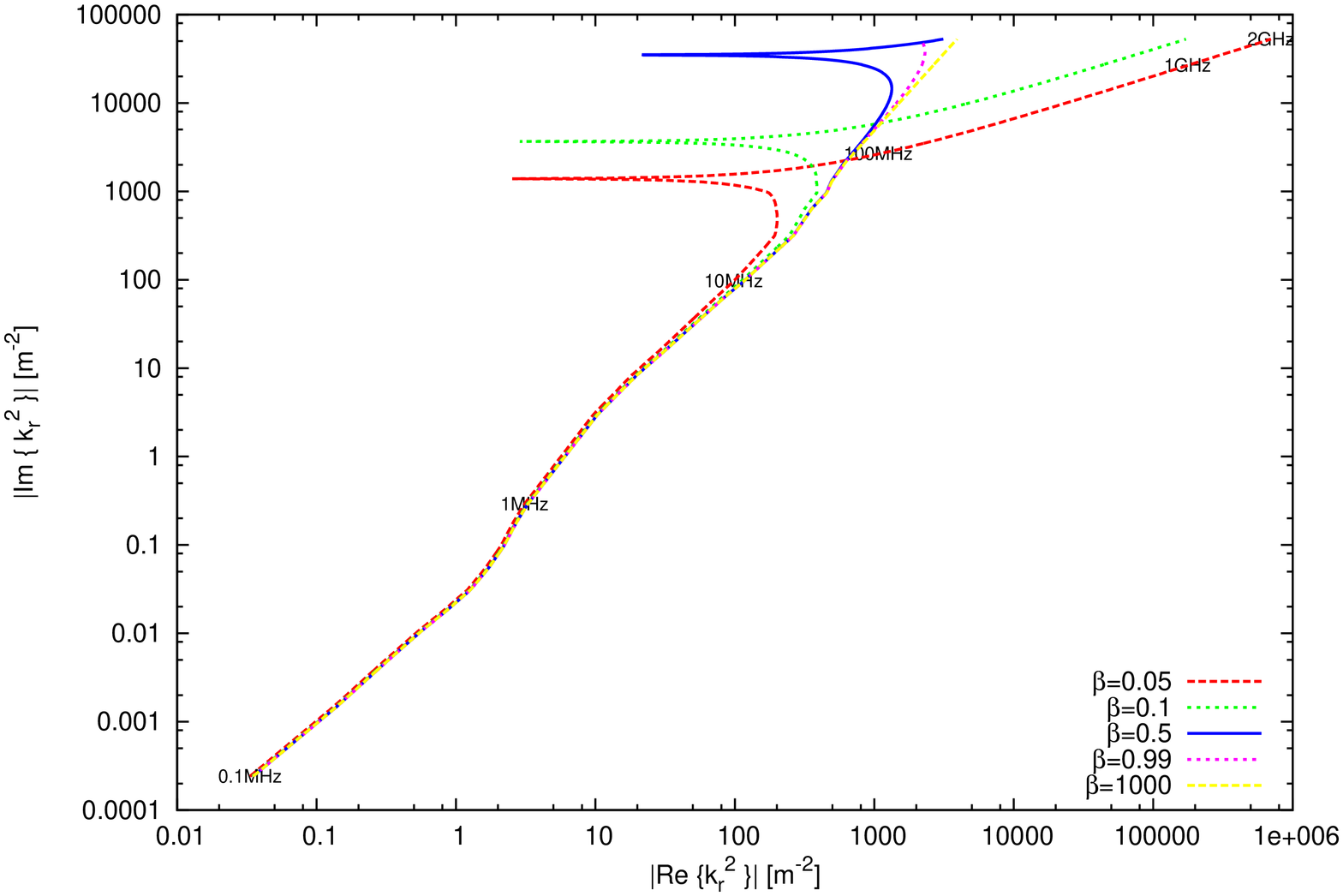}
		\caption{Transverse complex wavenumber in normal (see Fig. \ref{fig:kPerpSquare}) and loglog display for the material presented in Fig. \ref{mischung43}}
	\label{fig:kPerpSquare}
\end{figure}
The transverse wavenumber as calculated by Eq. \ref{separationEq} is plotted in Fig. \ref{fig:kPerpSquare} where one can see that the $\beta$-dependence is small if $\beta>0.5$ and $f<100\;$MHz.
This motivates again the radial model, i.e. neglecting the $\beta$ dependence entirely.
\begin{figure}[ht]
		\includegraphics[width=0.3\textwidth]{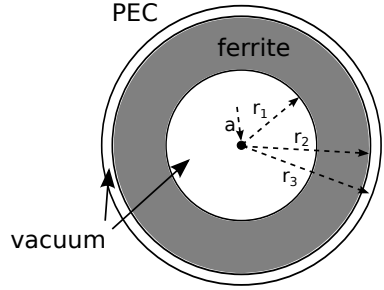}
	\caption{Ferrite ring for benchmarking the measurement setup. Dimensions: $	r_1=\rm{1.78}{cm}\, ; \; r_2=\rm{3.05}{cm}\, ; \; r_3=\rm{3.3}{cm}\, ; \; L=\rm{2.54}{cm}.$}
	\label{AnalyticSetup}
\end{figure}
For simple analytical treatment due to $k_\perp=k_r$, we will focus on a concentrical cylinder setup, as shown in Fig.\;\ref{AnalyticSetup}.
\begin{figure}[ht]
	\centering
		\includegraphics[angle=-90, width=.47\textwidth]{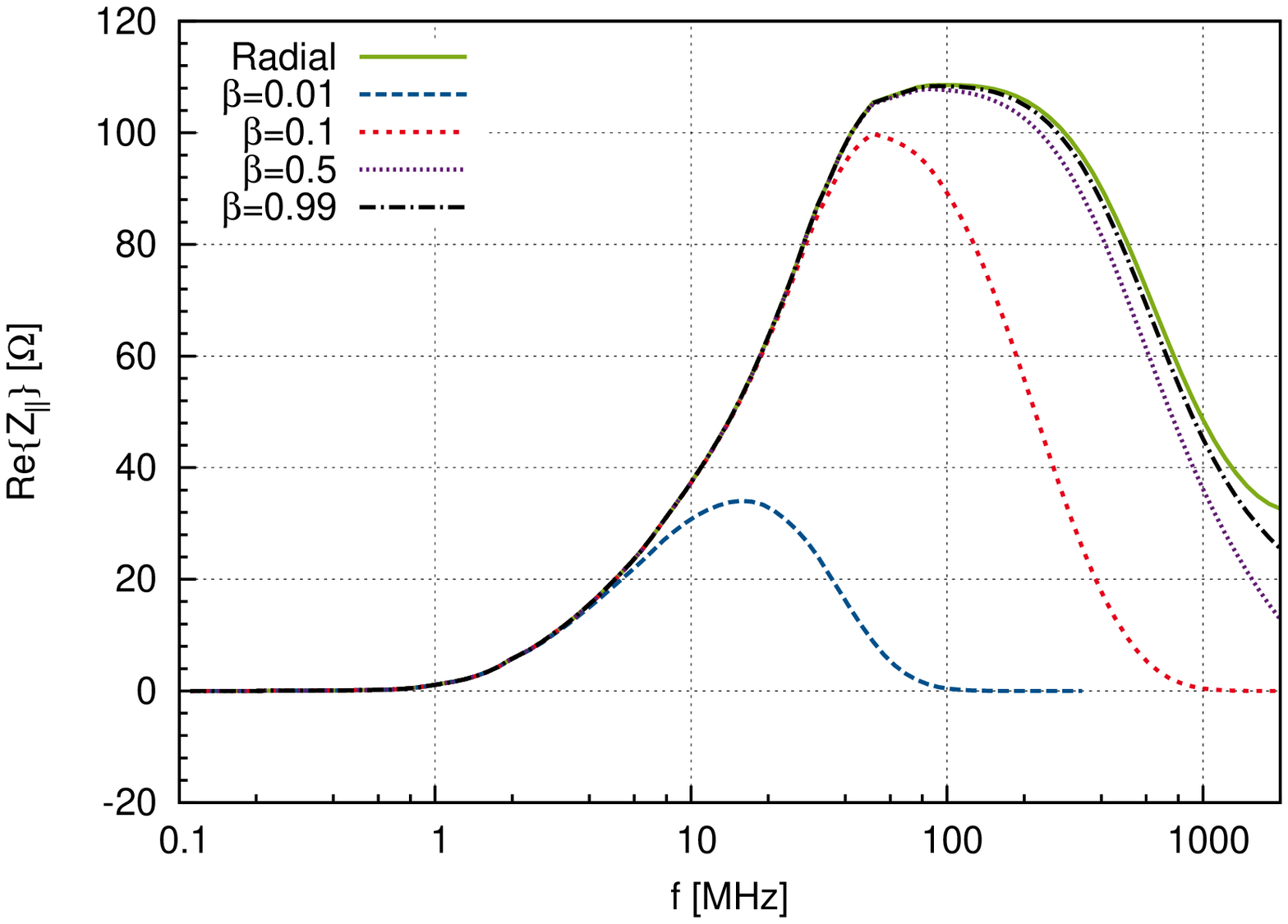}
		\includegraphics[angle=-90, width=.47\textwidth]{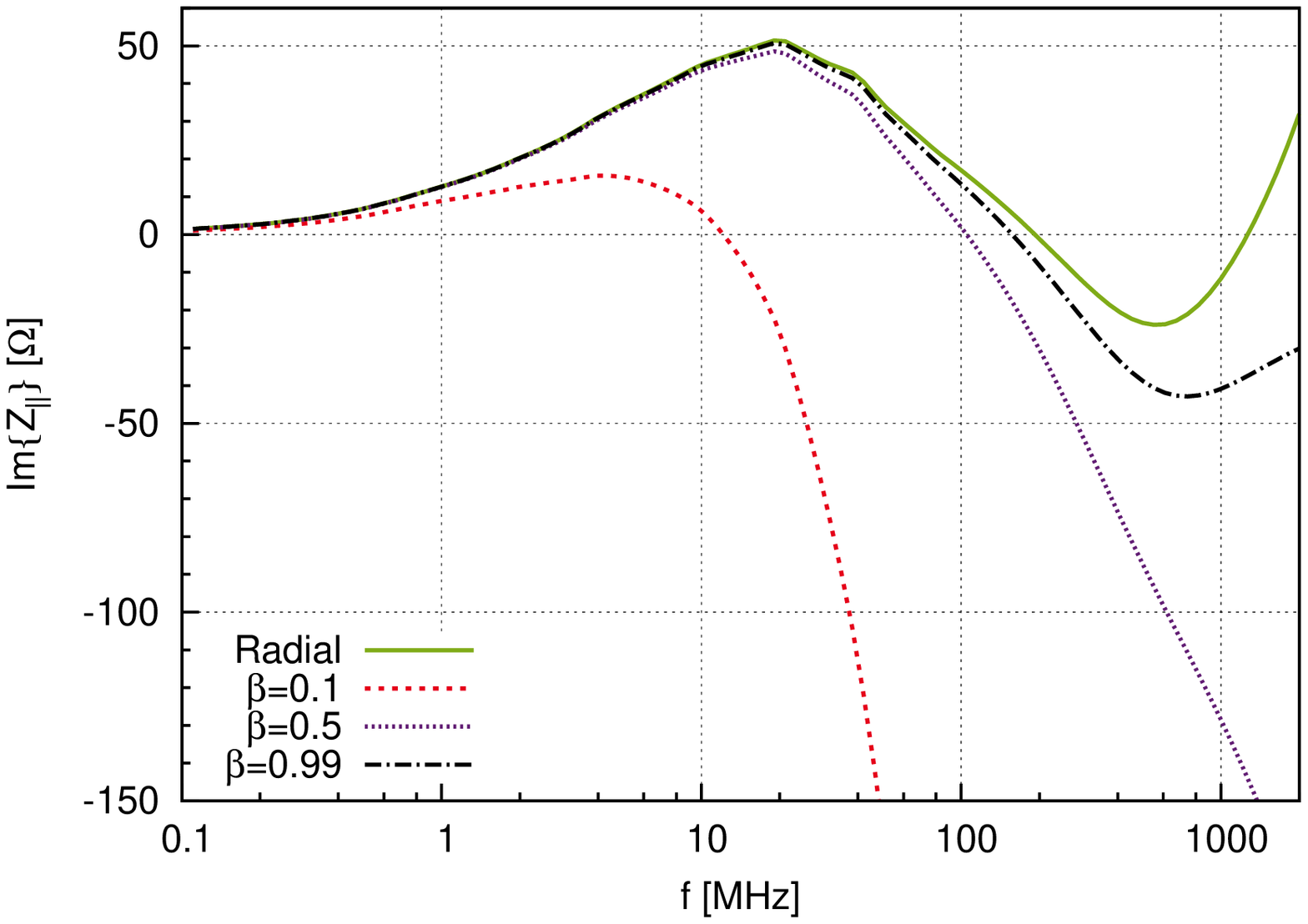}
	\caption{Beam model: Longitudinal impedance for different beam velocity}
	\label{analytic_beam}
\end{figure}

For all three models a solution is found from the ansatz
\begin{align}
		E_z=
			\left\{
				\begin{array}{ll}
					 \left(A_0+A_1 J_m(k_r r)\right)e^{-i k_z z} &r<a\\
					\left(B_1\cdot J_m(k_r r) + B_2\cdot N_m(k_r r)\right) e^{-i k_z z} & a\leq r <r_1\\
					\left(C_1\cdot J_m(k_r^F r) + C_2\cdot N_m(k_r^F r)\right) e^{-i k_z z} & r_1\leq r <r_2\\
					D_1\left(J_m(k_r r) - \frac{J_m(k_r r_3)}{N_m(k_r r_3)}N_m(k_r r)\right) e^{-i k_z z} & r_2\leq r \leq r_3
				\end{array}
			\right.
			\label{ansatz}
\end{align}
where the wavenumbers in radial direction are distinguished by $k_r$ for vacuum and $k_r^F$ inside the Ferrite.
Note that this ansatz is invalid for the coaxial model in case of no losses, since $E_z=0$ for the pure TEM mode.
For the coaxial model one applies $A_0=A_1=0$ and $B_1$ defines an arbitrary amplitude (Eigenvector scaling). The beam model requires additionally a particular solution without boundary conditions, i.e. 
\begin{align}
			E_z=A_0\cdot H(r-a),
\end{align}
satisfying Eq. \ref{equ_helmholz} with
\begin{align}
			A_0=\frac{iq}{\omega \eps_0 \pi a^2}.
\end{align}
Note that since $A_0$ is independent of $\beta$, the $\beta$ dependence in the general ansatz Eq. \ref{ansatz} is given entirely through $k_r$ and $k_r^F$. Therefore, since the impedance originates from the ferrite, the relativistic $\beta$ enters similar as a material property. Also one cannot expect to find a general impedance scaling law with $\beta$ since the impact of $\beta$ on $k_r$ and $k_r^F$ is different which means that the total impact depends on the geometry.

For solving the equation system \ref{ansatz} one has to determine $5$ constants in the coaxial line model ($B_2,C_1,C_2,D_1,k_z$) from $5$ matching conditions and $6$ constants in the beam model ($A_1,B_1,B_2,C_1,C_2,D_1$) from $6$ matching conditions. The matching conditions are
\begin{align}		
		E_z|_{r_i+}&=E_z|_{r_i-}\\
		H_\phi|_{r_i+}&=H_\phi|_{r_i-},
\end{align}
with
\begin{align}
		H_\phi=-\frac{i\omega \epsilon}{k_r^2}\cdot \frac{\partial E_z}{\partial r}.
\end{align}
in all models, obtained from component-wise rearranging Maxwell's equations.

In the Coaxial Line model one obtains a nonlinear transcendent Eigenvalue Equation, that has the Eigenvalue $k_r=(\omega^2\muc\epsc-k_z^2)^{1/2}$ in the arguments of the Bessel functions. 
For the simplified case of $r_2=r_3$ the Eigenvalue equation reads
\beq
\frac{\eps}{k_r}\frac{J_0'(k_r r_1)N_0(k_r a)-J_0(k_r a)N_0'(k_r r_1)}{J_0(k_r r_1)N_0(k_r a)-J_0(k_r a)N_0(k_r r_1)}
=\frac{\eps^F}{k_r^F}\frac{J_0'(k_r^F r_1)N_0(k_r^F r_2)-J_0(k_r^F r_2)N_0'(k_r^F r_1)}{J_0(k_r^F r_1)N_0(k_r^F r_2)-J_0(k_r^F r_2)N_0(k_r^F r_1)}.
\eeq
This can be solved only numerically and solution is a Quasi-TEM mode, having a small $E_z$-component but no cut-off frequency. The complex $k_z$ is shown in Fig \ref{fig-dispersion} and determines the transmission by $S_{21}=exp(-ik_z l)$. The impedance is then found by a conversion formula described in the next chapter.
In the beam model one finds longitudinal impedance (m=0) from  
\beq
			Z_\parallel(\omega)=-\frac{1}{q^2}\int_{\mbox{\small beam}}\vec{E}\cdot \vec{J}^*\dV=-\frac{l}{q}\bigg(\underbrace{\frac{2 J_1(k_r a)}{k_r a}}_{\approx 1}\cdot A_1+A_0\bigg) \label{exc_imp}.
\eeq
The longitudinal impedance is shown in Fig. \ref{analytic_beam} for different $\beta$. As already expectable from Fig. \ref{fig:kPerpSquare}, the beam model agrees with the radial model for LF and not too small $\beta$. 

Before we discussing the wire technique we shortly summarize some parameters important for the comparison of the models:
The wave impedance is defined as $Z_{wave}=S_z/|\Hv_\perp|^2$ and the (measurable) characteristic impedance is 
\beq
Z_0=\frac{\int_0^{r_3}\Ev_\perp\cdot\dsv}{\oint \Hv \cdot \dsv}.
\eeq
The longitudinal space charge impedance, as it will be dominating in Fig. \ref{analytic_beam} for very low $\beta$, can also be deduced from the characteristic impedance (electric part) and the image current inductance (magnetic part), i.e.
\beq
E_z=-\partial_z(Z_0I)-\partial_t(\mu_0 I \frac{g_b}{2\pi}).
\eeq
Subsequently, one obtains for a perfectly conducting circular beam pipe
\beq
Z_\parallel^{spch}=-i\omega\frac{\eta}{c}l\frac{g_b}{2\pi} \frac{1}{\beta^2\gamma^2}.
\eeq
In the radial model one has only the magnetic part since the transverse electric field is zero.
Table \ref{Ztable} shows an overview of intrinsic parameters of the models.
Note that the geometry factor for the beam and the coaxial line model are different due to the presence of fields within the beam.

\begin{table}[htb]
\begin{tabular}{||l||l|l|l|} \hline\hline
															& Beam Model 																														& Radial Model 															& Coaxial Line Model 							 \\  \hline\hline
$k_z$ 												& $\frac{\omega}{\beta c}$ 																							& $0$																				&  Eigenvalue  											\\  \hline
$k_r$ (vacuum)								& $\frac{i\omega}{\beta \gamma c}$																			& $\frac{\omega}{c}$												&  $\sqrt{(\omega/c)^2-k_z^2}$  		\\  \hline
$Z_{wave}^{REF}$ 							& $\eta/\beta$ 	\;\;		(!)																							& $0$																				&	 $\eta$														\\  \hline
$Z_{wave}^{DUT}$ 							& $\frac{k_z}{\omega\epsc}=\frac{1}{\beta c \epsc}$ 										& $0$																				&	 $\frac{k_z}{\omega\epsc}$					\\  \hline
$Z_0$ (vacuum)								& $\frac{g_{b}}{2\pi}Z_{wave}^{REF}$																		& $0$																				&	$\frac{g_{c}}{2\pi}Z_{wave}^{REF}$ \\  \hline
$Z_\parallel^{spch} (vacuum)$	& $-ik_z l Z_{0}/\gamma^2$																							& $i\omega \mu_0 l\frac{g_b}{2\pi}$ 				& $0$																				\\  \hline
$g$-factor										& $g_b=\frac{1}{2}+\ln\frac{r_3}{a}$																		& $g_b$																			& $g_c=\ln\frac{r_3}{a}$			\\  \hline
cut-off $\omega_c$						& $\approx\frac{\beta\gamma c}{a}\sqrt\frac{2}{g_b}$										& 	--																			& $\approx\frac{2c}{\pi(a+r_3)}$			\\  \hline
\end{tabular}
\caption{Overview of properties in the different models ($\eta=\sqrt{\mu_0/\eps_0}=377\Omega$)}
\label{Ztable}
\end{table}


\section{Wire Measurement Technique}
\label{sect_meas_tech}
The classical wire technique is based on a coaxial setup, where the device under test (DUT) can be seen as an additional complex impedance added in the coaxial line replacement circuit. Figure \ref{coax} shows the setup and the replacement circuit model of an infinitely short piece of it. 
Usually the measurement is performed with respect to a reference line, which can be either a piece of beam pipe or the vacuum vessel of the DUT. There are also approaches to obtain the reference signals analytically, especially for plain beam pipes.
An important parameter in the analysis is the electrical length in units of radians, defined by 
\beq
\Theta=2\pi \frac{l}{\lambda}=kl
\eeq
where the wavelength $\lambda=2\pi/k$ can have different values in longitudinal and transverse direction and in different materials.
There is also an important distinction between a lumped impedance, i.e.
\beq
\frac{\partial Z_\parallel (\omega,z)}{\partial z}=Z_\parallel^{total}(\omega) \delta(z-z_0)
\eeq
and a distributed impedance,
\beq 
\frac{\partial Z_\parallel (\omega,z)}{\partial z}=\frac{Z_\parallel^{total}(\omega)}{l}.
\eeq
In practice, one has neither of the two but something in between. The impedance jump (geometric impedance) at the beginning of the DUT is always lumped, while the body of the DUT (resistive wall) is almost equally distributed. The modeling of lumped impedances is just an impedance element in longitudinal direction, while distributed impedances are represented by a TEM-line with an impedance element $Z_\parallel/l$ equally distributed to each infinitely short transmission line element.

\subsection{Distributed Impedance}

\begin{figure}[ht]
	\centering
	\begin{minipage}[t]{0.47\textwidth}
	\centering
		\includegraphics[width=4.5cm]{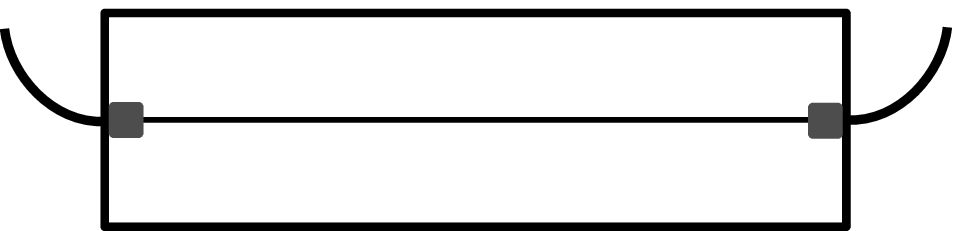}
	\end{minipage}
	\begin{minipage}[t]{0.47\textwidth}
		\centering
		\includegraphics[width=4.5cm]{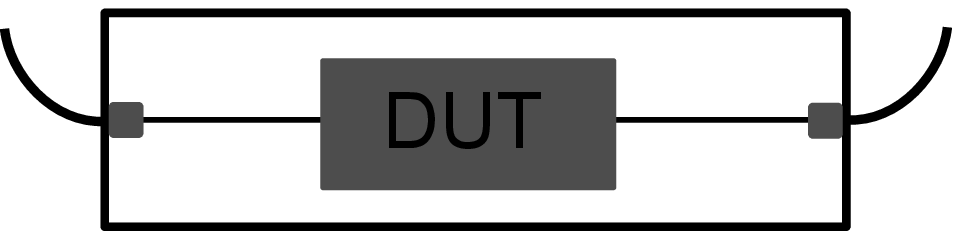}
	\end{minipage}
	
	\begin{minipage}[t]{0.47\textwidth}
	\centering
		\includegraphics[width=4.5cm]{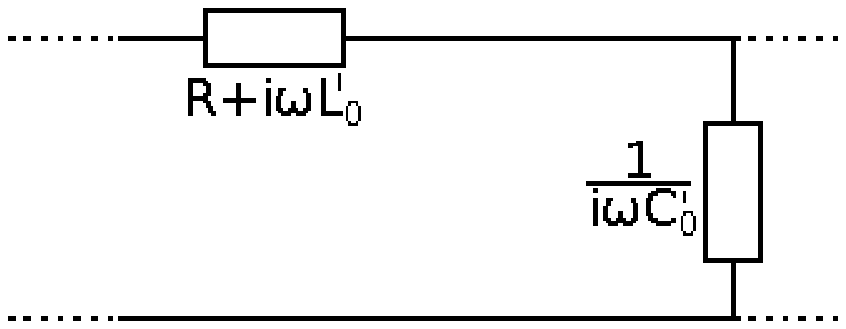}
	\end{minipage}
	\begin{minipage}[t]{0.47\textwidth}
		\centering
		\includegraphics[width=4.5cm]{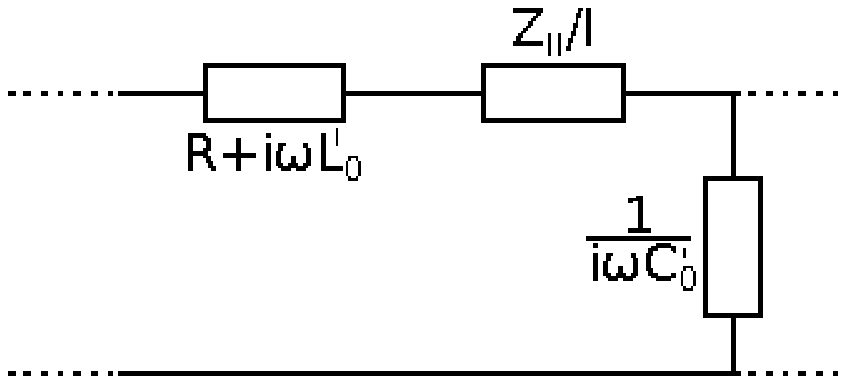}
	\end{minipage}
	\caption{Transmission line replacement circuit for distributed impedance}
	\label{coax}
\end{figure}
For equally distributed impedance sources the complex wave numbers in the setup shown in Fig. \ref{coax} are given by \cite{Pozar1998}
\beq
	k_z^{DUT}=\omega \sqrt{C'_0 L'_0} \sqrt{ 1-i\frac{R'_0+Z_\parallel/l}{\omega L'_0}} \hspace{1cm}
	k_z^{REF}=\omega \sqrt{C'_0 L'_0} \sqrt{ 1-i\frac{R'_0}{\omega L'_0}}
\eeq
\beq
	Z_{0}^{(REF)} =\sqrt{\frac{R'_0 +i\omega L'_0}{i\omega C'_0}}
\eeq
which can be solved as 
\beq
	Z_\parallel=i Z_0 \frac{(k_{z,DUT}^2-k_{z,REF}^2)\cdot l}{k_{z,REF}} = i Z_0l\cdot (k_{z,DUT}-k_z^{REF})\cdot \left( 1+ \frac{k_z^{DUT}}{k_z^{REF}}\right) 
	\label{Zlongfromk}
\eeq
These wavenumbers can be obtained from the scattering matrix measured by the VNA.
The scattering matrix of a piece of transmission line of length $l$ and characteristic impedance $Z_{0,d}$ in an environment of characteristic impedance $Z_0$ is given by \cite{Vaccaro1994}
\begin{align}
	S=\left(
	\begin{matrix}
	S_{11} & S_{12}\\
	S_{21} & S_{22}
	\end{matrix}\right)	
	=\frac{	\left(
	\begin{matrix}
	(Z_{0,d}^2-Z_{0}^2)\sin(k_zl) & -2iZ_{0,d}Z_{0}\\
	-2iZ_{0,d}Z_{0} & (Z_{0,d}^2-Z_{0}^2)\sin(k_zl)
	\end{matrix}
	\right)}
	{(Z_{0,d}^2+Z_{0}^2)\sin(k_zl)-2iZ_{0,d}Z_{0}\cos(k_zl)}
 \label{Sdist}
\end{align}
In case of no reflections at DUT, i.e. $Z_{0,d}\simeq Z_0$, Eq. \ref{Sdist} simplifies to  
\beq
	S_{21}=S_{12}=e^{-ik_z l} .
	\label{equ_S_assumption}
\eeq
Otherwise one has to introduce a corrected $S_{21}$ parameter $S_{21}^C:=\exp(-ik_z l)$ that can be obtained by solving Eq. \ref{Sdist} for $\cos(k_z l)$. 
The quadratic equation for $S_{21}^C$ is called Wang-Zhang\cite{Wang2000}-formula,
\begin{align}
	(S_{21}^C)^2+\frac{S_{11}^2-S_{21}^2-1}{S_{21}}S_{21}^C +1 =0 \;\;\rm{with}\; |S_{21}^C|<1   \label{WangZhang}
\end{align}
and requires knowledge of the $S_{11}$-parameter.
The wavenumber $k_z$ is found from the complex logarithm of Eq. \ref{equ_S_assumption} with either original or corrected $S_{21}$. It can be inserted into \ref{Zlongfromk} to obtain 
\beq
	Z_\parallel=Z_0 \cdot \ln \left(\frac{S_{21}^{REF}}{S_{21}^{DUT}}\right)\cdot \left[1+\frac{\ln (S_{21}^{DUT})}{\ln (S_{21}^{REF})}\right]. \label{improved-log-formula}
\eeq
In the literature this is called (Vaccaro\cite{Vaccaro1994}-Jensen\cite{Jensen2000}-) improved-log formula.
Although this formula is exact, it is in some cases disadvantageous since it is very sensitive to statistical errors of subsequent DUT and REF measurements.
Its approximation under the assumption of small $|S_{21}^{DUT}-S_{21}^{REF}|$ , i.e. $\ln(S_{21}^{DUT})\approx \ln(S_{21}^{REF})$ is the more robust but less accurate (Walling \cite{Walling1989}-) log-formula,
\beq
Z_\parallel=2\cdot Z_0 \cdot \ln \left(\frac{S_{21}^{REF}}{S_{21}^{DUT}}\right) .
\label{log-formula}
\eeq

\subsection{Lumped Impedance}
For purely lumped impedances, i.e. an impedance circuit $Z_d$ element squeezed between two reference lines wit characteristic impedance $Z_0$, one finds \cite{Pozar1998}
\begin{align}
	S=\left(
	\begin{matrix}
	S_{11} & S_{12}\\
	S_{21} & S_{22}
	\end{matrix}\right)	
	=\frac{1}{2Z_0+Z_d}	\left(
	\begin{matrix}
	 Z_{d} & 2Z_0\\
	 2Z_0 & Z_{d}
	\end{matrix}
	\right)
 \label{S_lumped}
\end{align}
resulting in the Hahn-Pedersen \cite{Hahn1978} formula,
\beq
   Z_\parallel=2 Z_0\frac{S_{21}^{REF}-S_{21}^{DUT}}{S_{21}^{DUT}}.
	\label{HPformula}
\eeq
This is an improvement of the original Sands and Rees \cite{Sands1974} formula
\beq
   Z_\parallel=2 Z_0\frac{S_{21}^{REF}-S_{21}^{DUT}}{S_{21}^{REF}}.
	\label{SRformula}
\eeq
Both Eqs. \ref{HPformula} and \ref{SRformula} can be obtained from Taylor expansion of the positive/negative logarithm in Eq. \ref{log-formula}. 
Note that the reflection $S_{11}$ does not play a role for the determination of purely lumped impedances.

\section{Discussion of Analytical Results}
\label{sect_ana_res}
\begin{figure}[ht]
	\centering
		\includegraphics[angle=-90, width=.47\textwidth]{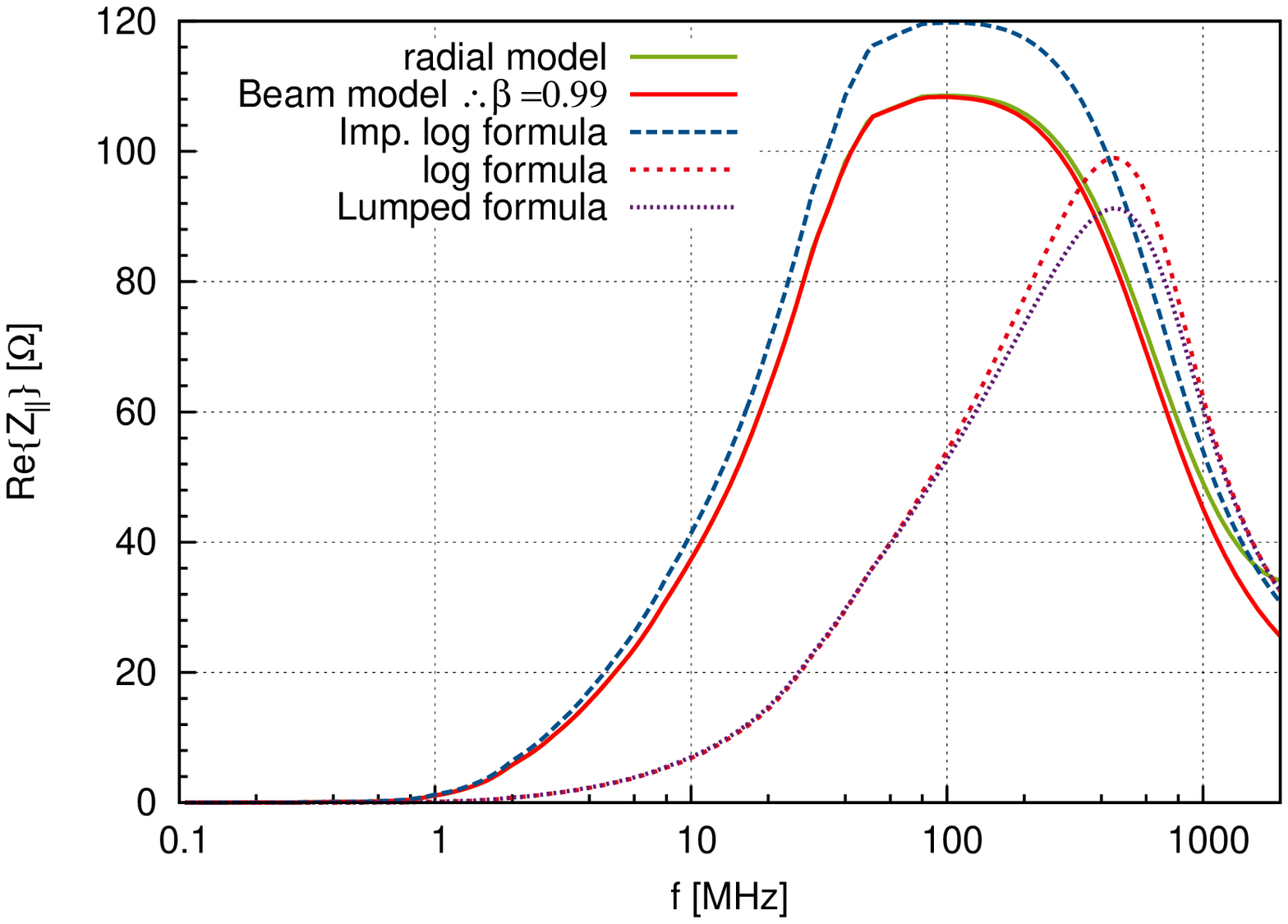}
		\includegraphics[angle=-90, width=.47\textwidth]{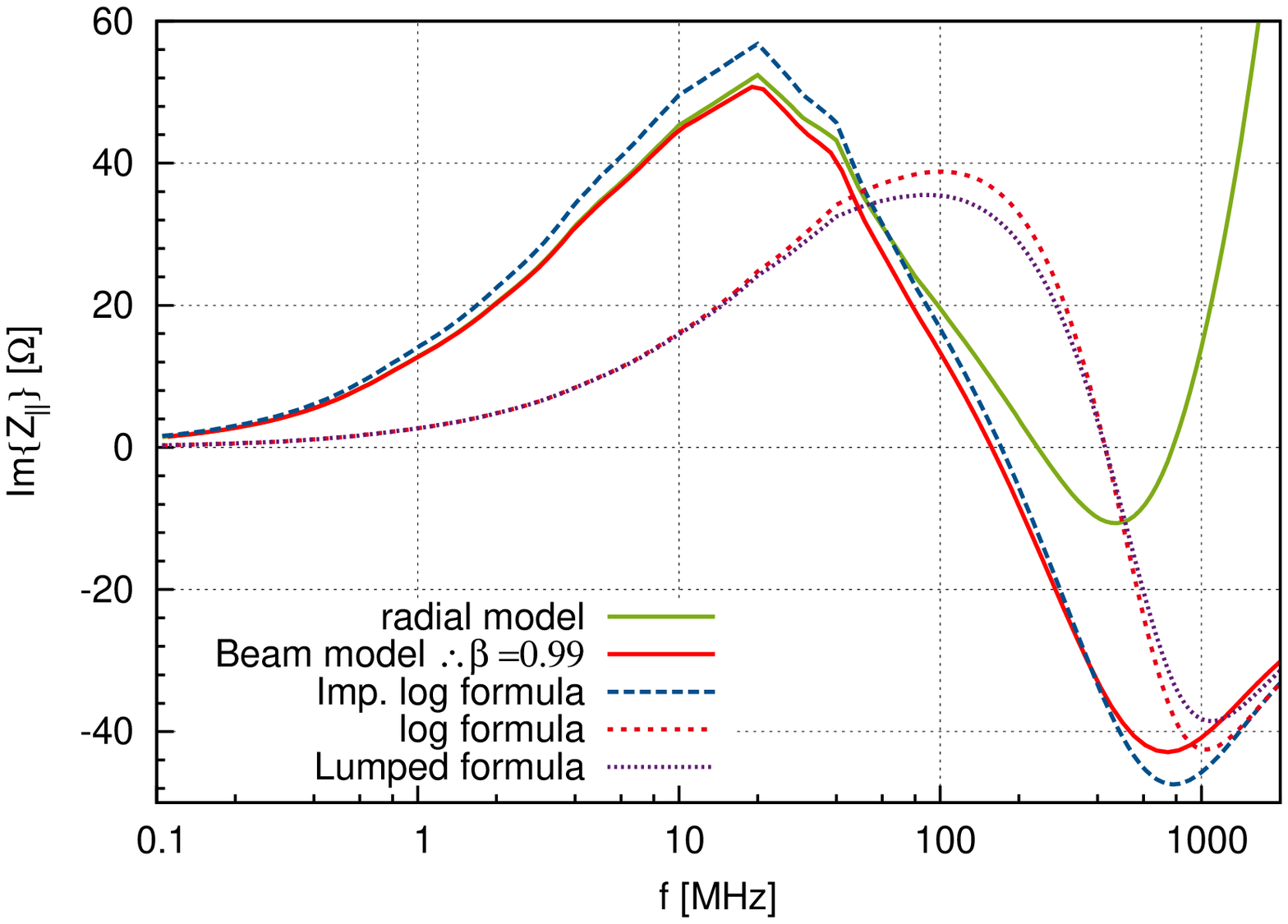}
	\caption{Analytic coaxial wire method with different $S_{21}\rightarrow\Zc_\parallel$-conversion formulas vs. beam and radial model. The wire radius has been chose as $r_0=0.225$ mm as it was the smallest practically achievable.}
	\label{analytic_wiremethod_formulas}
\end{figure}
\begin{figure}[thb]
	\centering
		\includegraphics[angle=-90, width=.47\textwidth]{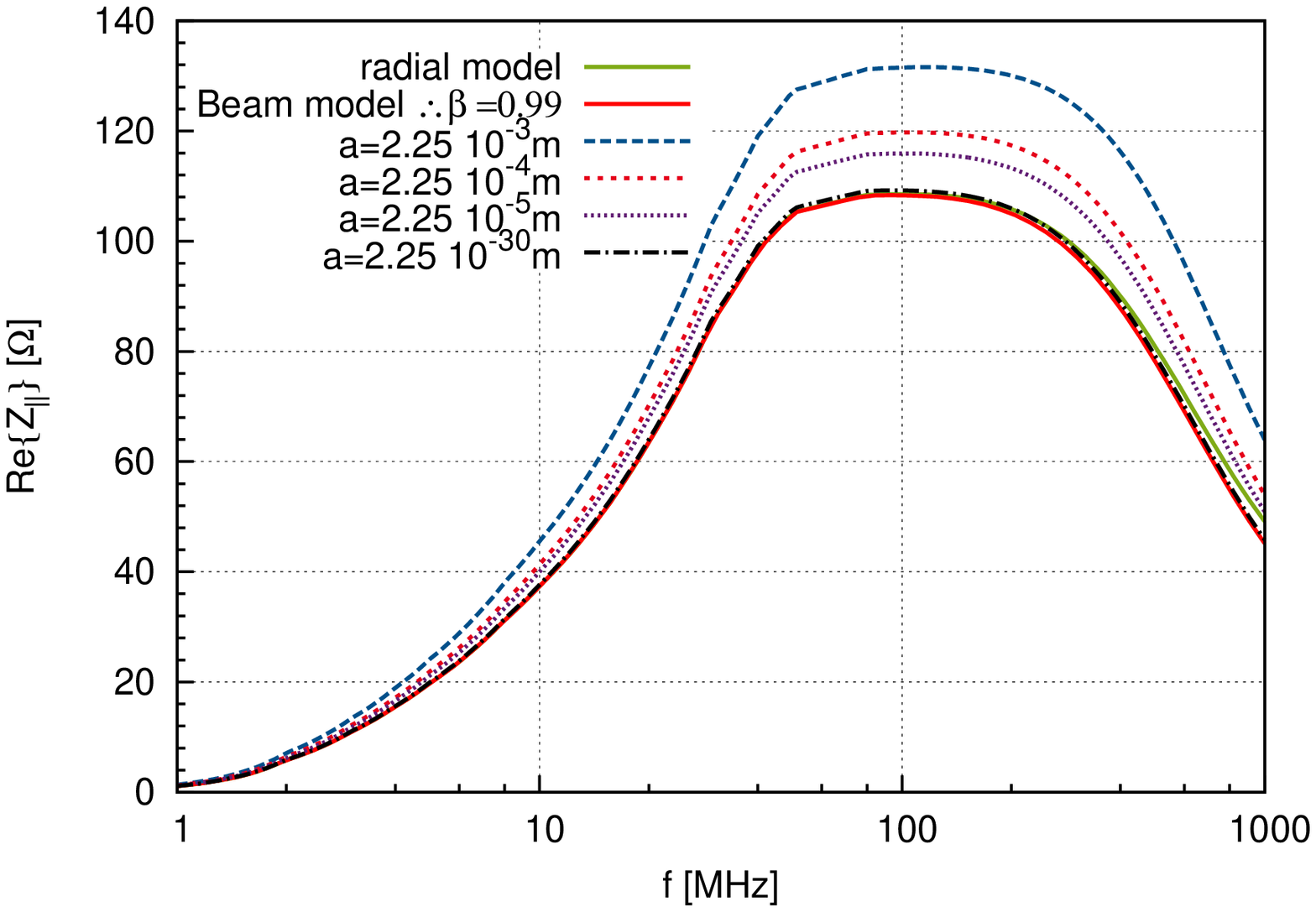}
		\includegraphics[angle=-90, width=.47\textwidth]{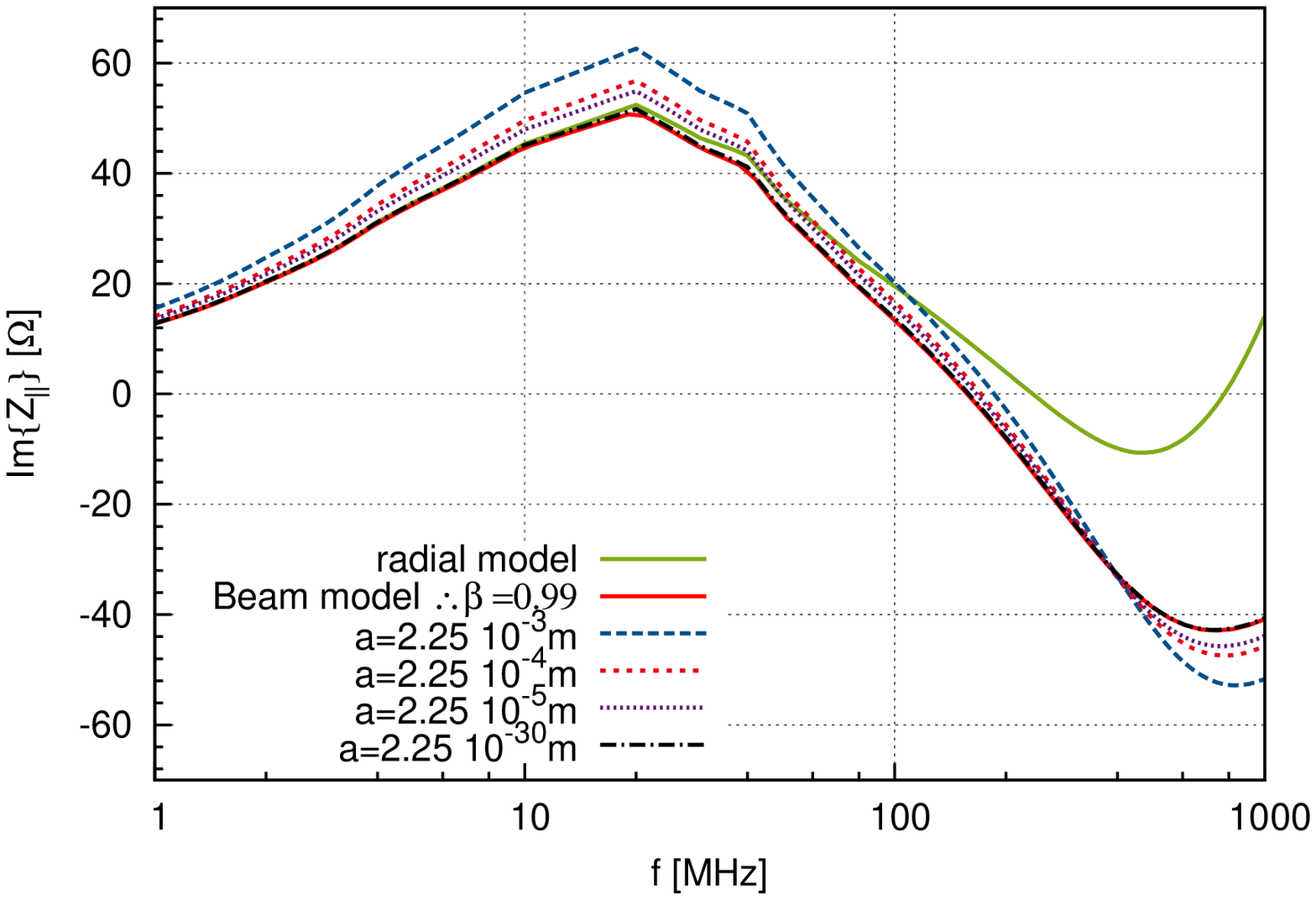}
	\caption{Analytic coaxial wire method with (exact) improved-log-formula vs. radial model}
	\label{analytic_wiremethod_improvedlogformula}
\end{figure}
\begin{figure}[htb]
	\centering
		\includegraphics[angle=-90, width=.47\textwidth]{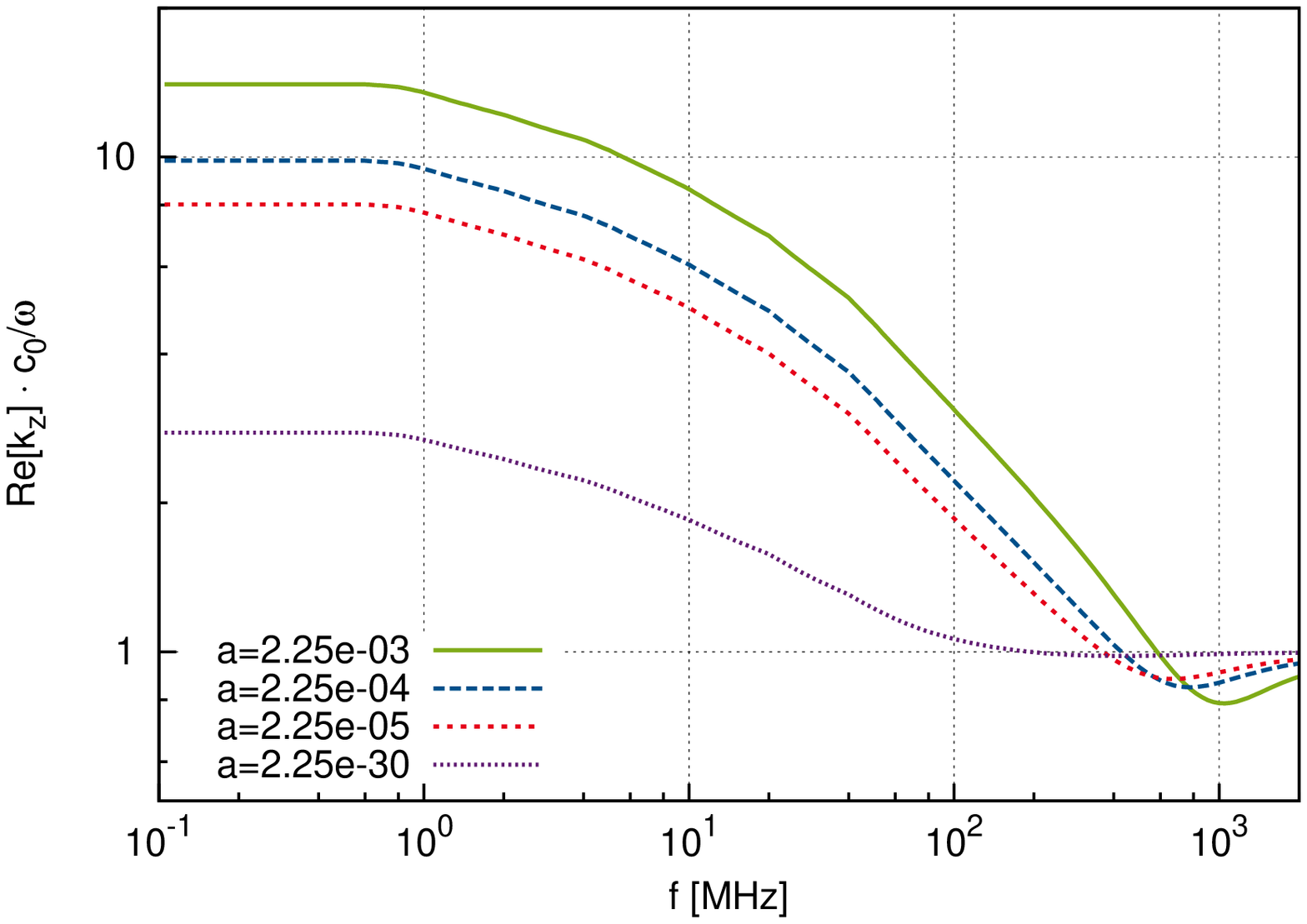}
		\includegraphics[angle=-90, width=.47\textwidth]{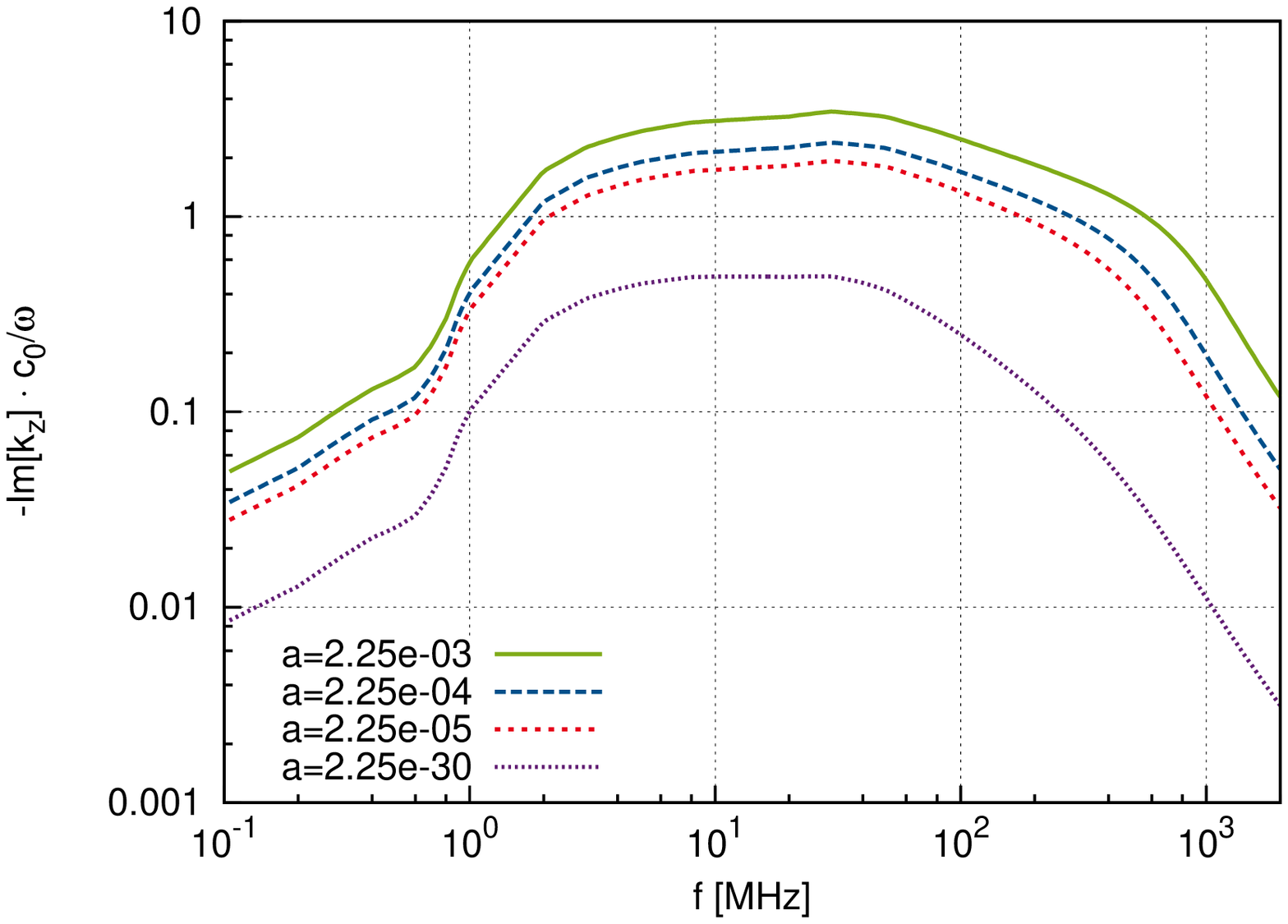}
		\includegraphics[angle=-90, width=.47\textwidth]{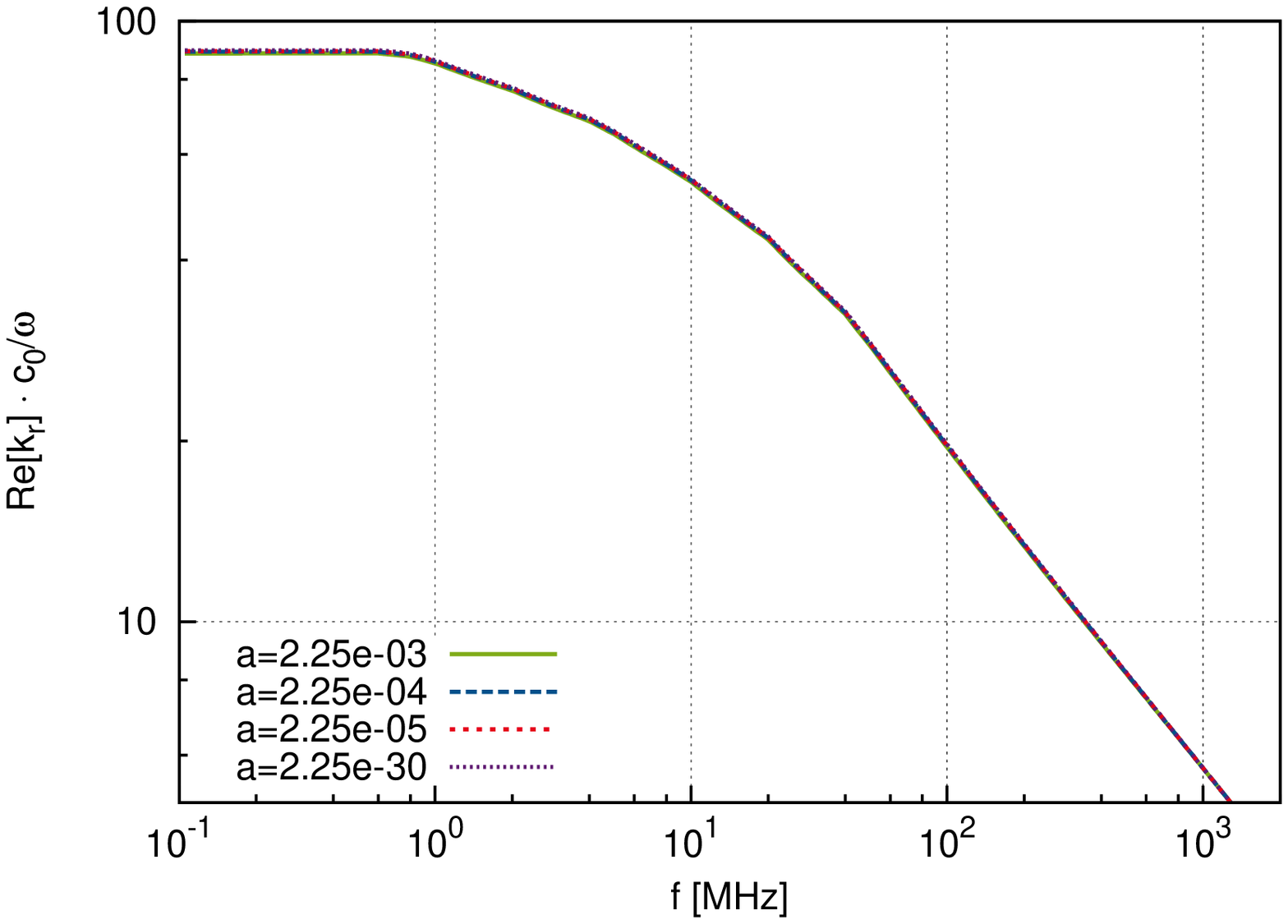}
		\includegraphics[angle=-90, width=.47\textwidth]{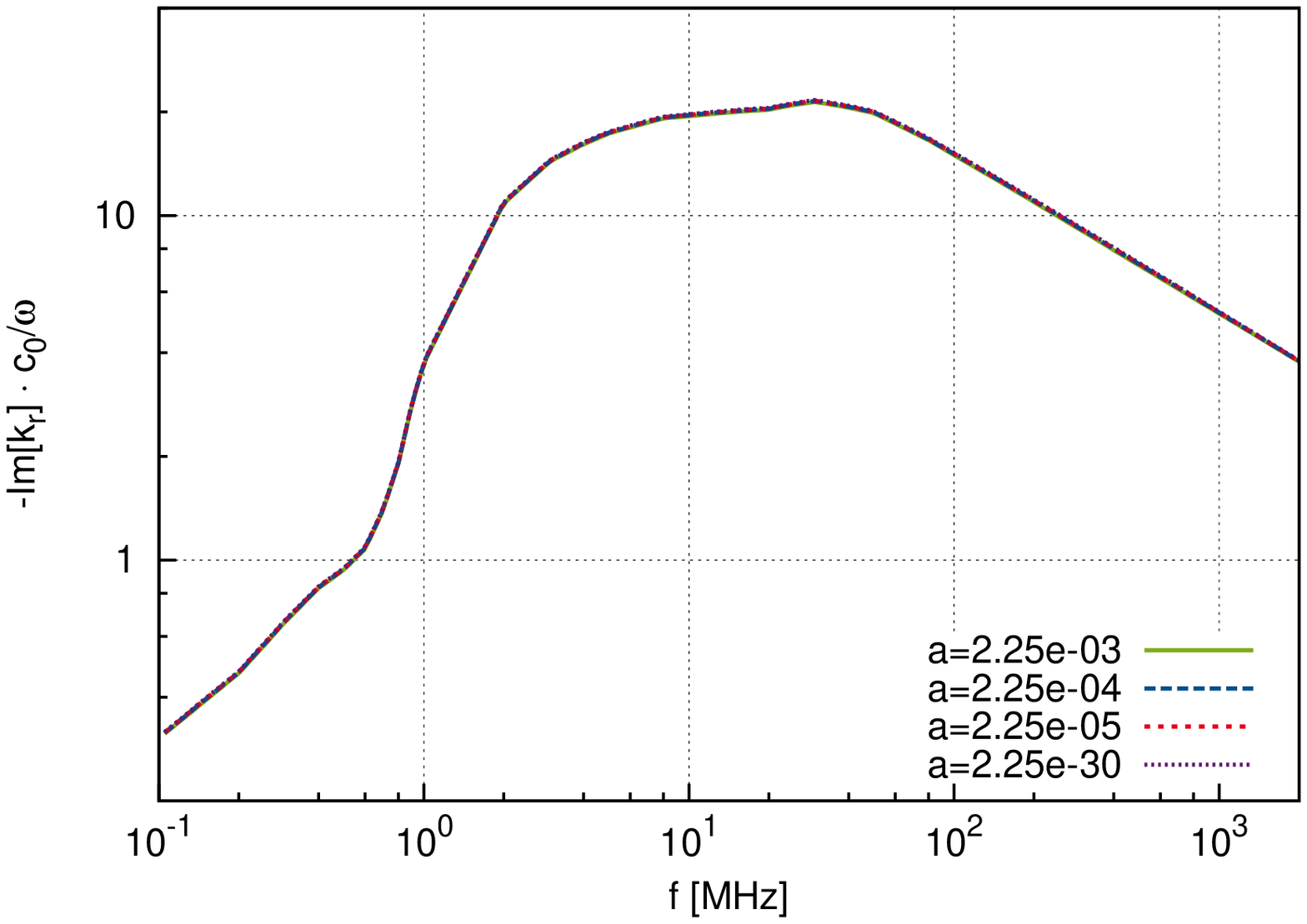}
	\caption{Wavenumbers in Ferrite for the coaxial model}
	\label{fig-dispersion}
\end{figure}
The impedance of the ferrite ring in Fig. \ref{AnalyticSetup} is determined from the Eigenvalue $k_z$ and the formulas  \ref{improved-log-formula}, \ref{log-formula}, and \ref{HPformula}. Figure \ref{analytic_wiremethod_formulas} shows a comparison of the Eigenvalue impedances and the impedances from the beam and current (radial model) excitation. One can see that the beam and the radial model fit well for the real part, but at high frequency the imaginary part deviates due to longitudinal phase shift. The improved-log impedance deviates only slightly from the highly relativistic beam impedance whereas the lumped- and log-formula deviate strongly. As visible in Fig. \ref{analytic_wiremethod_improvedlogformula} the deviation for the improved-log-formula can be accounted to the finite wire thickness. 
When the wire becomes very thin (practically not possible), the Eigenvalue $k_z$ approaches the the plane-wave wavenumber $\omega/c$ (see Fig. \ref{fig-dispersion}) and therefore the transmission $S_{21}=\exp(-ik_zl)$ is equal to the one for the ultrarelativistic beam. 

This means that the improved-log-formula has to give the same impedance as calculated in the beam-excited model by Eq.\ref{exc_imp}.
Further one can see in Fig. \ref{fig-dispersion} that the radial wavenumber in the ferrite depends only very little on the wire radius $a$. The losses enter the $S_{21}$-parameter and the impedance via the imaginary part of $k_z$, which depends on the wire radius. Nonetheless this error enters the distributed impedance only logarithmically.
The convergence of the measured impedance for $a\rightarrow 0 $ is also discussed in \cite{Argan1999}.

\section{Numerical Modelling}
\label{sect_num}
Beam coupling impedances can be obtained from time domain simulations and FT of the wake potential. Also the S-parameters obtained in bench measurements can be numerically simulated in both FD and TD. The advantage of time domain simulations is that one directly obtains broadband results. Frequency Domain methods use the (interpolated) material data as given directly in FD, whereas in TD an impulse response, i.e. a rational transfer function, approximated to a certain order, is required. Details can be seen e.g. in \cite{Gutschling2000}.
\begin{figure}[ht]
	\centering
		\includegraphics[angle=0, width=.47\textwidth]{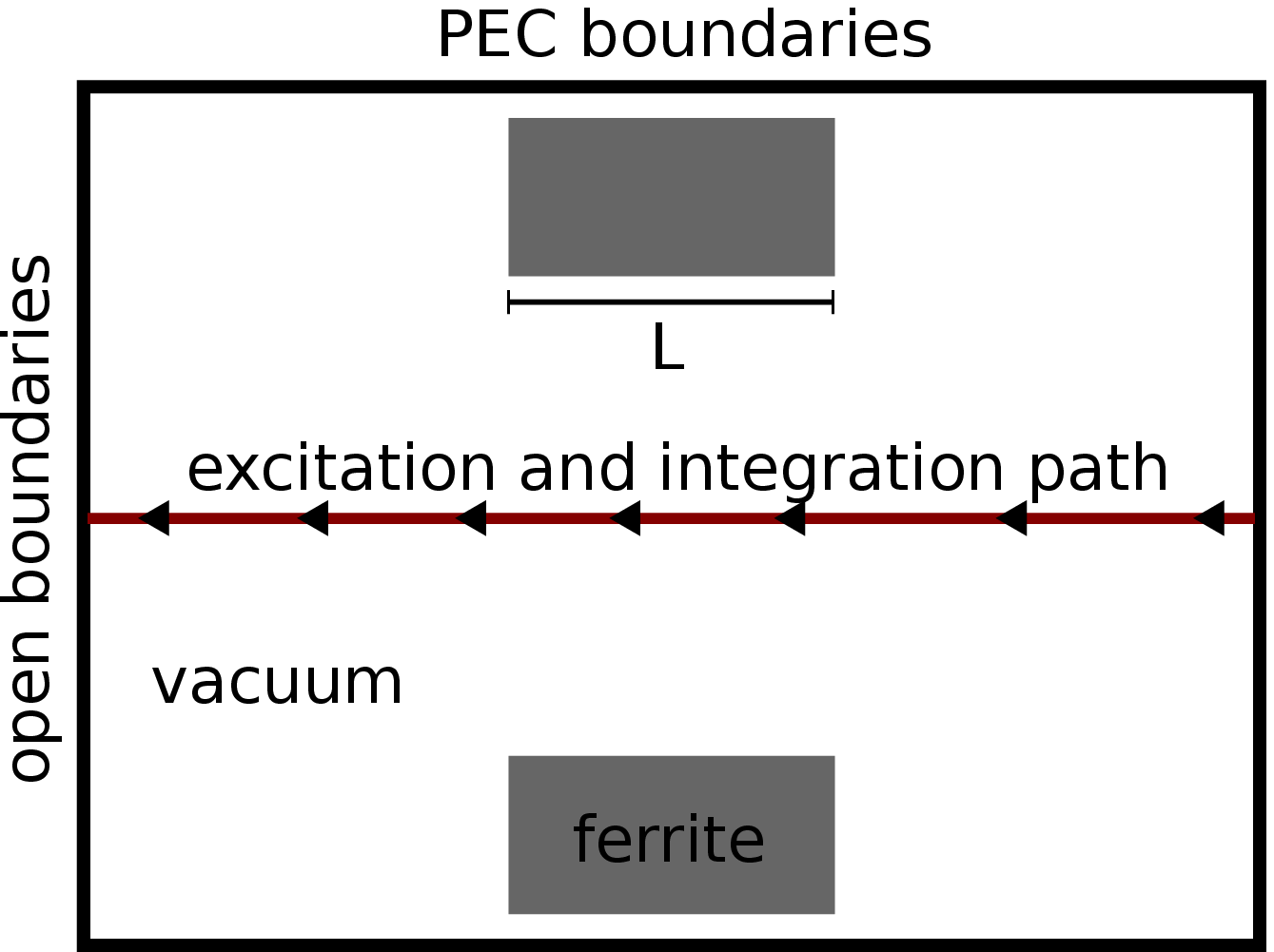}
		\includegraphics[angle=0, width=.40\textwidth]{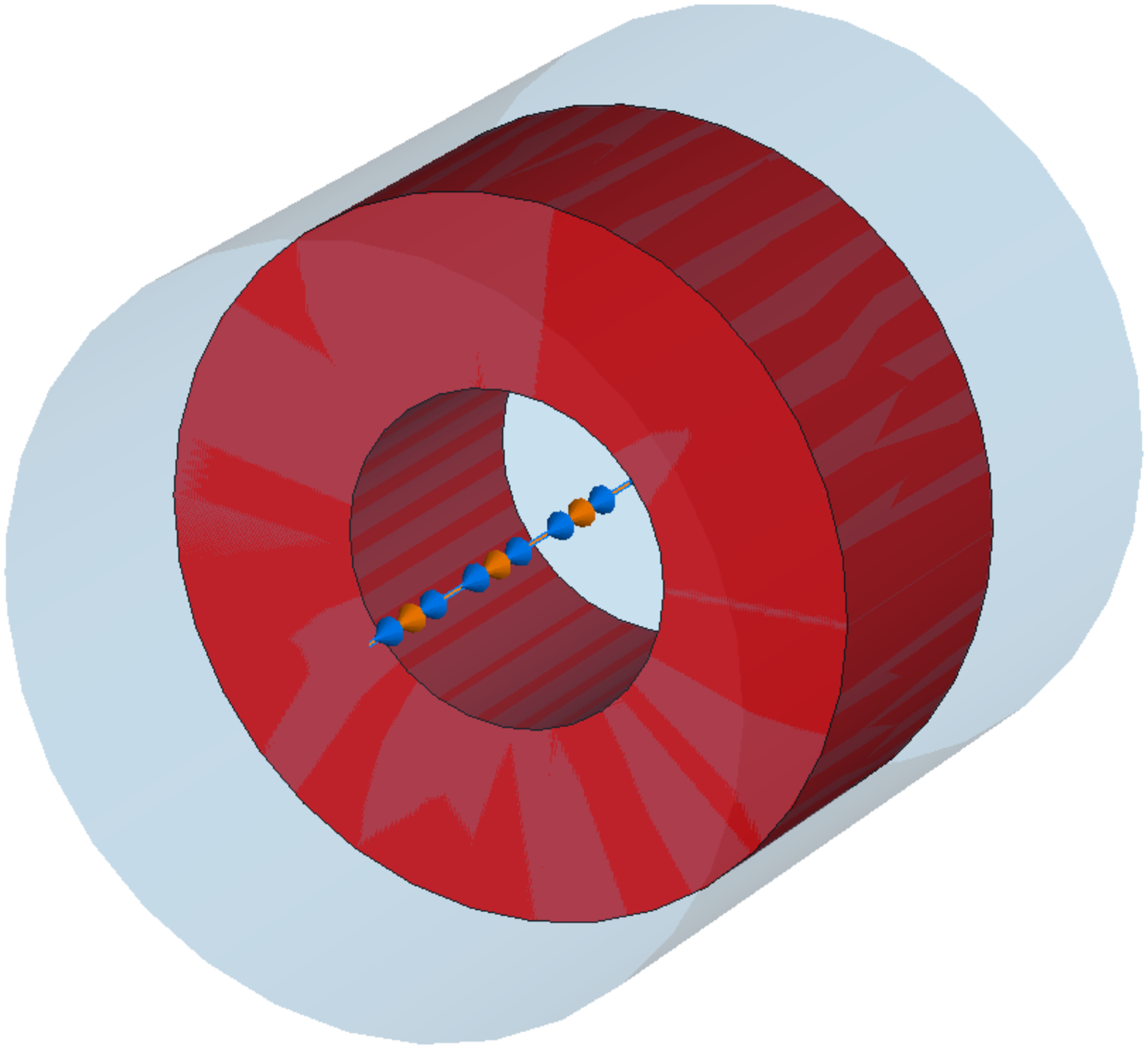}
	\caption{Longitudinal cut of the ferrite ring model and CST model}
	\label{num_sketch}
\end{figure}
The following will show both wake simulations using CST Particle Studio (PS) \cite{CST} and S-parameter simulations in TD and FD using CST Microwave Studio (MWS) \cite{CST}. 
Figure \ref{num_sketch} shows the setup, where open boundaries or waveguide-ports are used for beam/waveguide entry and exit planes.

\subsection{Impedance from wake field calculation}
The beam in the wakefield-simulation is taken as infinitely (practically one mesh cell) thin and with a Gaussian longitudinal profile with $\sigma=$10.5 cm.
The integrated wakelength is 50m. The mesh has 180,000 cells leading to a computation time of less than 1 hour.

The practical limitations of the wakefield solver arise from the required long wakelength for low frequencies and the small time step required for stability.
The wakefield solver operates (explicitly) in TD and is therefore subject to the Courant-critereon,
\beq
\delta t \leq \min_{i,j,k} \left(c\sqrt{\frac{1}{\delta x_i^2}+\frac{1}{\delta y_j^2}+\frac{1}{\delta z_k^2}}\right)^{-1}
\eeq
i.e. the spacial mesh determines the maximum stable timestep.
For low frequencies, the accuracy is also subject to the (K\"upfm\"uller- \cite{KM}) uncertainty principle,
\beq
\Delta f \geq\frac{1}{\Delta t}=\frac{c}{\Delta l}
\eeq
where $\Delta f$ is frequency-uncertainty of a given quantity (e.g. the impedance) and $\Delta l$ is the wakelength.
Via the discrete Fourier transform, $\Delta f$ is proportional to the frequency resolution of the impedance.
For low frequencies this way of computing impedances becomes inapplicable since $\Delta l$ is proportional to the total computation time.
A small relief to this limitation is obtained for low-$Q$ structures by zero-padding before applying the FFT. 
A frequency domain solver \cite{Doliwa2007a} \cite{Niedermayer2012b}, or an implicit time domain solver, would not be limited by this.

\begin{figure}[htb]
	\centering
		\includegraphics[angle=-90, width=.47\textwidth]{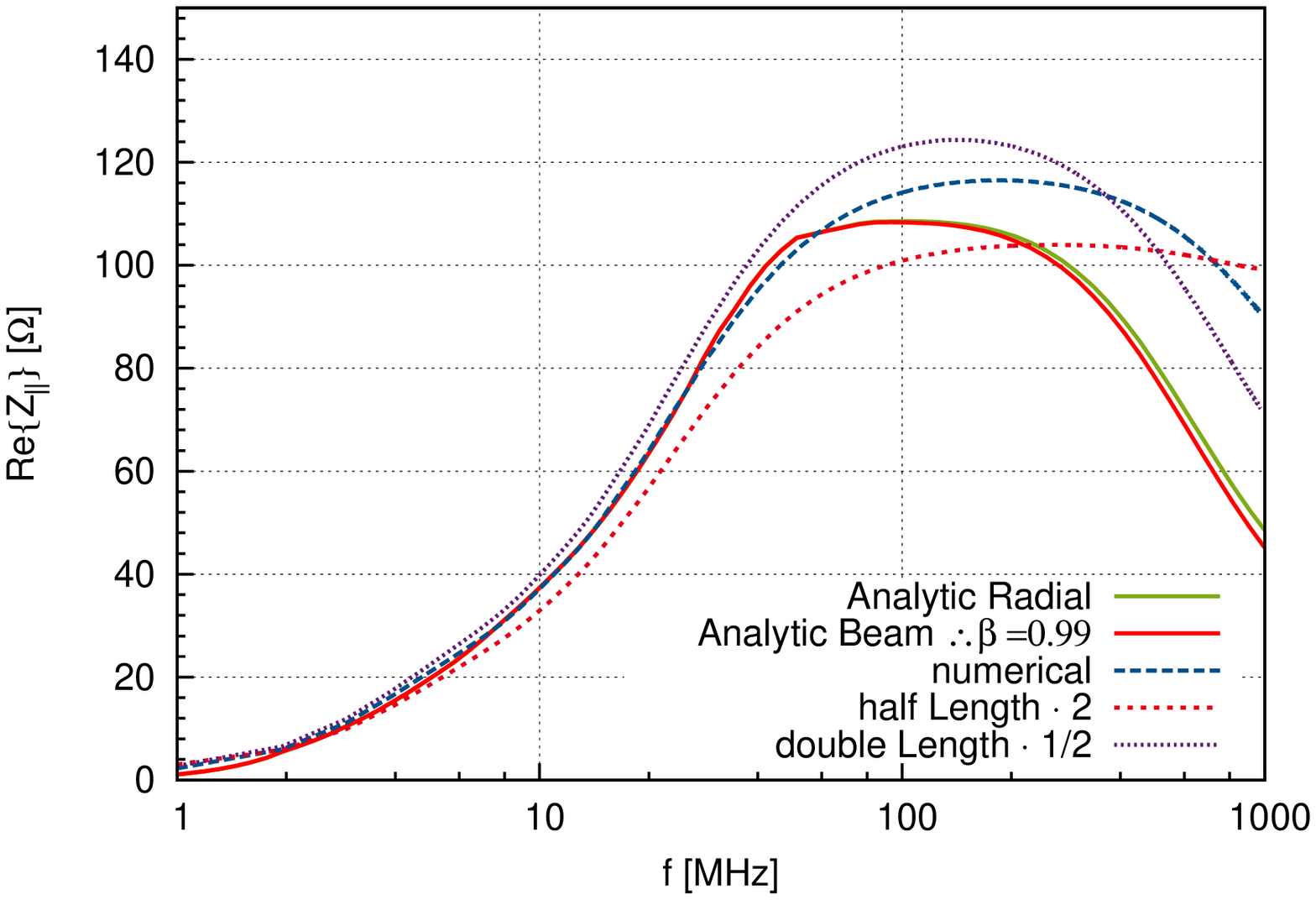}
		\includegraphics[angle=-90, width=.47\textwidth]{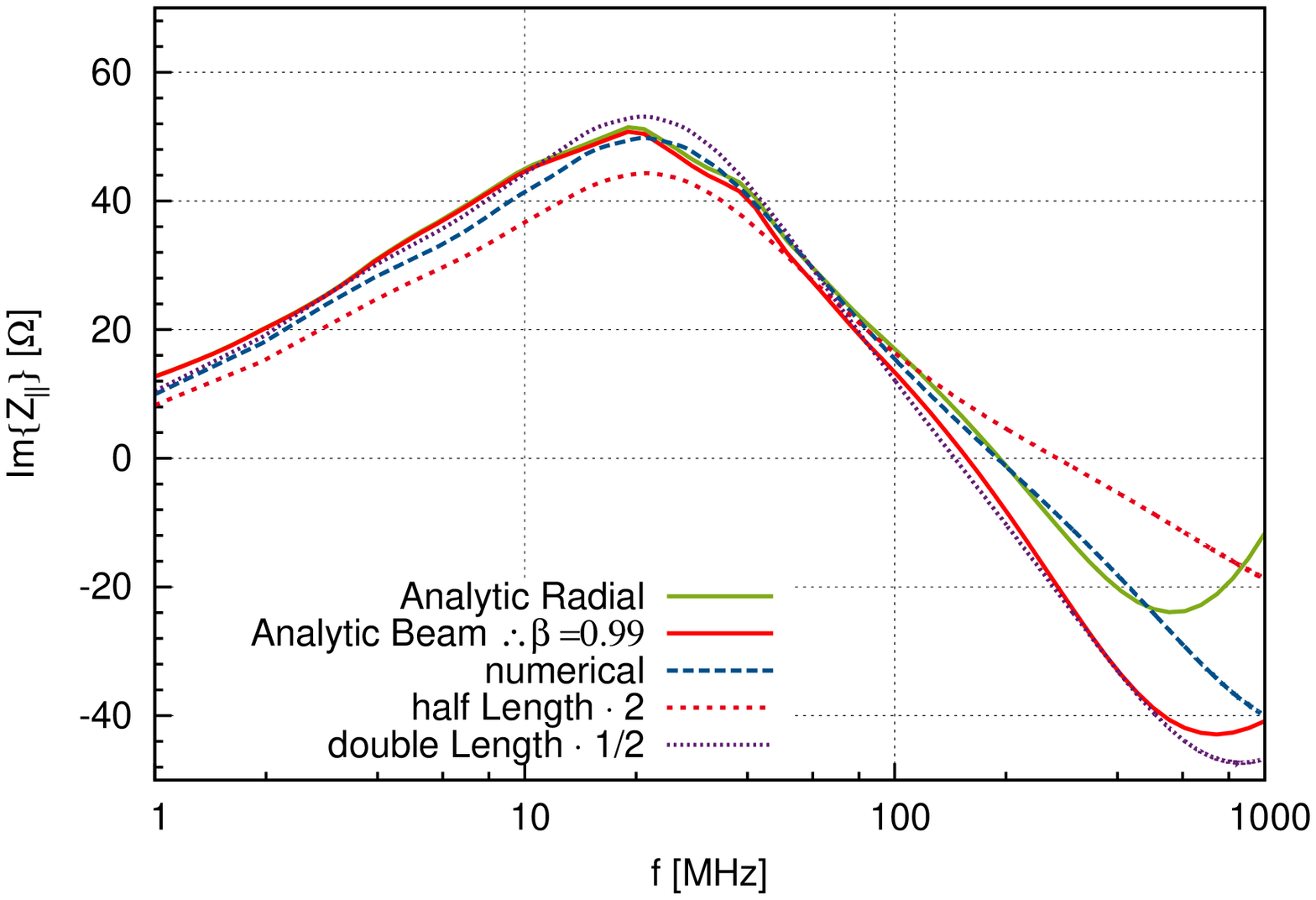}
	\caption{CST PS simulation vs. analytical 2D beam and radial model }
	\label{compare_ps_analytic_re}
\end{figure}
Figure \ref{compare_ps_analytic_re} shows the simulation results. Note that slight discrepancies arise from the fitting of the material data on some rational transfer function ansatz. 
The simulation has been rerun for different lengths to check the scaling. As visible Fig. \ref{compare_ps_analytic_re}, the simulation curves roughly approach the analytical ones for longer DUTs, i.e. fulfillment of the 2D assumption.

\subsection{Simulation of the Measurement Process}
\begin{figure}[htb]
	\centering
		\includegraphics[angle=-90, width=.47\textwidth]{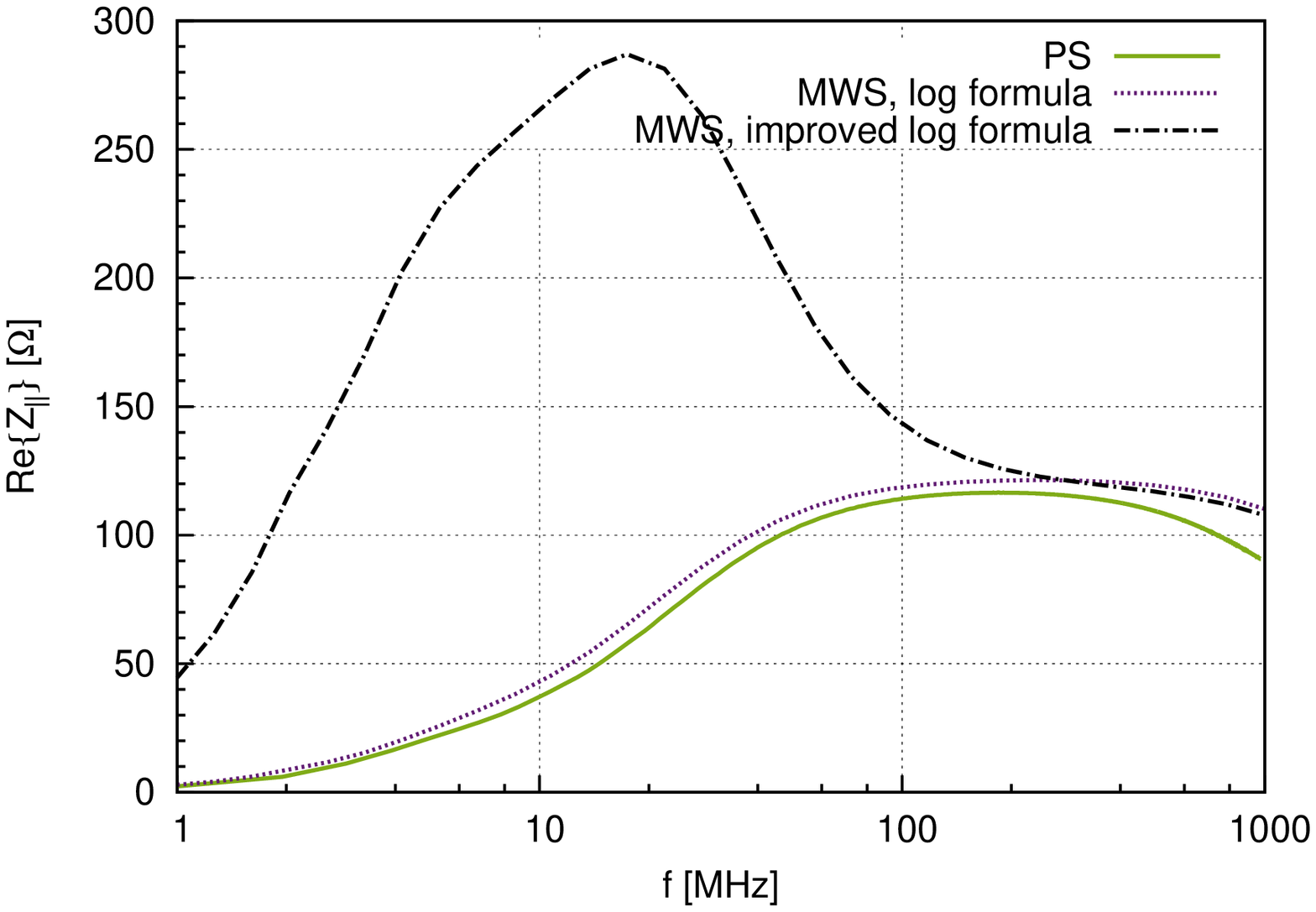}
		\includegraphics[angle=-90, width=.47\textwidth]{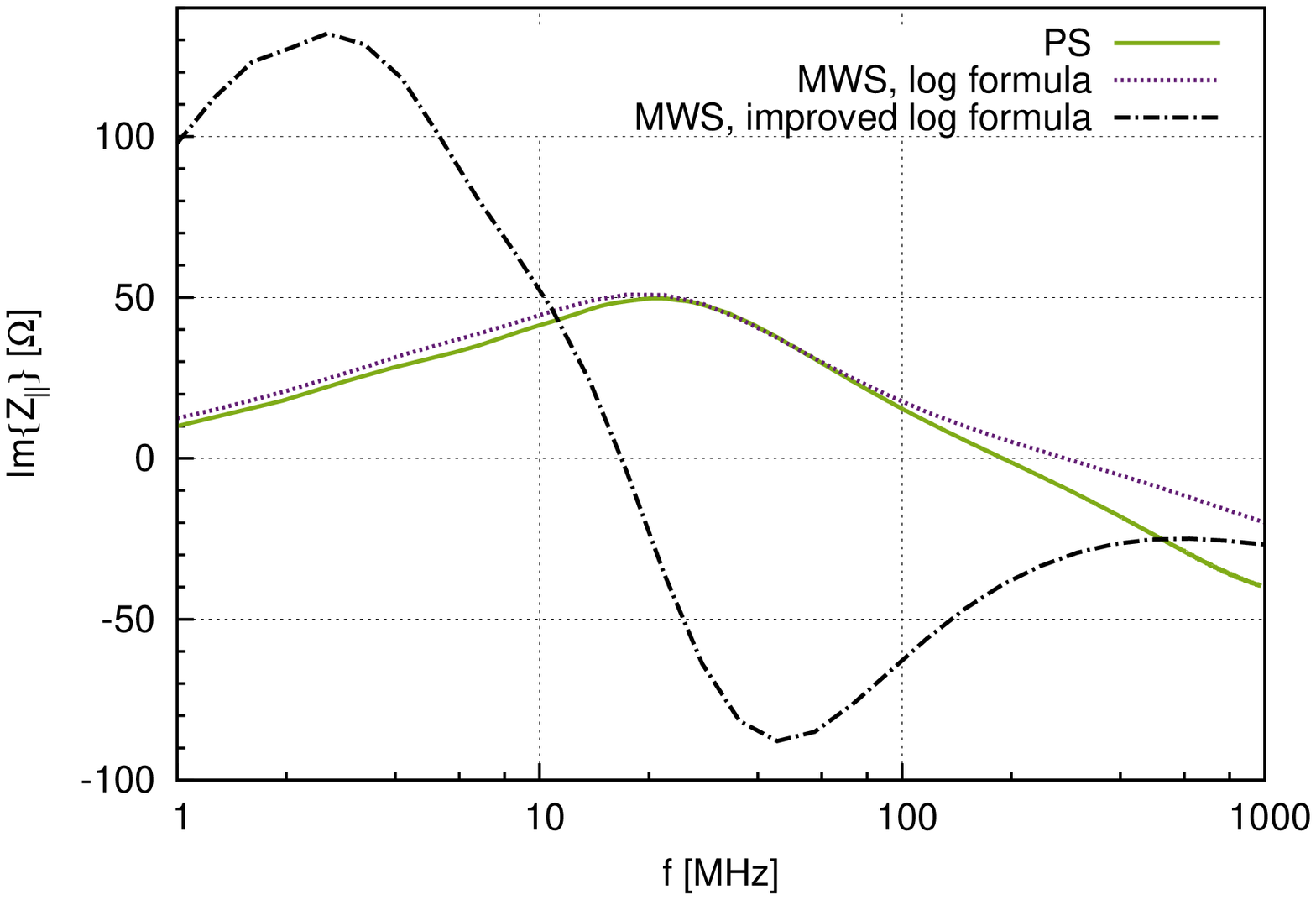}
		\caption{MWS S-parameter simulation with different conversion formulas vs. PS-solution}
	\label{mws_different_formulas}
\end{figure}

\begin{figure}[htb]
	\centering
		\includegraphics[angle=-90, width=.47\textwidth]{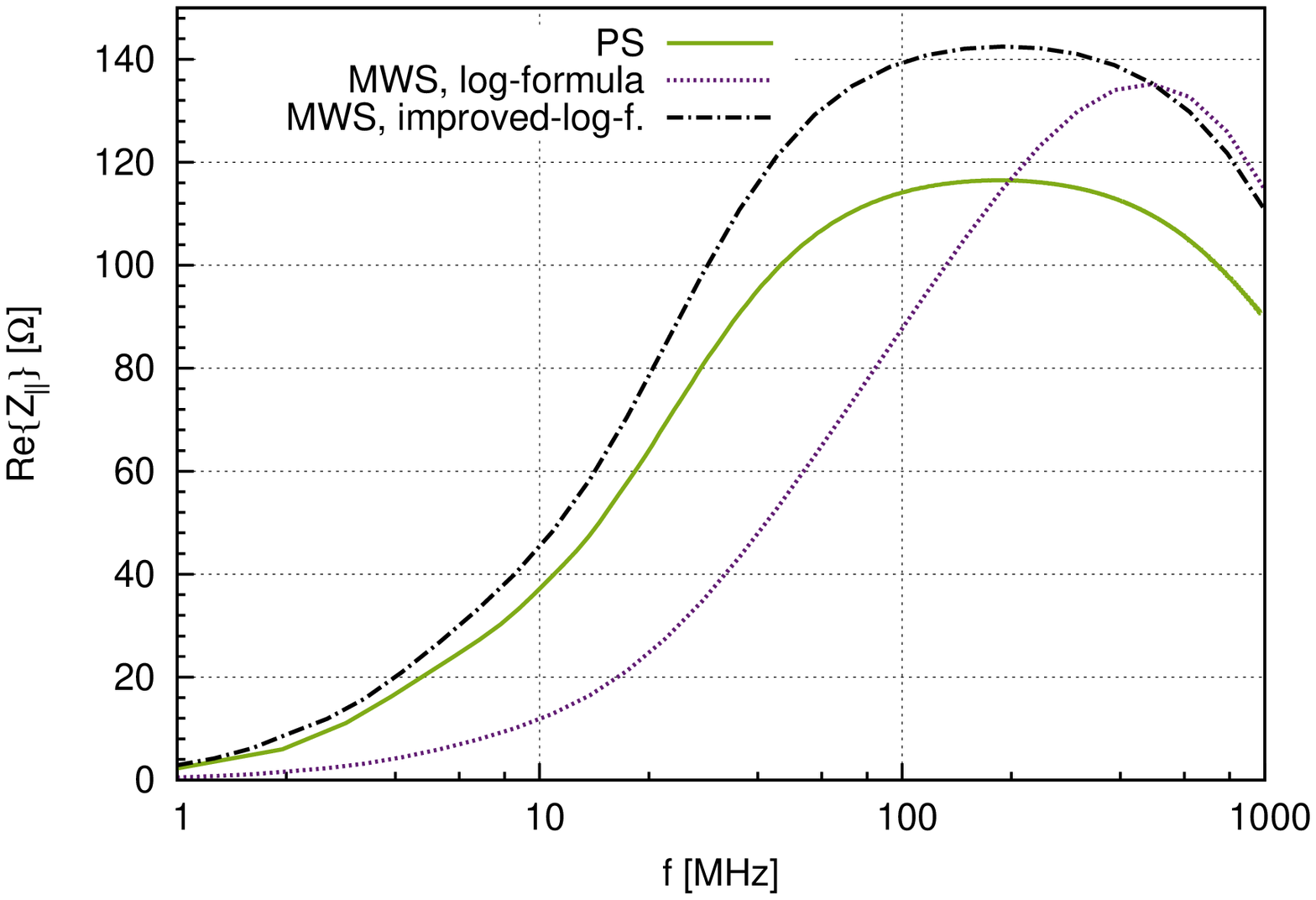}
		\includegraphics[angle=-90, width=.47\textwidth]{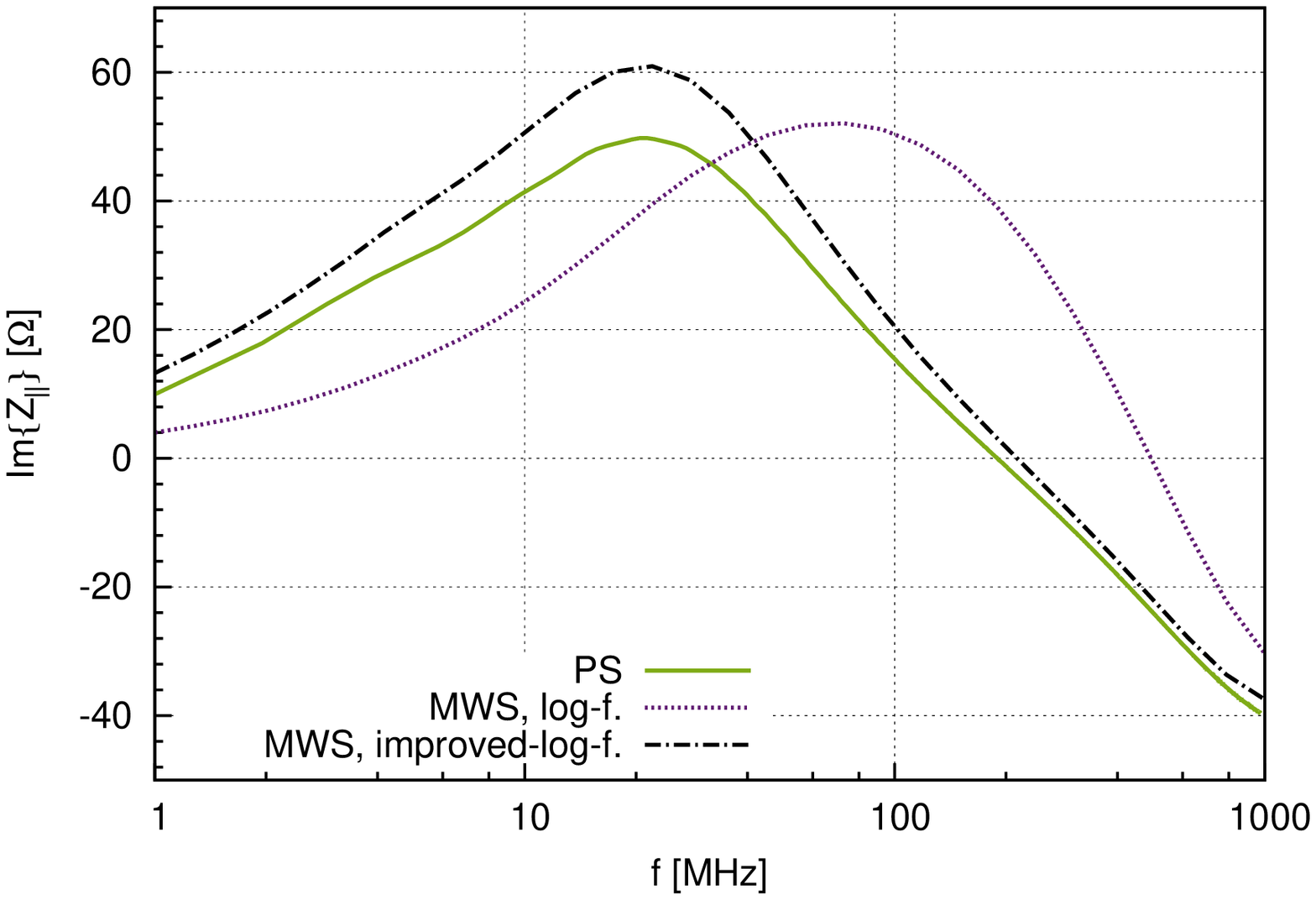}
		\caption{MWS S-parameter simulation with reflection correction vs. PS-solution}
	\label{mws_different_formulas_corrected}
\end{figure}

The measurement process has been simulated using CST MWS. In order to obtain higher accuracy by avoiding the material data fitting error, the FD solver has been employed. Ports with 20 waveguide modes serve as boundary condition. 
The longitudinal impedance calculated from the $S_{21}$-parameter is shown in Fig. \ref{mws_different_formulas}. The curve for the improved-log formula shows a strong resonance, which is accounted to the reflection at the edge of the DUT. This can be corrected using the Wang-Zhang-formula \ref{WangZhang},	 
providing new transmission parameters to insert into Eq. \ref{log-formula}, \ref{improved-log-formula} or \ref{HPformula}. The corrected results are visible in Fig. \ref{mws_different_formulas_corrected}.
The match between the log-formula and the PS-curve is purely by chance. After reflection correction the improved-log-formula matches the PS simulation within a deviation of about 20\%. This can be accounted to the finite wire radius (see also Fig. \ref{analytic_wiremethod_improvedlogformula}). Note that many mesh cells are required to resolve thin wires in S-parameter simulations.

\section{Measurement data evaluation}
\label{sect_meas}
In order to conclude on setup-independent properties of the DUT, the measurement has been performed for two different setups shown in Fig. \ref{setup}. A copper wire of 0.225mm diameter has been chosen because of its small thickness, good conductivity and low susceptibility to deformations. In the large setup the wires have been stretched by tightening the screws of the end-plates about 3mm on the inner side of the box. In the small setup, the fixation was done using orthogonal PCBs, soldered together under tension of the wire. 
\begin{figure}[htb]
	\centering
		\includegraphics[angle=0, width=.47\textwidth]{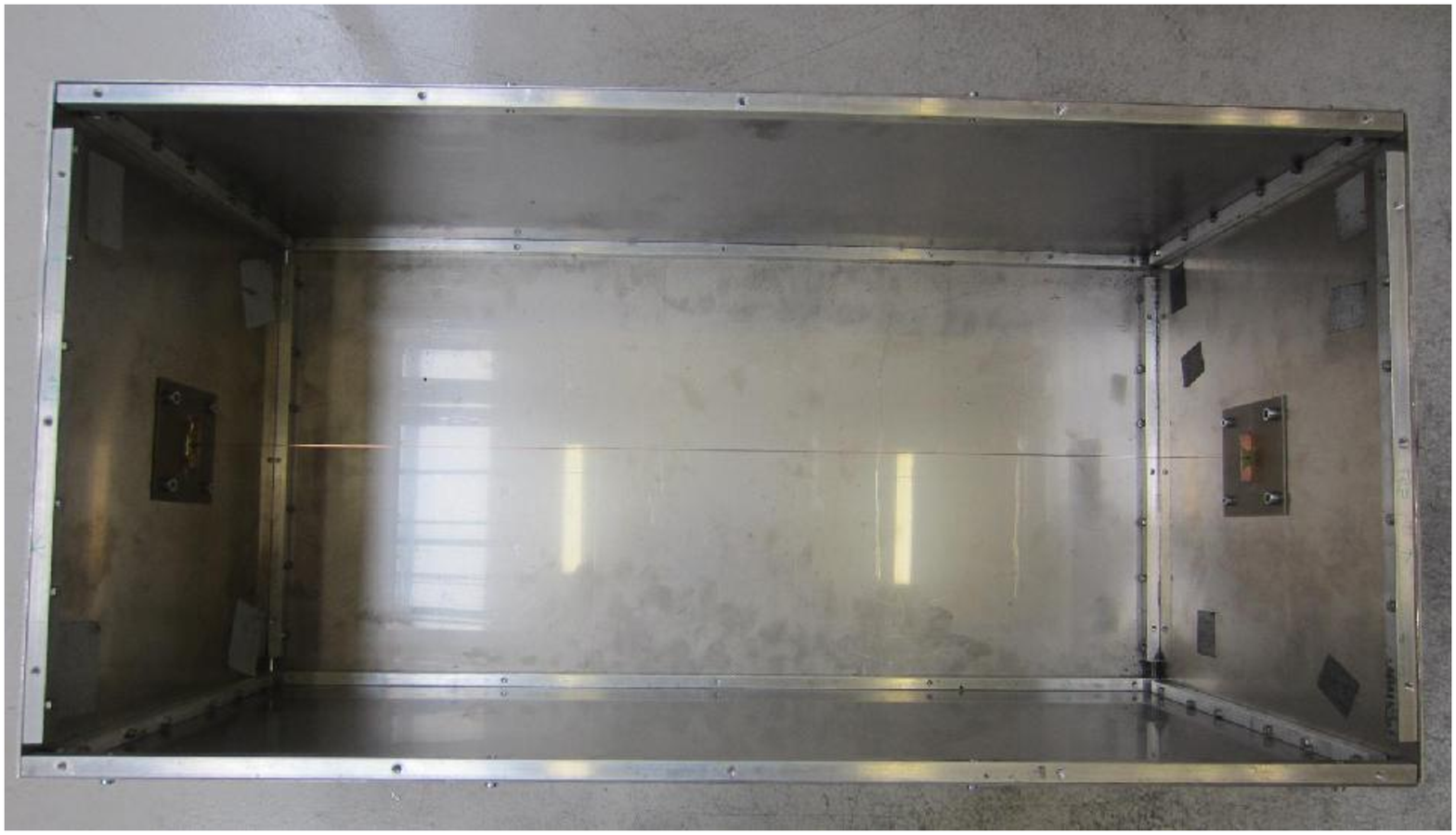}
		\includegraphics[angle=0, width=.47\textwidth]{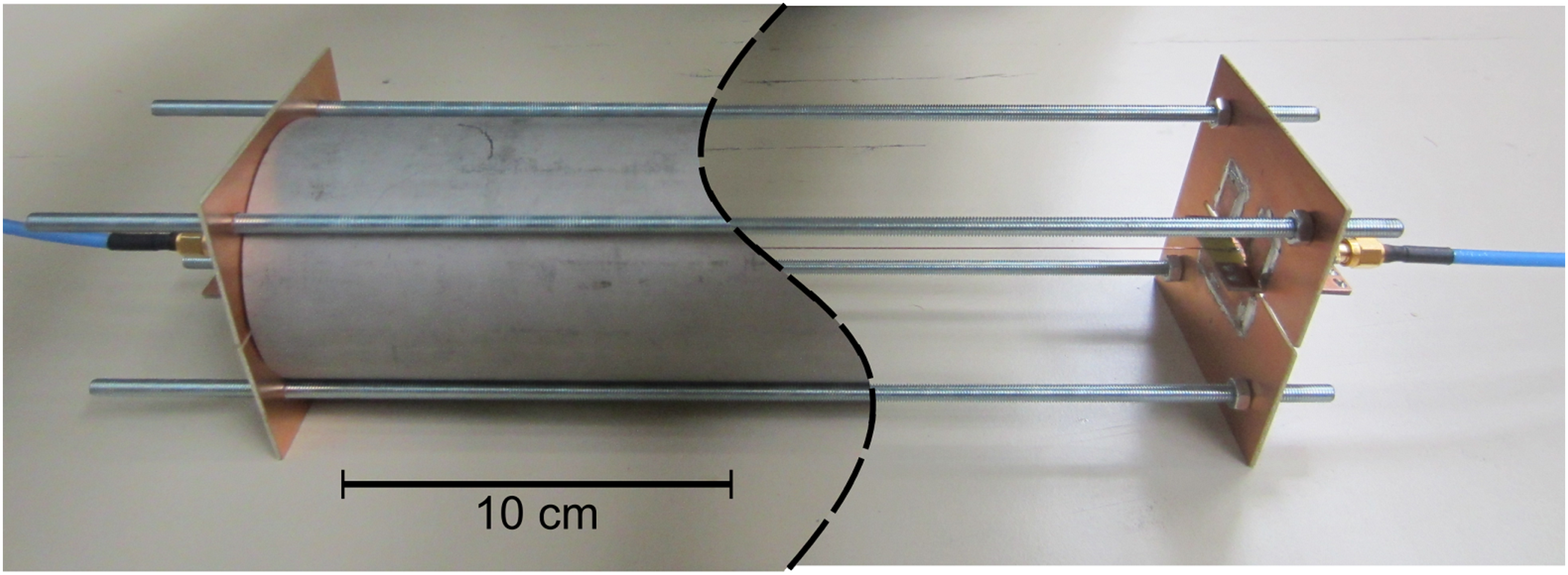}
	\caption{Different measurement boxes}
	\label{setup}
\end{figure}

\begin{figure}[htb]
	\centering
		\includegraphics[angle=-90, width=.47\textwidth]{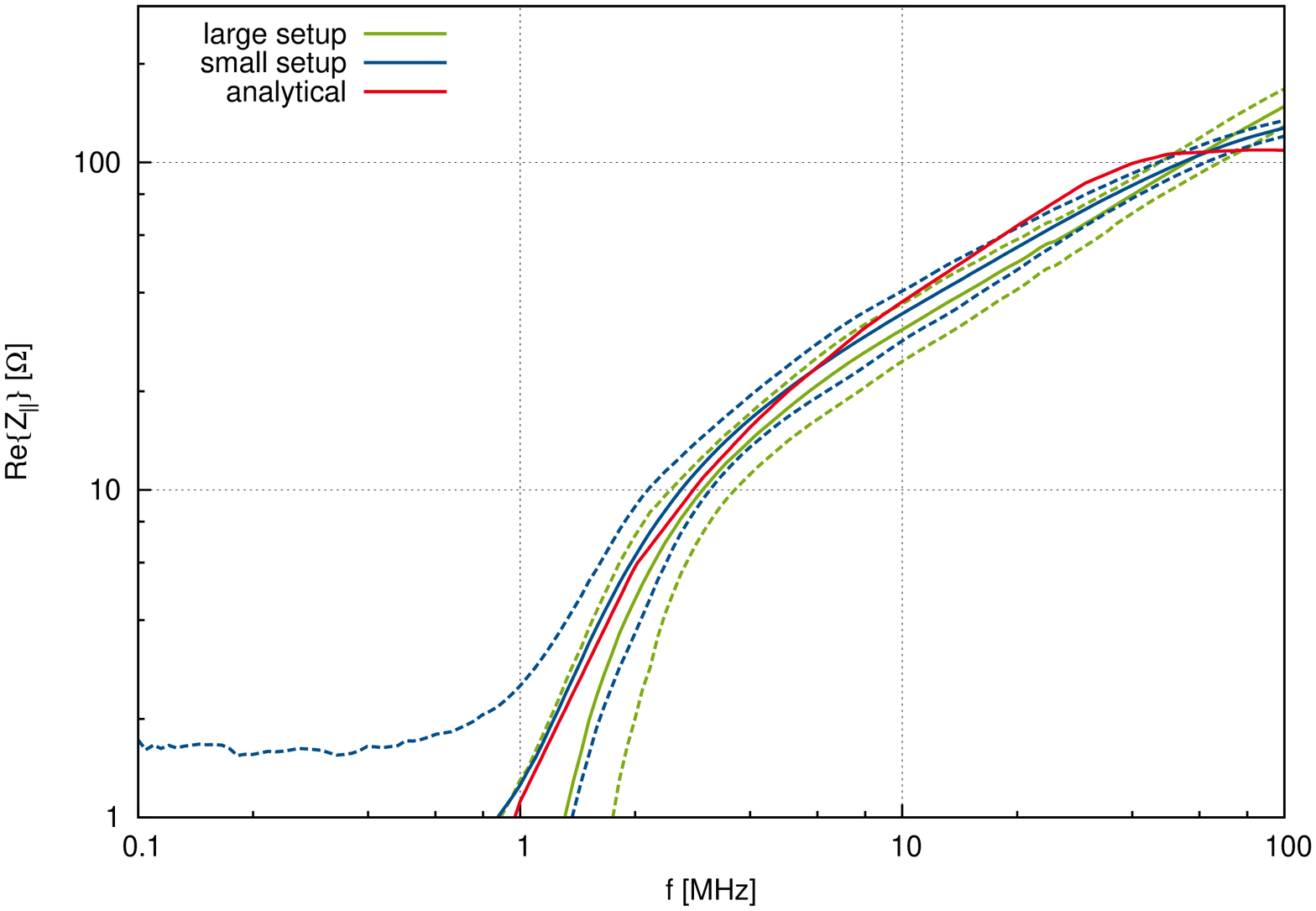}
		\includegraphics[angle=-90, width=.47\textwidth]{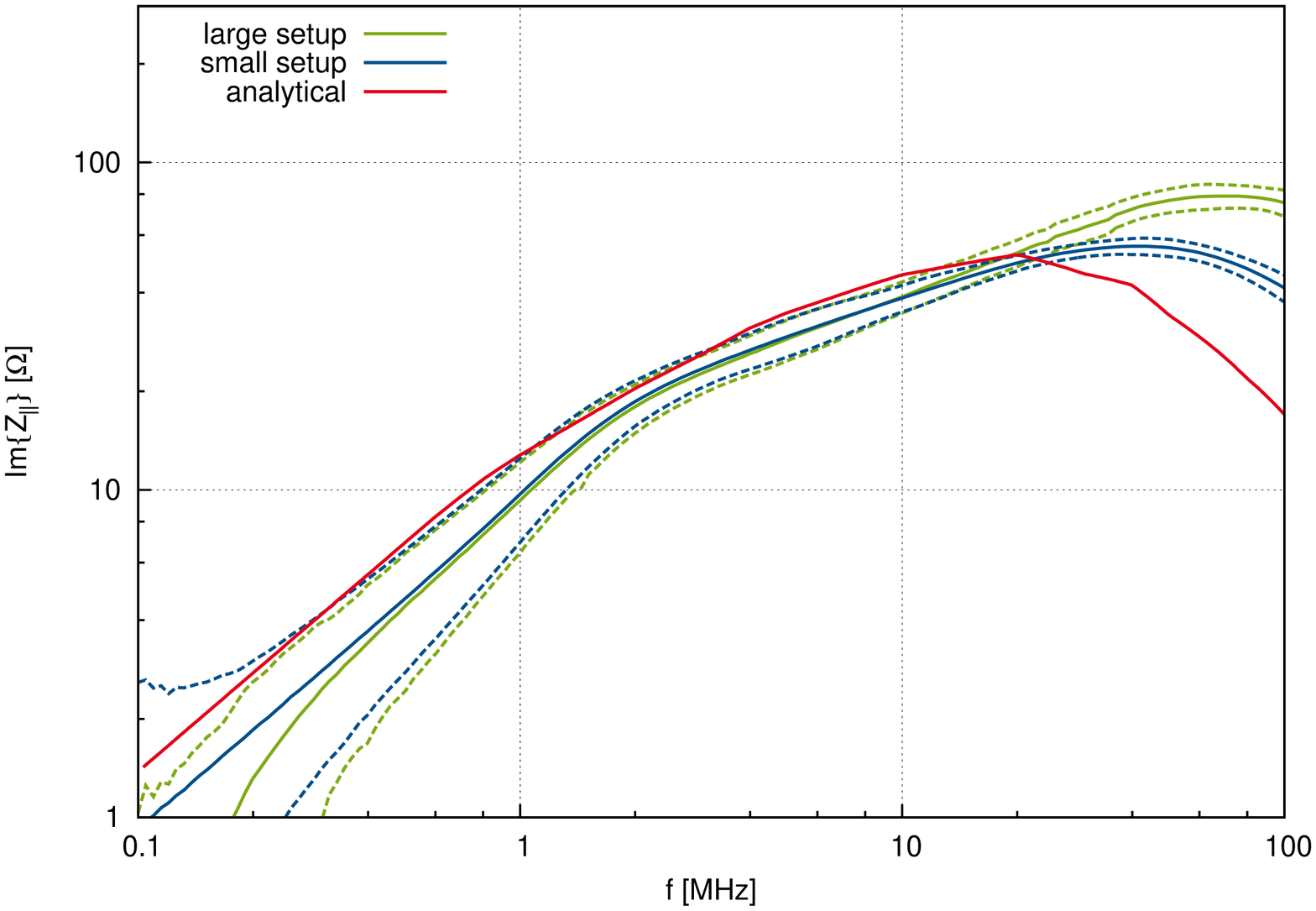}
	\caption{Wire measurements vs. anayltical (radial and beam agree for LF)}
	\label{meas_long_loglog}
\end{figure}

\begin{figure}[htb]
	\centering
		\includegraphics[angle=-90, width=.47\textwidth]{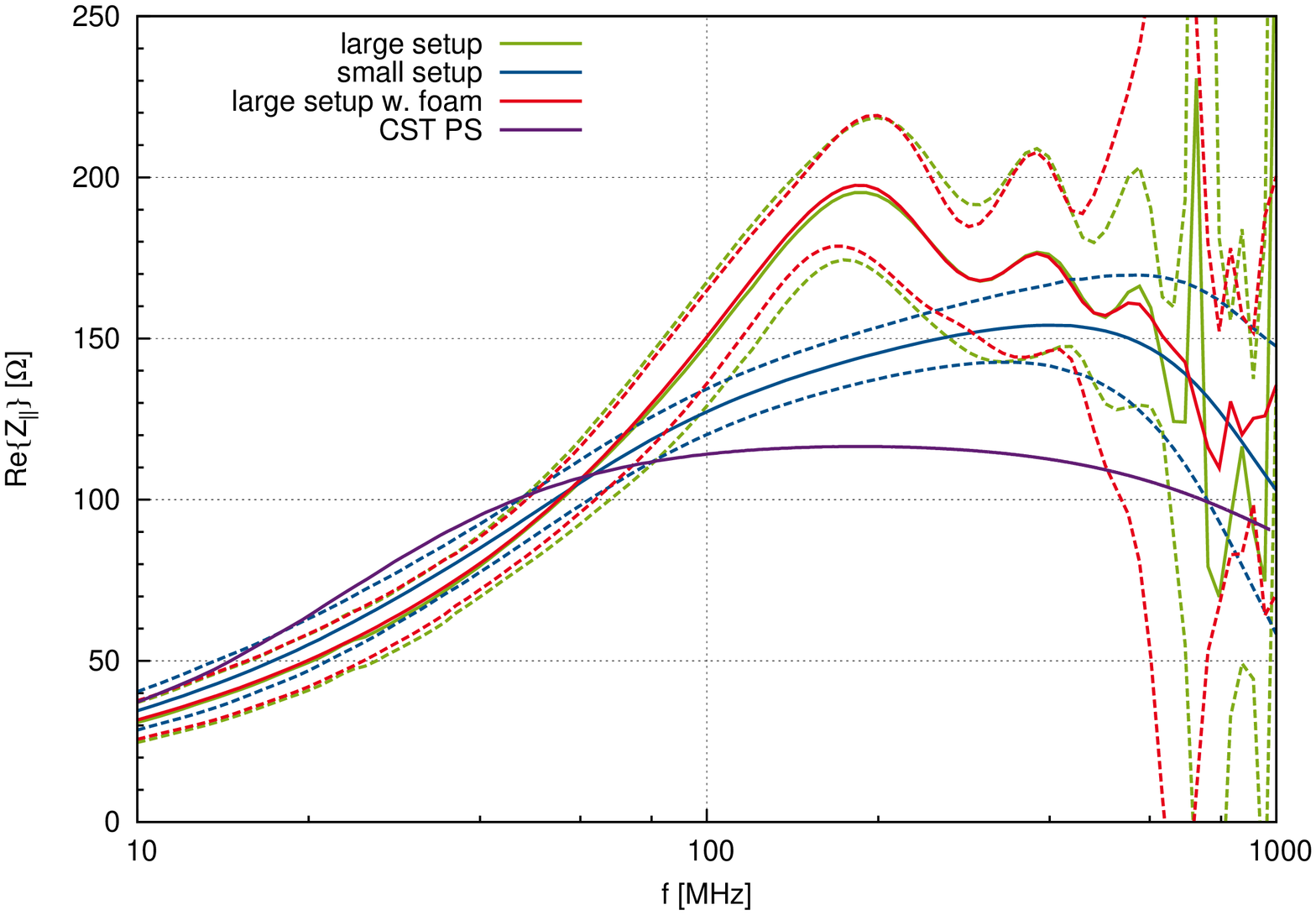}
		\includegraphics[angle=-90, width=.47\textwidth]{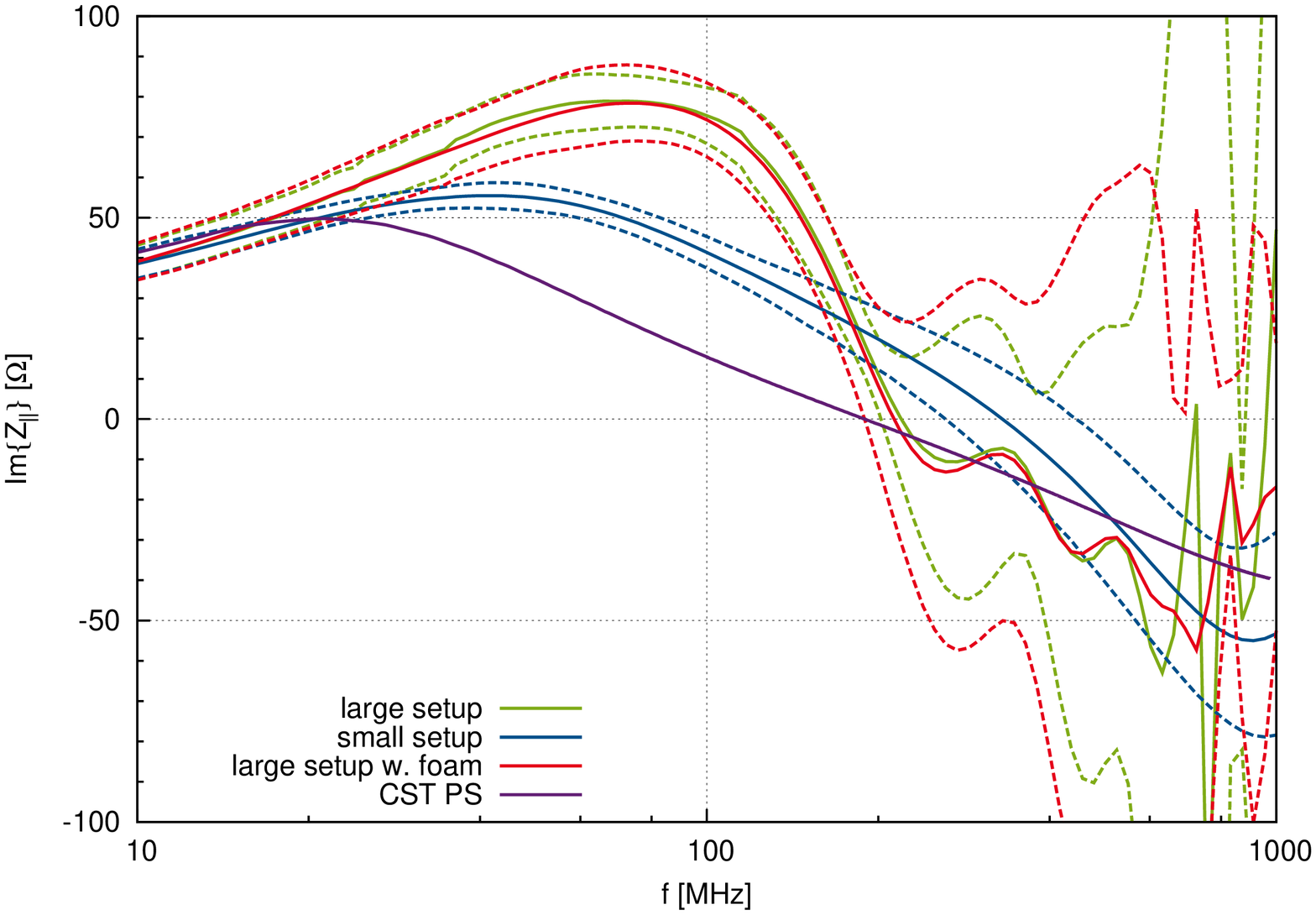}
	\caption{Measurements vs. wakefield simulation. At very high frequencies resonances of the setup could be damped by the foam. The 2D assumption of the analytical models is not valid here.}
	\label{meas_long_lin}
\end{figure}

The two setups are supposed to have such different properties, that agreement of results can be accounted to setup independent properties only.
Both measurements have been performed for the Ferrite ring with changing DUT and REF multiple times in order to obtain sufficiently well statistics. 
Due to the agreement for the simulations as visible in Fig. \ref{mws_different_formulas} the log-formula has been chosen for the evaluation since the improved-log-formula is supposed to show the strong resonance. The Wang-Zhang correction cannot be applied since the $S_{11}$-parameter cannot be measured due to multiple reflections between the matching section and the DUT. 
The results are show in Figs. \ref{meas_long_loglog} and \ref{meas_long_lin}. The dashed lines in the plots denote error bars.
\begin{figure}[htb]
	\centering
		\includegraphics[angle=0, width=.3\textwidth]{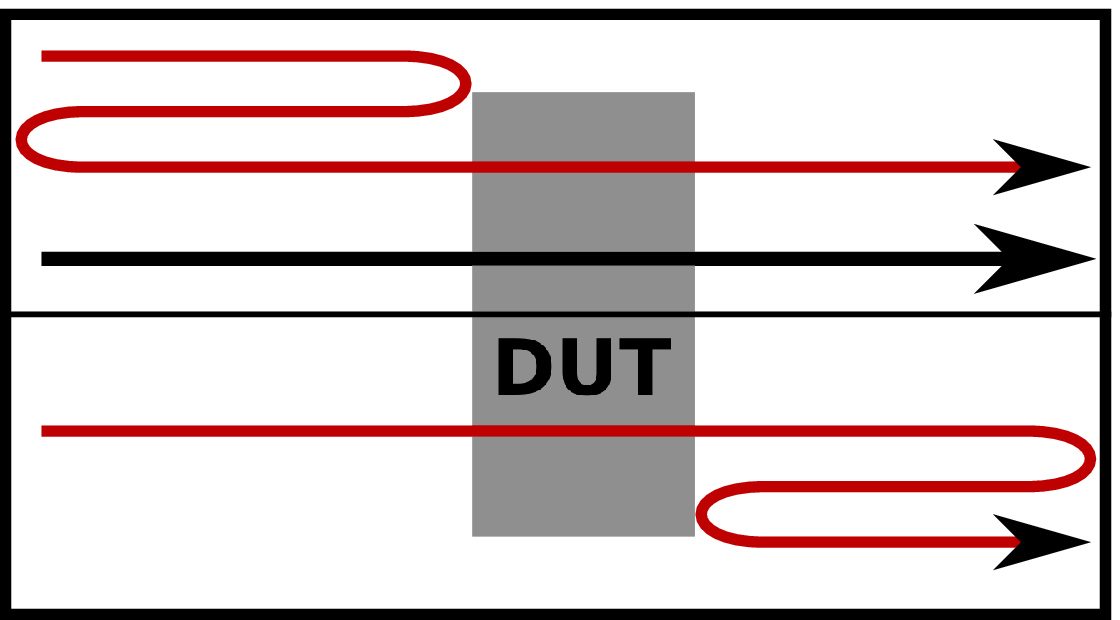}
		\includegraphics[angle=0, width=.3\textwidth]{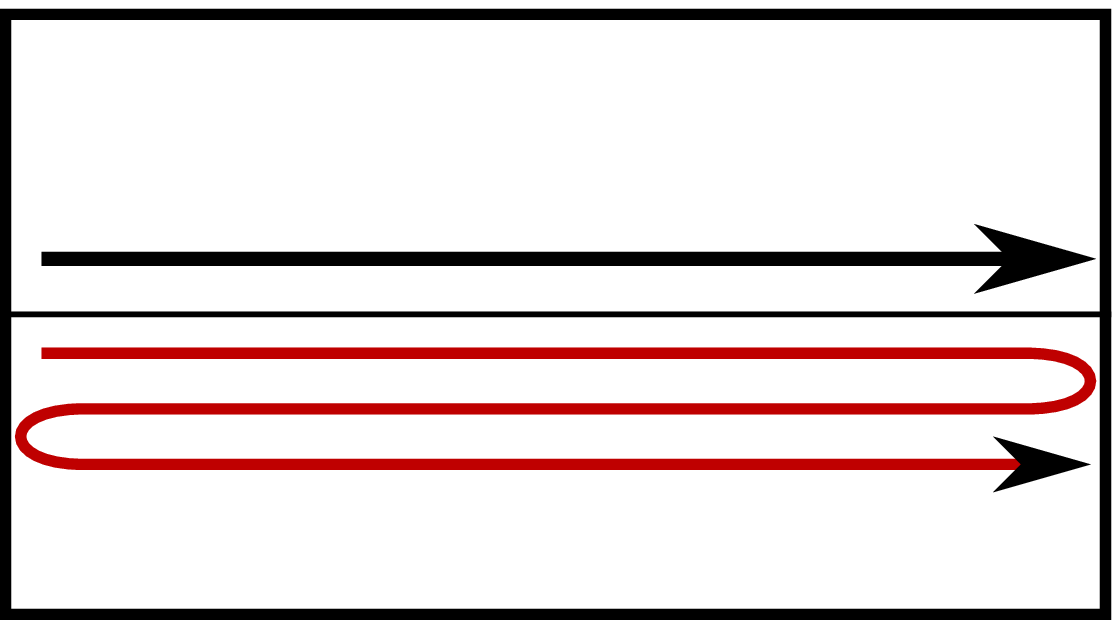}
	\caption{Dominating parasitic reflections for DUT and REF measurements}
	\label{reflections}
\end{figure}
\begin{figure}[htb]
	\centering
		\includegraphics[angle=-90, width=.47\textwidth]{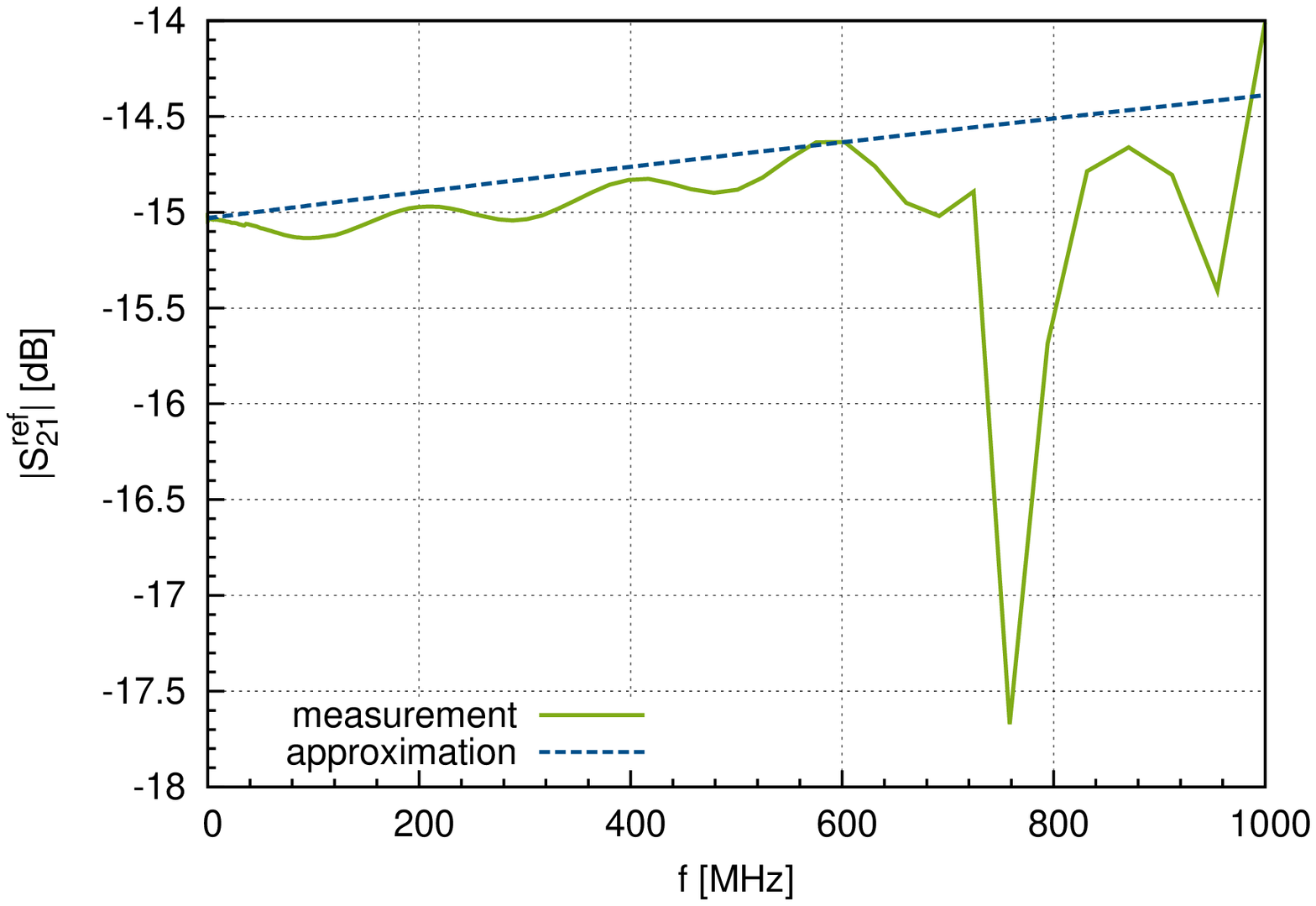}
		\includegraphics[angle=-90, width=.47\textwidth]{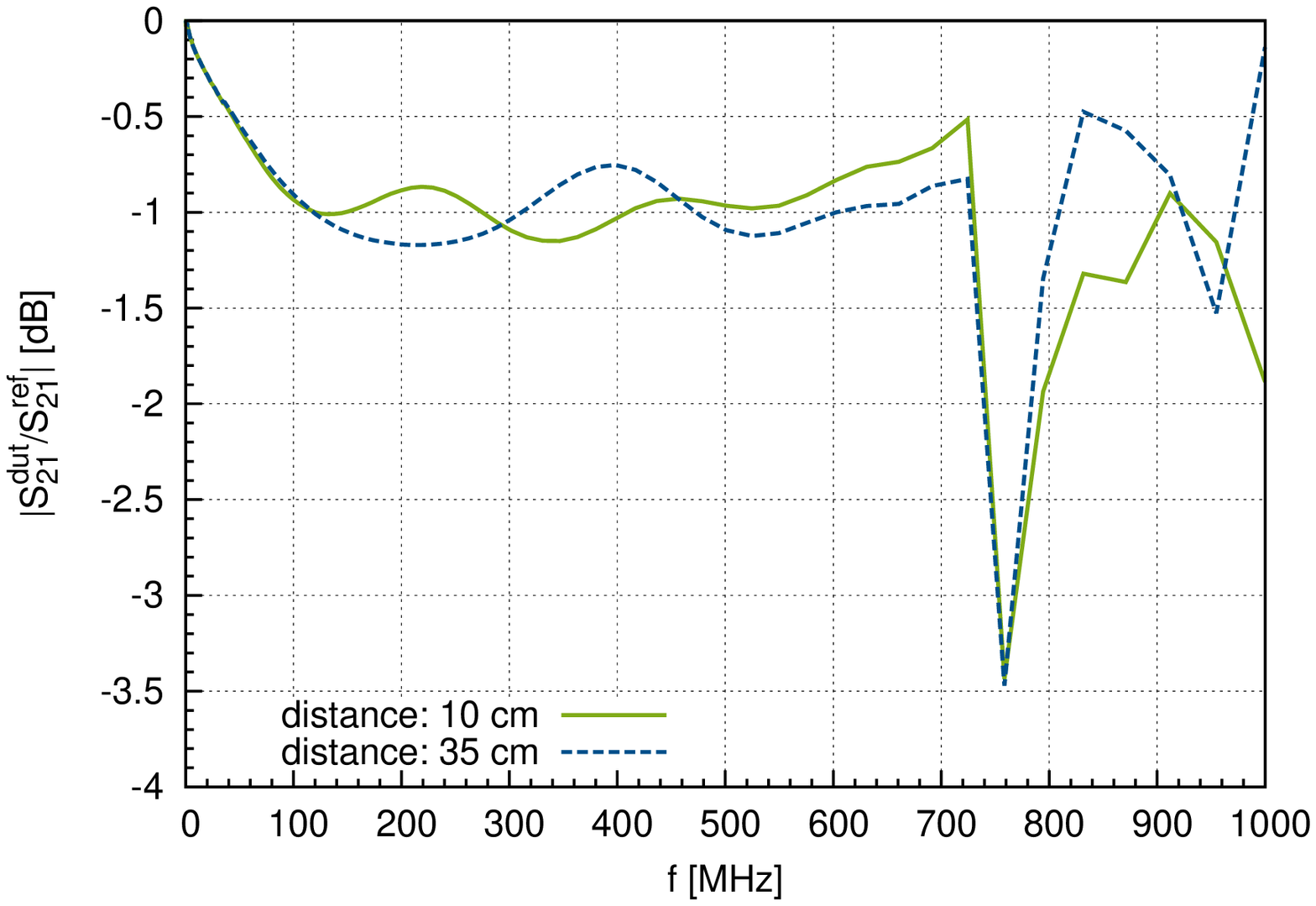}
	\caption{Reference measurement and smooth approximation. The DUT measurement depends on its longitudinal position.}
	\label{S21meas}
\end{figure}
They are obtained from independent consideration of systematic errors, such as geometry and characteristic impedance uncertainties, and statistical errors (standard deviation) such as noise, longitudinal shift and misalignment.
The error due to multiple reflections at the DUT (see Fig. \ref{reflections}) has been treated statistically for different positions of the DUT. In Fig. \ref{S21meas} one sees that the multiple reflections introduce a ripple on the measured $S_{21}$ which is position dependent. This is canceled by averaging over different positions. Also the REF measurement has been smoothed (note the scale in Fig. \ref{reflections}) to obtain similar smoothness as the averaged DUT signal.
  
The measurement results show that for low frequencies the agreement with the analytical calculation is well, while at larger frequencies discrepancies occur. At a first glance this can be accounted to resonances in the large measurement box, which can be partly damped by the RF attenuation foam. As always at high frequency, the smaller setup shows the better results. Its discrepancies with the CST-PS simulation can be accounted mostly to the material data fitting for TD simulation, the finite wire radius, and the uncertainty of the manufacturer's material data. For an estimation of the propagation of material data uncertainties see also Appendix \ref{A4}.

\section{Transverse Impedance}
\label{sect_trans}
\begin{figure}[htb]
	\centering
		\includegraphics[angle=0, width=.3\textwidth]{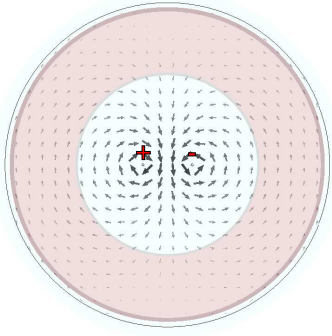}
		\caption{Magnetic field of dipole TEM Eigenmode obtained by multi-pin-portmode-solver}
	\label{cst_dipolefield}
\end{figure}

The dipolar transverse impedance can be measured by a two-wire setup, run on the differential mode. The magnetic field of such a mode can be seen in Fig. \ref{cst_dipolefield}. Note that the standard port mode solver in CST gives two arbitrary orthogonal TEM modes when there are two pins in the port. In order to select the differential mode one can apply a 'multi-pin-port' with predefined polarity of the wires.
\begin{figure}[htb]
	\centering
		\includegraphics[angle=-90, width=.47\textwidth]{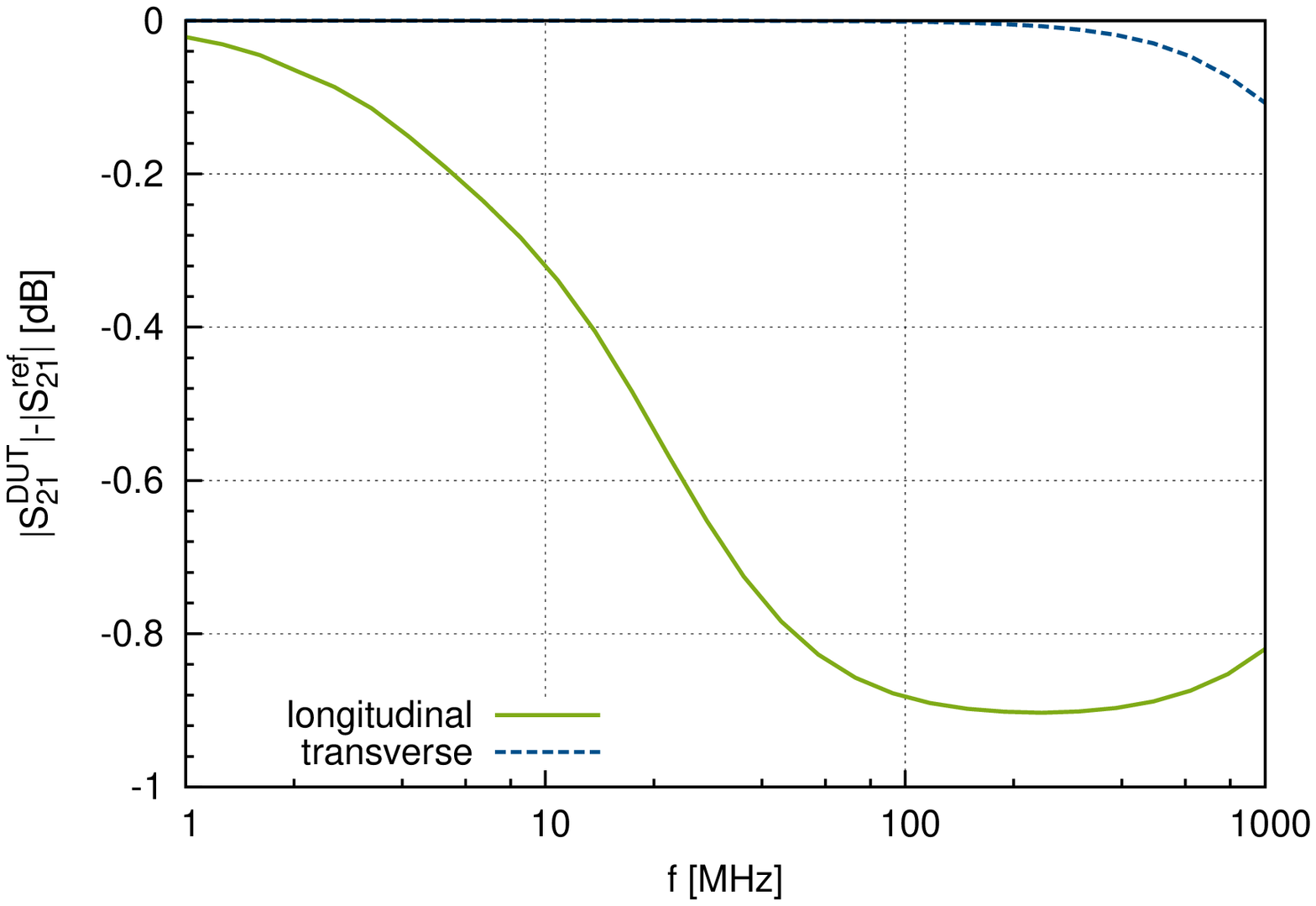}
		\includegraphics[angle=-90, width=.47\textwidth]{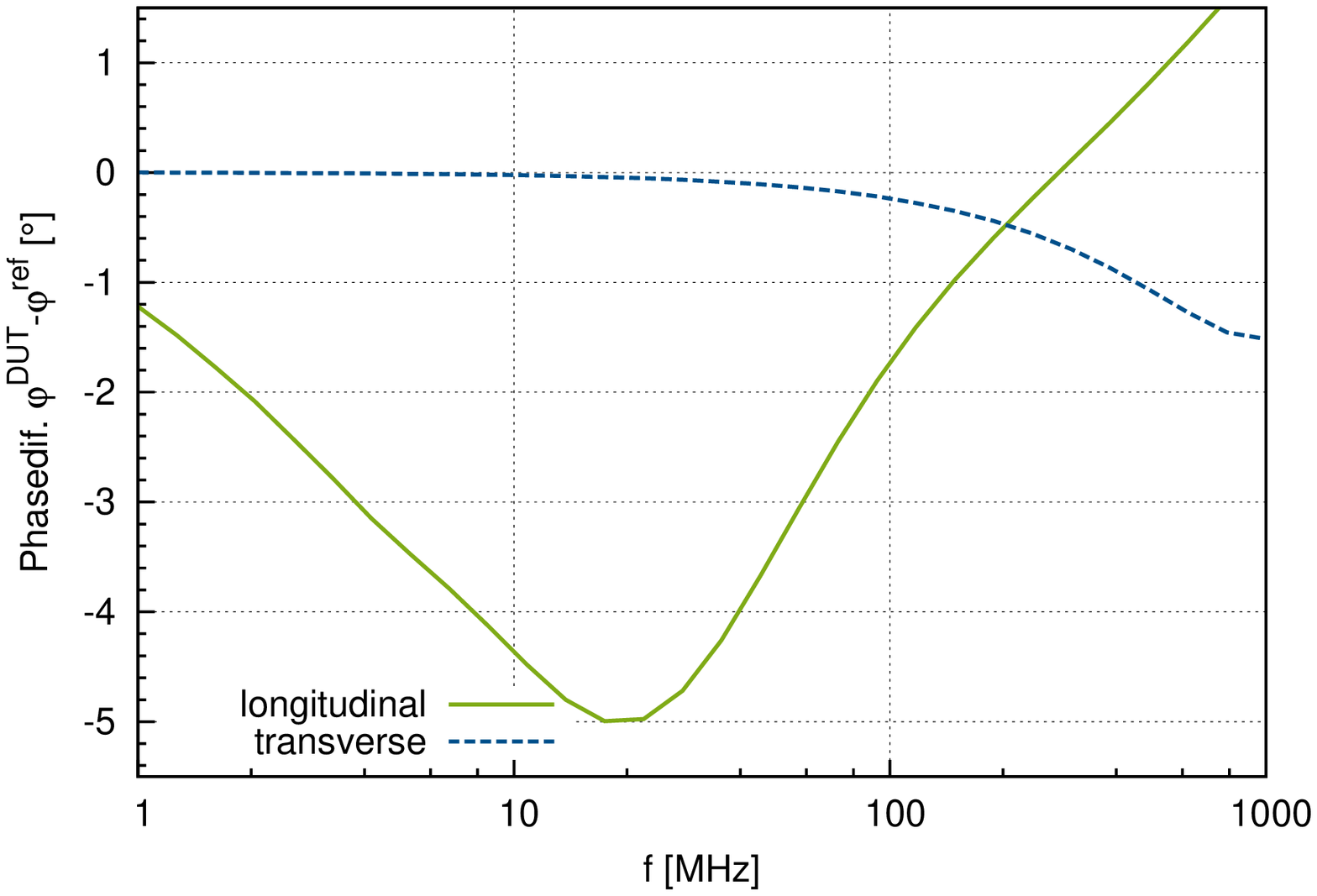}
		\caption{S-parameters for the monopole and dipole TEM mode (simulation)}
	\label{num_s21}
\end{figure}
Figure \ref{num_s21} shows the $S_{21}$-parameter of the simulation, as compared to the single-wire simulation. The magnitude and phase deviations (to REF) are much smaller for the dipole mode. One finds that the major difficulty in the dipolar measurement is the bad signal-to-noise ratio (SNR). The advantage of such a small $S_{21}$ is that the conversion formulas \ref{log-formula} and \ref{improved-log-formula}  can be linearized and agree with Eq. \ref{HPformula}, i.e. one does not have to distinguish between lumped and distributed impedances. Also the reflection at the DUT is negligible.
The characteristic impedance (REF) for the differential TEM mode is (see also \cite{Wang2004})
\begin{align}
		Z_0=\frac{\eta}{\pi}\ln\left(\frac{d+\sqrt{d^2-a^2}}{a}\frac{b^2-d \sqrt{d^2-a^2}}{b^2+d\sqrt{d^2-a^2}}\right) \label{equ_charimp_trans}
\end{align}
where $a$ is the wire radius, $b$ is outer radius and $2d=\Delta$ is the wire distance. 
The transverse impedance is defined as
\begin{align}
Z_x(\omega)&=\frac{i}{q \Delta}\int_{-l/2}^{l/2} (\Ev(\omega)+\vv\times\Bv(\omega))_x e^{i\omega z/v} \dz  \\
&=-\frac{v}{\omega q \Delta}\int_{-l/2}^{l/2} \frac{\partial E_z(\omega)}{\partial x} e^{i\omega z/v} \dz +\left[E_x e^{i\omega z/v} \right]_{-l/2}^{l/2}
\end{align}
with the second expression obtained from the Panofski-Wenzel \cite{Panofsky1956} theorem.
In good approximation one finds
\beq
Z_x(\omega)
\approx \frac{v}{\omega \Delta^2}\delta Z_\parallel,
\label{Zt}
\eeq
where $\delta Z_\parallel$ is the impedance obtained from the $S_{21}$ conversion formula for the differential mode.
\begin{figure}[htb]
	\centering
		\includegraphics[angle=-90, width=.47\textwidth]{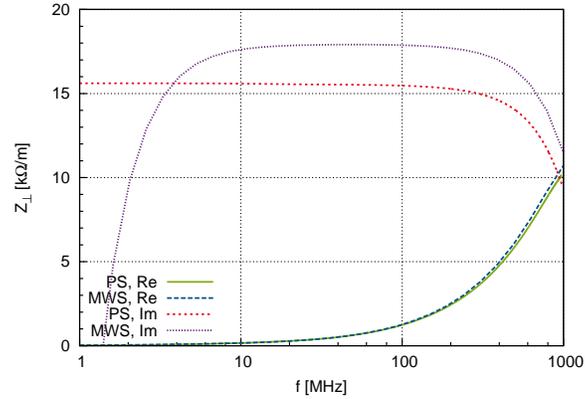}
		\caption{Transverse impedance PS vs. MWS}
	\label{mws_transversal}
\end{figure}
Figure \ref{mws_transversal} shows the same plot for the transverse impedance, also with good agreement for the real part. The disagreement for the imaginary part is accounted to extremely small change in the relative transmission, making it impossible to determine the phase of $S_{21}$ accurate enough. 

\begin{figure}[htb]
	\centering
		\includegraphics[angle=-90, width=.4\textwidth]{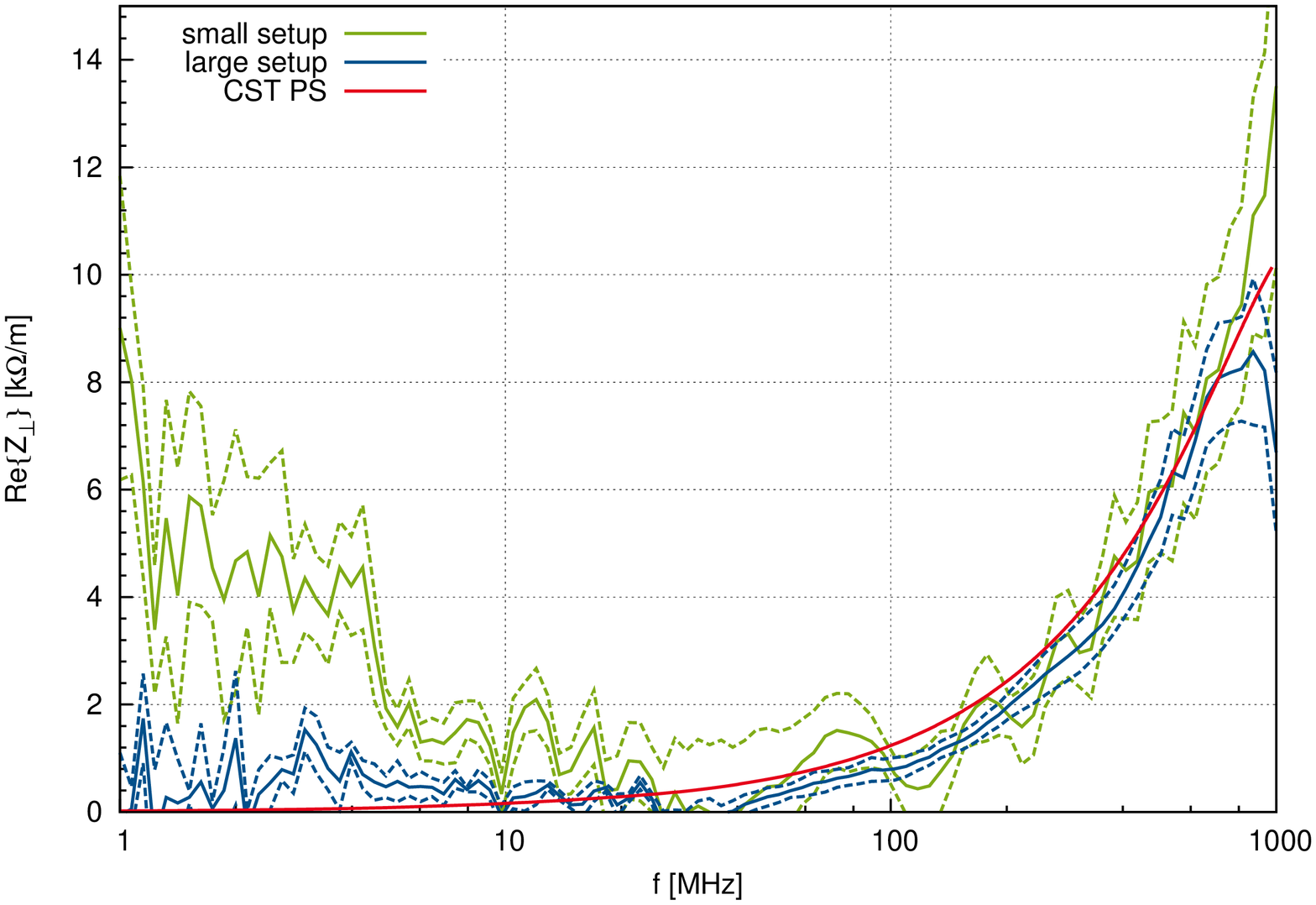}
		\includegraphics[angle=-90, width=.4\textwidth]{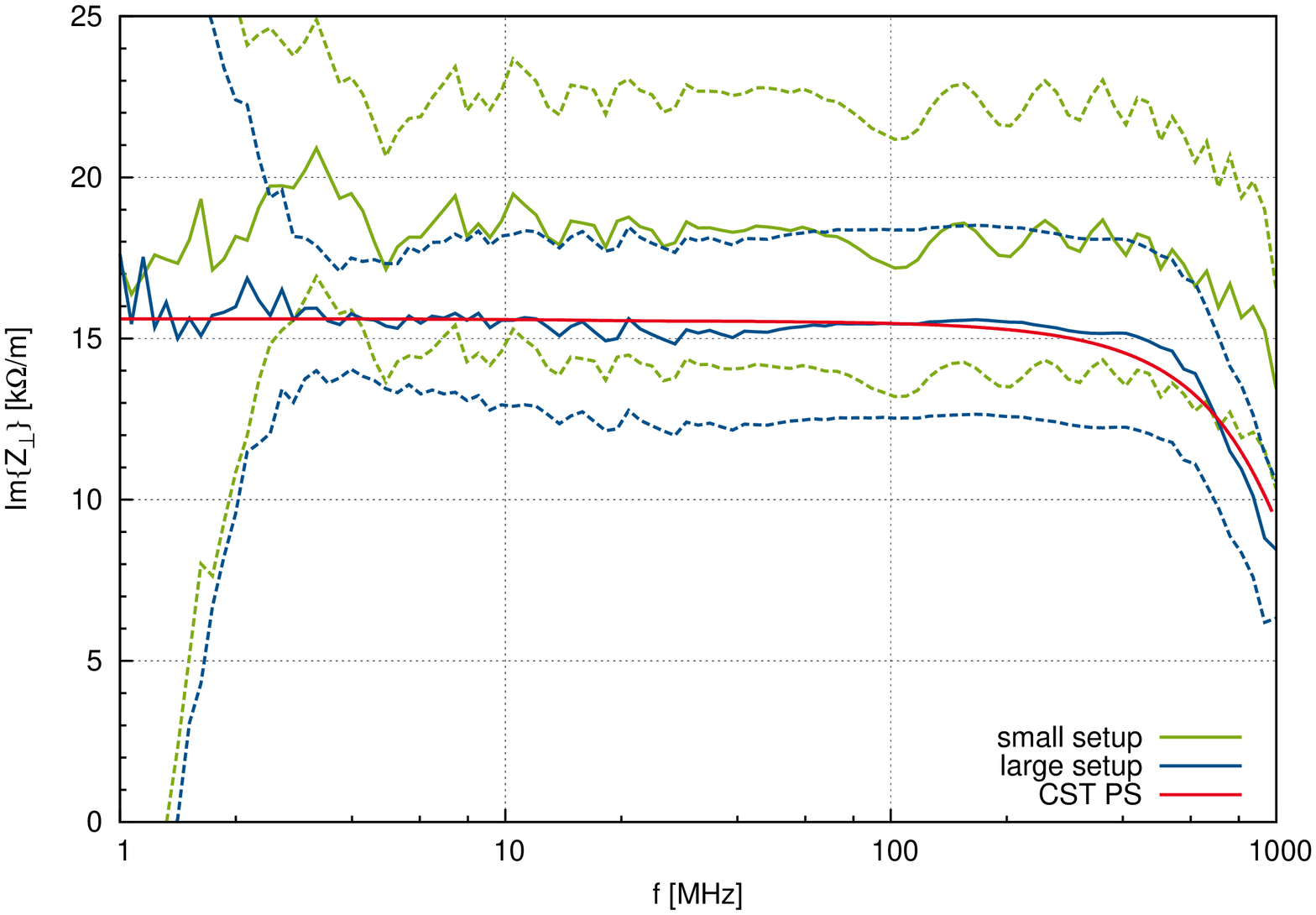}
	\caption{Transverse impedance: Measurement vs. wakefield simulation}
	\label{meas_trans}
\end{figure}



In a wakefield simulation the transverse impedance has been obtained by integrating the wake force on the beam axis and exciting the system by two particle beams. Those beams are off-centered by $\Delta/2$ and carry equal oppositely signed charge. The linear behaviour with $\Delta$ has been confirmed. Figure \ref{mws_transversal} shows a comparison for wakefield and S-parameter simulation. At low frequencies the S-parameter simulation becomes inaccurate, since the signal is smaller than the numerical errors. 

The measurements together with the error estimates are shown in Fig. \ref{meas_trans}. Note that for the two wire setup an autocal-kit can be recommended since otherwise 18 different connections have to be made which takes quite long and is quite susceptible to errors. Both the large and the small setup show good agreement with the wakefield simulation, but the error-bars become intolerably large at low frequency. This can be improved using the coil measurements, see \cite{Roncarolo2009} and Appendix \ref{A3}.

\section{Conclusion}
\label{sect_conclusion}
A generalized two-dimensional approach to the longitudinal impedance for a bench measurement, using transmission line quasi-TEM eigenmodes, and for a particle beam has been presented. 
It was found that the beam velocity enters the impedance calculation in close relation to the material properties. 
Therefore simple scaling laws with $\beta$ only exist in the case of  
frequency independent material properties, see e.g. \cite{Niedermayer2012}.

From the dispersion relation (Eq. \ref{dispersion_tandelta}) follows that for low frequency and velocities close to the speed of light, the radial model can be employed, i.e. the limit $\beta\rightarrow\infty$ can be applied. The radial model is used for simplified measurements, i.e. the coil method, or for impedance simulations using the power dissipation method \cite{Roncarolo2009} \cite{Niedermayer2012}.
Another important issue originating from the dispersion relation is that for very low $\beta$ one requires a dense transverse mesh in numerical simulations.

The interplay between simulations and bench measurements has been outlined: On the one hand simulations are needed to crosscheck the 'a priori' assumptions in the measurements. In particular, the proper de-embedding of the measurement box has to be checked by simulations. On the other hand measurements are needed to validate simulations, which can then be performed for arbitrary $\beta$.
Note that the wire bench measurements are incapable of resembling $\beta<1$ since the wave impedance for the real beam is $Z_{wave}=\eta/\beta$ while a TEM wave in vacuum always has $Z_{wave}=\eta$.

For the determination of the distributed impedance from $S_{21}$ measurements the 'improved-log-formula' has been re-derived. It was found that for a perfectly uniformly distributed impedance, i.e. when the 2D assumptions are exactly fulfilled, the formula recovers the impedance from the scattering parameter exactly, provided the wire radius tends to zero. Note that this convergence is very slow (logarithmic), such that in practice always an error of about 10-20\% remains.
The 'log-formula' and the 'lumped-formula' have been compared for the example ferrite ring with the analytical $S_{21}$ and found too inaccurate. 
For the simulation of the measurement setup the 'log-formula' showed an approximate agreement to the wakefield simulation while the 'improved-log-formula' showed a parasitic resonance. This could not be explained completely, but it is accounted to the 'log-formula' being less sensitive. This was also observed in the practical measurements, when errors due to subsequent changing of DUT and REF measurements propagated through the 'improved-log-formula' but not through the 'log-formula'.
The parasitic resonance in the simulation of the measurement evaluated by the 'improved-log-formula' could be removed by applying the Wang-Zhang reflection correction. This works very well in the simulation but in the real measurement $S_{11}$ cannot be determined properly due to multiple reflections between the DUT and the matching resistors.

For the transverse impedance impedance it does not matter which $S_{21}\rightarrow Z$ formula is applied since the measurement signal is extremely small. When linearizing the $S_{21}\rightarrow Z$ formulas for $S_{21}^{DUT}\simeq S_{21}^{REF}$, they all agree with each other. The limiting property of the two-wire measurement is the signal-to-noise-ratio (SNR) which becomes poor, particularly at low frequencies. For those low frequencies the coil method is a well-working alternative. 

\begin{table}[htb]
\begin{tabular}{|l|l|l|l|l|} \hline
Priority & \multicolumn{2}{|c|}{\textbf{Longitudinal}} & \multicolumn{2}{c|}{\textbf{Transverse}}\\
\hline
1 & $S_{21}\rightarrow Z$ & a priori 																											& Noise &  $\rightarrow$Averaging  		\\  \hline
2 & Reflections &  \parbox[t]{3cm}{$\rightarrow$Average \\DUT position} 							& \parbox[t]{3cm}{Random setup\\ modification} 	 &  $\rightarrow$Averaging  		 \\ \hline
3 & Wire thickness & a priori 		 																										& \parbox[t]{3cm}{Wire distance \\ \& thickness} &  a priori 		  \\ \hline
4 & Noise &  $\rightarrow$Averaging  																									& $S_{21}\rightarrow Z$ &  a priori 		 \\ \hline
5 & \parbox[t]{3cm}{Random setup\\ modification} 	 &  $\rightarrow$Averaging  				& Reflections & \parbox[t]{3cm}{$\rightarrow$average \\DUT position} 		 \\ \hline
6 & Misalignment & $\rightarrow$Averaging 																						& Misalignment & $\rightarrow$Averaging  \\ \hline
\end{tabular}
\caption{Prioritization of error sources in the measurements and their diminishment}
\label{tab:error}
\end{table}

An overview of the measurement error sources is given in Tab. \ref{tab:error}. Statistic errors can be diminished by averaging over e.g. DUT position or many DUT/REF setup changes, provided the SNR is reasonably high. 


%
%

%

\begin{acknowledgments}
UN and LE wish to thank Elias Metral, Fritz Caspers, and Manfred Wendt for the hospitality at CERN and inspiring discussions.
\end{acknowledgments}

\appendix
\section{Other Geometries and Material properties for the Example Setup}
\label{A1}
An open boundary condition (radiation condition) can be applied in Eq. \ref{ansatz} by exchanging the bracket after the $D_1$ constant by the Hankel function $H_m^{(2)}(k_r r)$.
\begin{figure}[ht]
	\centering
		\includegraphics[angle=-90, width=.47\textwidth]{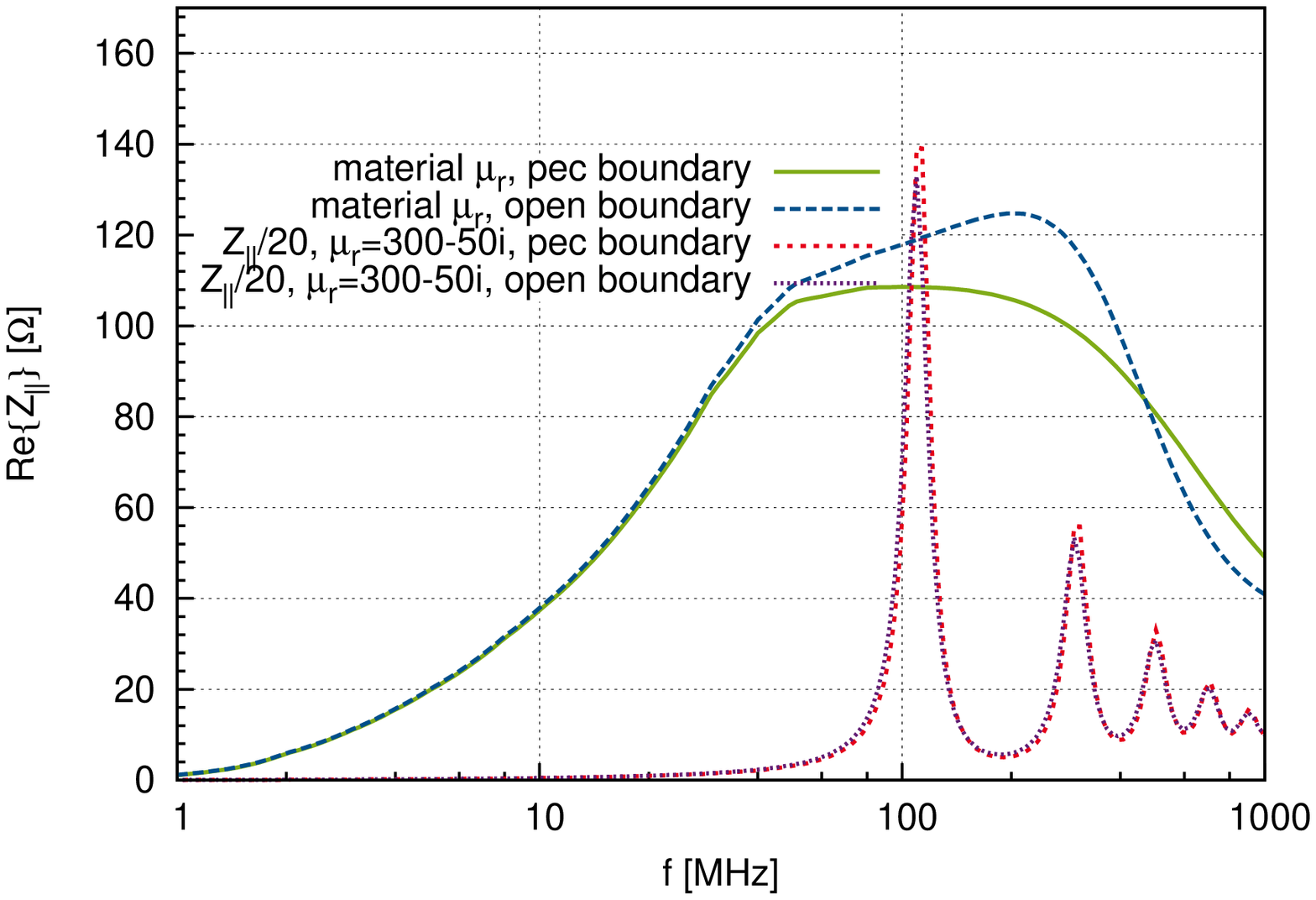}
		\includegraphics[angle=-90, width=.47\textwidth]{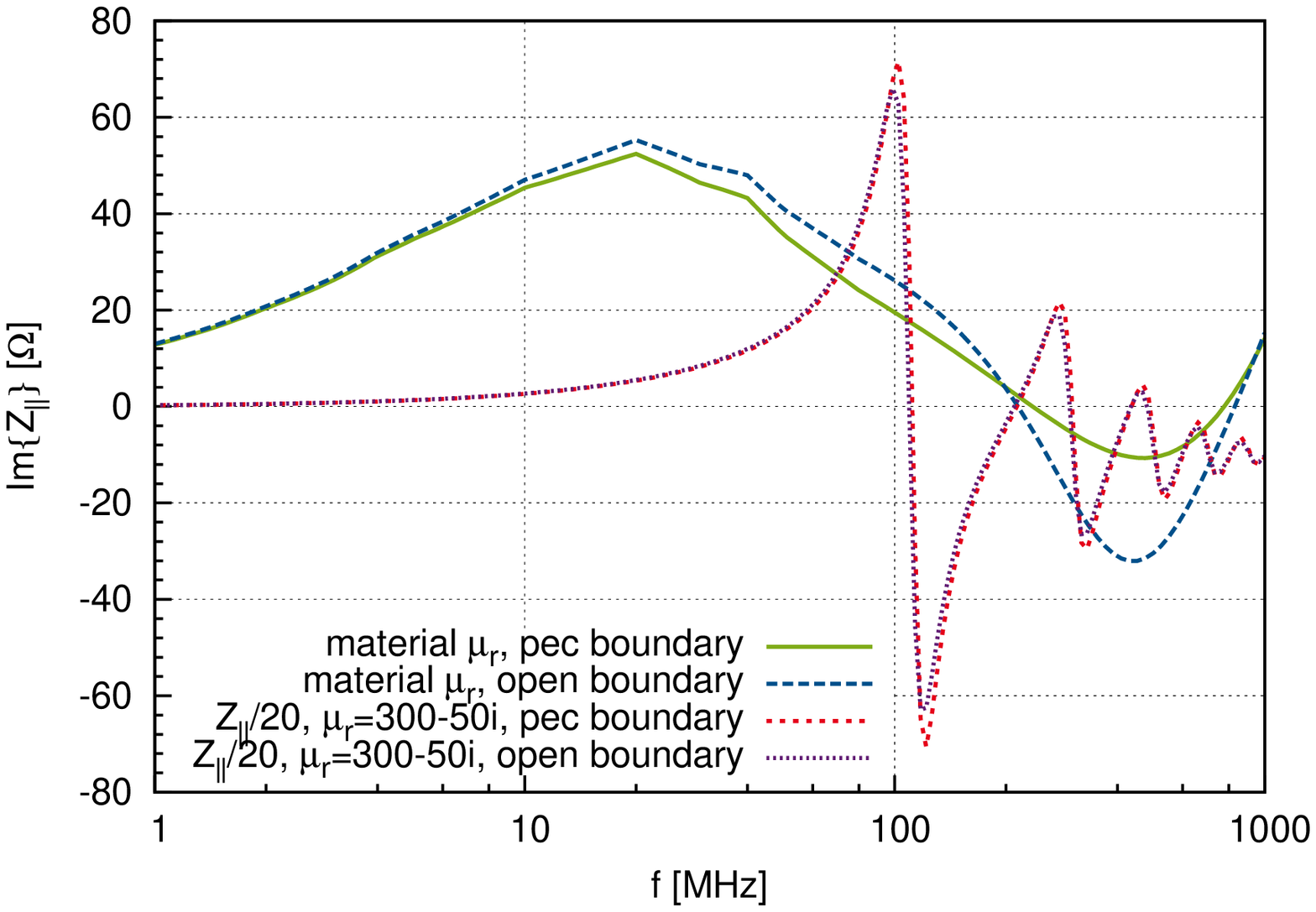}
	\caption{Radial model: Comparison of real material parameters to dispersion-free (artificial) material, open and closed boundaries. Only in the dispersion-free case geometrical resonances are visible.}
	\label{analytic_radial}
\end{figure}
This, and the impedance for an artificial material with constant complex permeability is shown in Fig. \ref{analytic_radial}. Without the dispersion the geometric resonances become visible. Relevant for the measurement is that even in the large box the electrical lenght between the ferrite and the boundary is much smaller than the electrical lenght of the ferrite itself. This motivates neglecting the effect of the boundary, especially at low frequency.

\section{Technical issues of the measurement setup}
\label{A2}
\begin{figure}[htb]
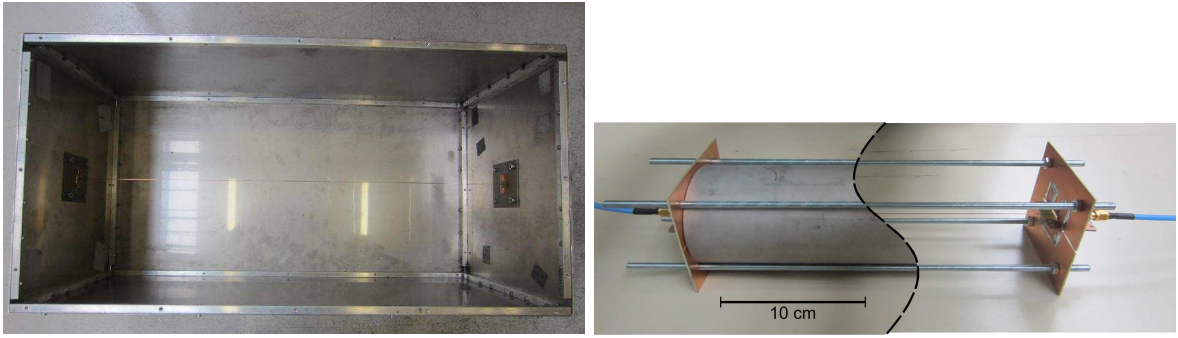

	\centering
		\includegraphics[angle=0, width=.47\textwidth]{large_setup}
		\includegraphics[angle=0, width=.47\textwidth]{small_setup}
	\caption{Different measurement boxes}
	\label{setup}
\end{figure}
The cables connecting the box with the VNA have to be phase-stable, even in the case of manipulating them for subsequent DUT and REF measurements. Standard SMA cables have been tested and found insufficient. Of course, precision measurement cables could do the job, but they are very expensive. A cost effective alternative is found by semi-rigid SMA-cables. Due to only small movements during setup changes, phase deviations are tolerably small. Note that also the calibration of the VNA is made at the end of the SMA cables. SMA-N adapters can be used for N-calkits since their electrical length can be neglected below 1GHz.


For frequencies below roughly 50\;MHz resistive matching is the method of choice. It is based on building a resistive network that makes each side see its own characteristic impedance. 
On the NWA side it makes sense to use a commercially available attenuator piece instead, since its $\Pi$ or T-bridge network has very linear frequency and phase response and can therefore easily be accounted in the REF measurement. On the measurement box side a longitudinal resistor has to be used which is involved to optimize. The low-pass cut-off of real resistors determines the maximum frequency of the resistively matched setup. Different end pieces for the wire(s) have been tried out:
\begin{enumerate}
	\item Orthogonal PCBs with SMD metal film resistors
	\item $90\deg$ SMA flange with carbon or metal film resistors
\end{enumerate}
The SMD resistors can be precisely mounted, nonetheless they show (dependent on type) a bad high frequency behaviour. Similarly bad behaviour is found for the metal film resistors. Comparably good rf-behaviour is found for particular carbon resistors, so called 'grounding resistors'. They keep their purely real resistance up to about 30 MHz. Nonetheless they are specified with a tolerance of 20\%, which requires measuring each resistor with a precise Multimeter and choosing a proper combination.
\begin{figure}[htb]
	\centering
		\includegraphics[angle=0, width=.47\textwidth]{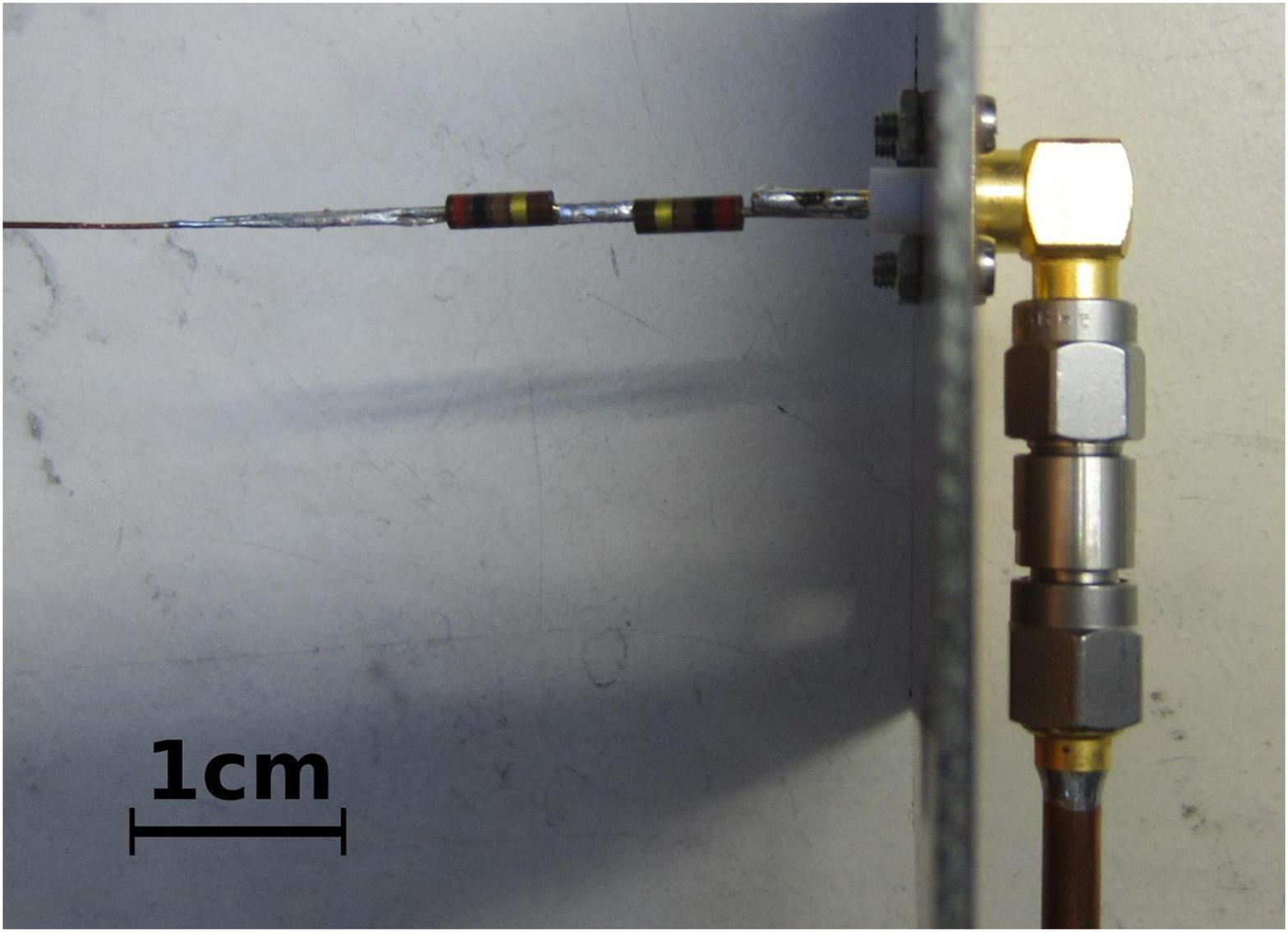}
		\includegraphics[angle=0, width=.47\textwidth]{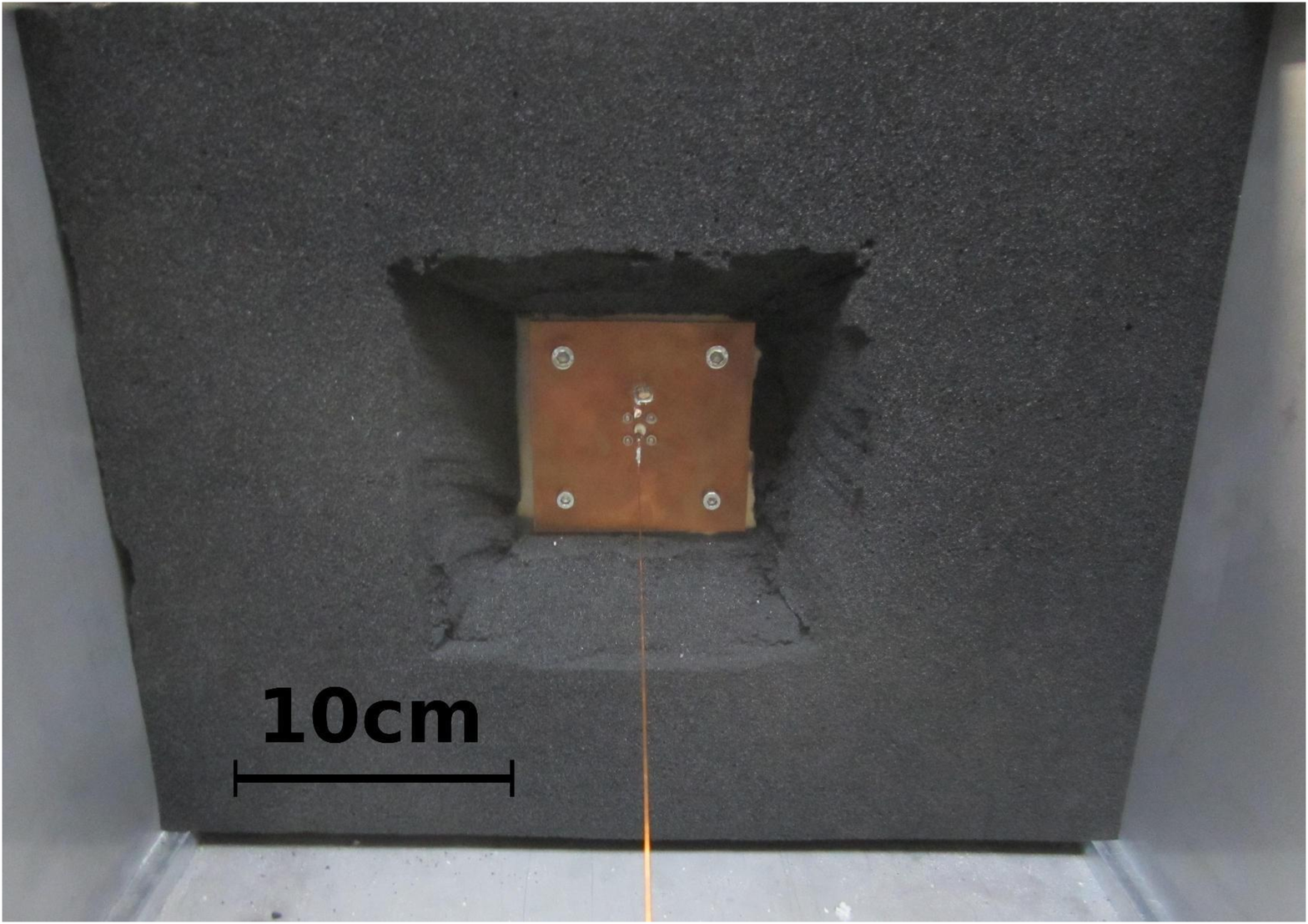}
	\caption{Matching resistors with 10dB attenuator and absorber foam}
	\label{matching}
\end{figure}
For frequencies above 30\;MHz reflections on the resistive matching section occur. They can be damped using RF-attenuation foam. Nonetheless, the changing of DUT/REF without changing the properties of the foam is technically involved.

\section{Transverse Impedance Coil Measurements}
\label{A3}
In order to enhance the extremely small signals in the two-wire method for low frequency, a multiturn coil can be used\cite{Roncarolo2009}.
Both the flux and the induced voltage are amplified by the number of turns $N$, and one finds instead of Eq. \ref{Zt}
\begin{align}
	Z_\perp=\frac{c\cdot \delta Z}{\omega \cdot \Delta^2 \cdot N^2} \label{equ_coilDet}.
\end{align}
\begin{figure}[htb]
	\centering
		\includegraphics[angle=0, width=.4\textwidth]{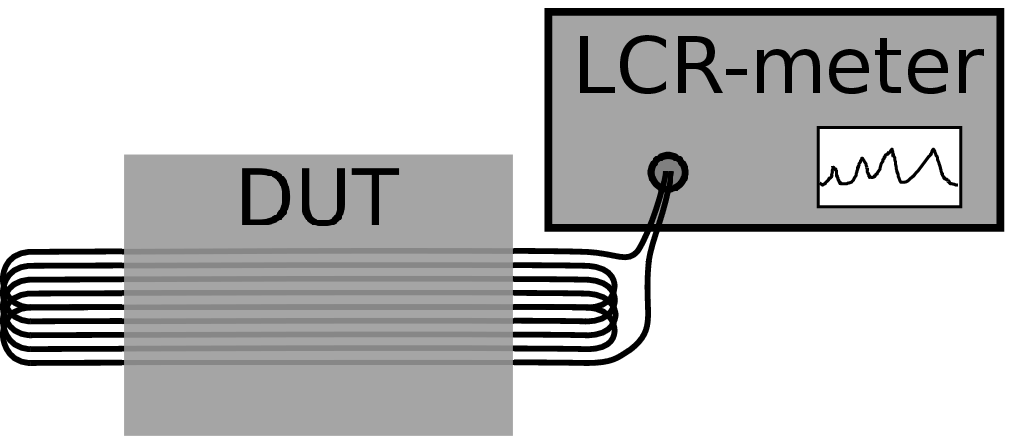}
		\includegraphics[angle=0, width=.4\textwidth]{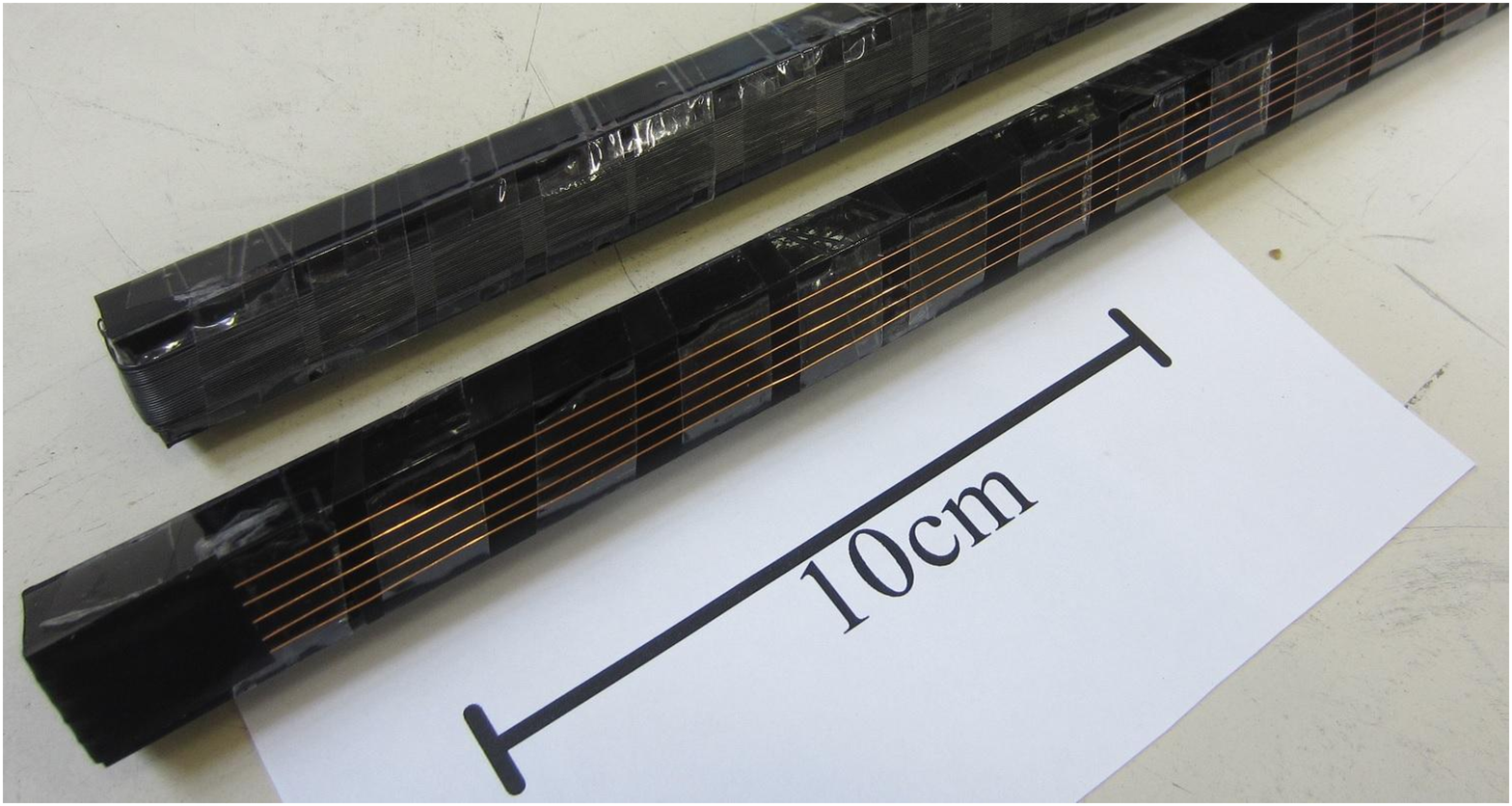}
	\caption{Transverse impedance measurement for very low frequency}
	\label{fig_coil}
\end{figure}
Since ferrite structures usually have only small transverse impedance contributions at such low frequencies, the method is benchmarked using a metal pipe of 2\;mm wall thickenss.
Figure \ref{fig_coil} shows the measurement setup, in which the coil impedance change $\delta Z=Z^{DUT}-Z^{REF}$ is determined by a LCR-meter. 
\begin{figure}[htb]
	\centering
		\includegraphics[angle=-90, width=.4\textwidth]{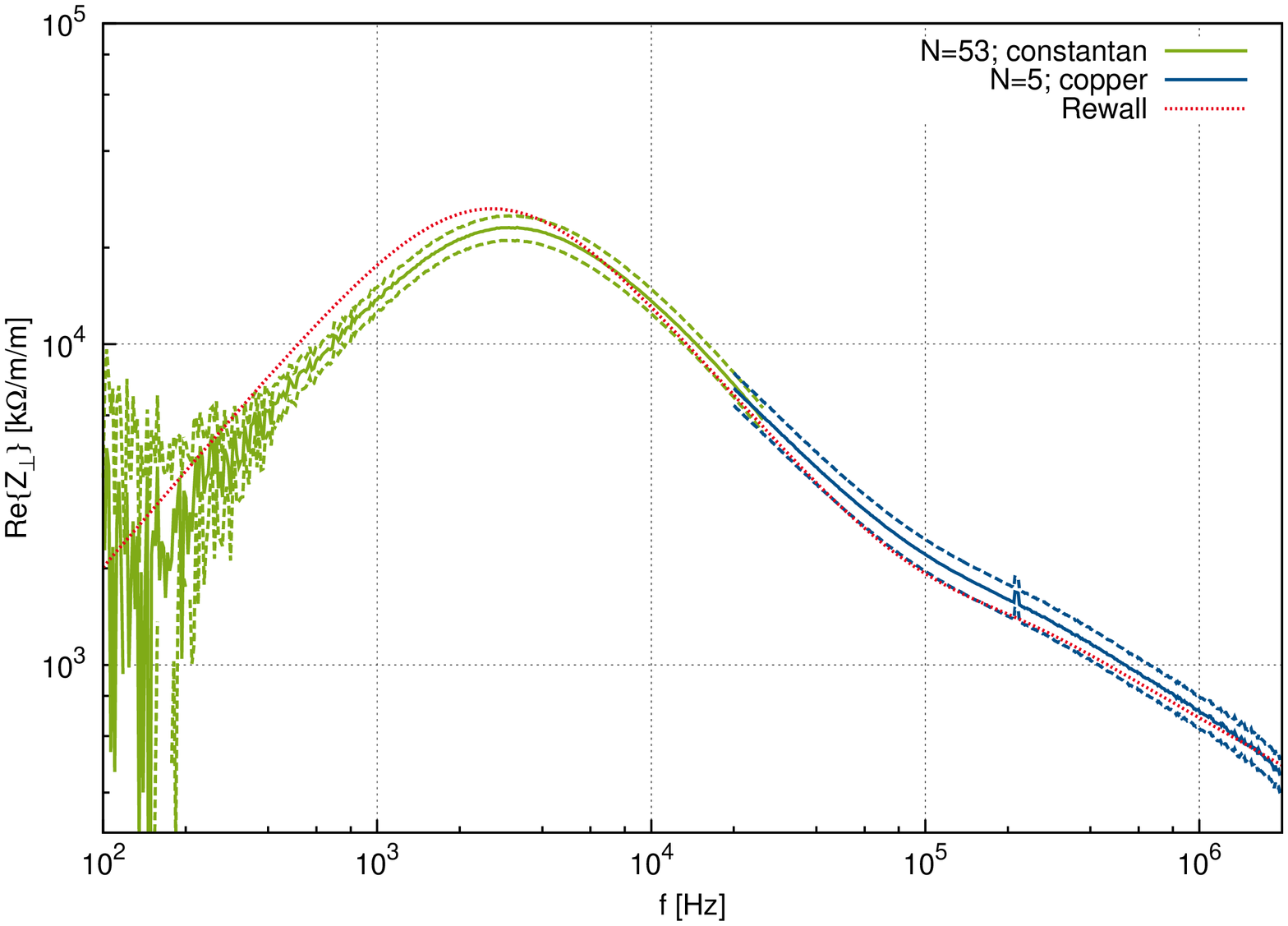}
		\includegraphics[angle=-90, width=.4\textwidth]{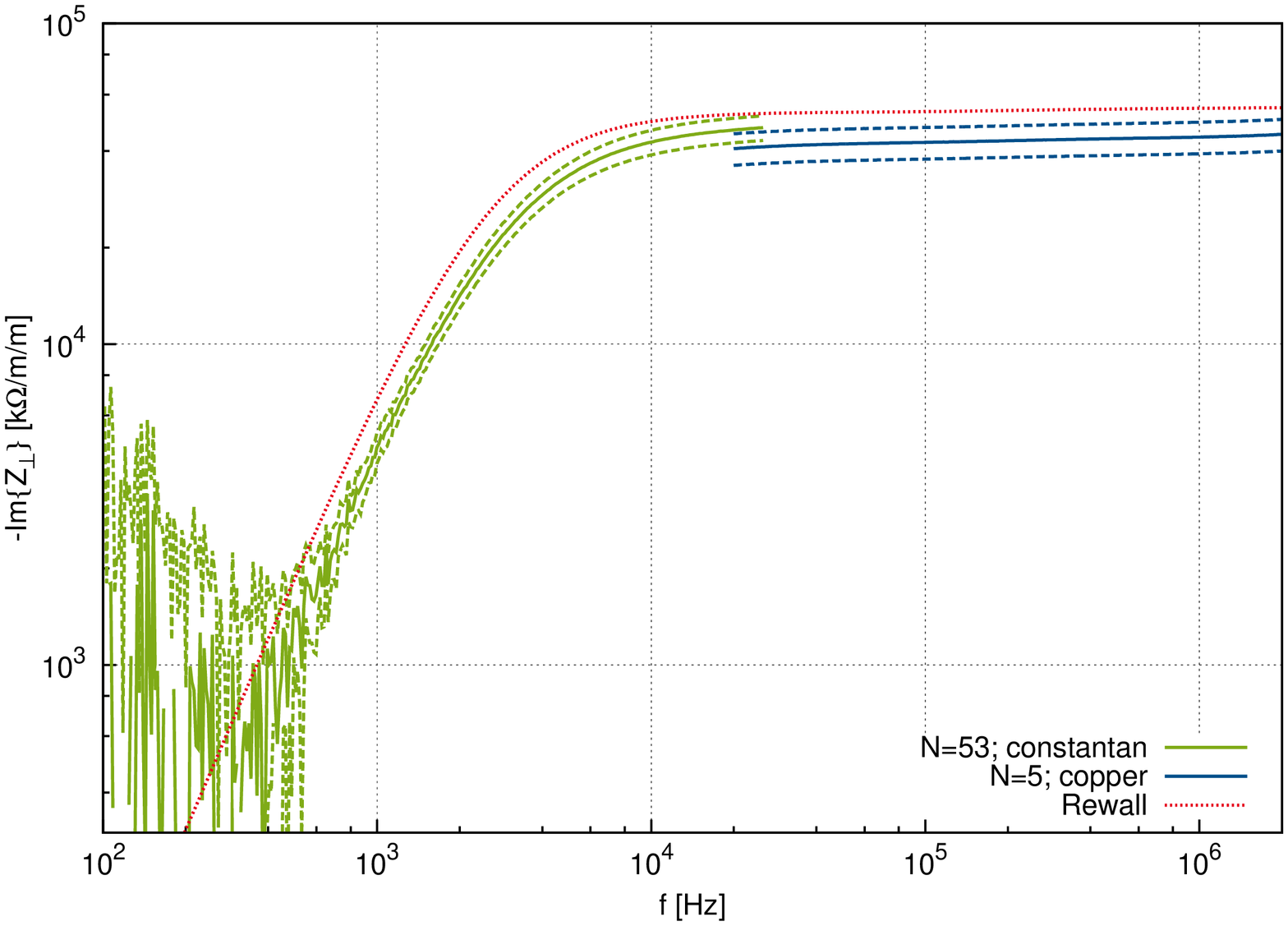}
	\caption{Transverse impedance at LF: Coil measurement vs. analytical calculation by ReWall\cite{Mounet2009}. The dashed lines indicate standard deviation.}
	\label{coilmeasurement}
\end{figure}
The coil-method has an upper frequency limit, given by the coil resonance. It can be increased by taking fewer turns and increasing the turn distance (decreasing the inter-turn capacitance).
At very low frequency the accuracy limitation comes from the instrument noise ($\delta Z\propto \omega$) and from temperature drift of the coil, i.e.
\begin{align}
			R(T)=\frac{L}{\pi r^2}\rho(T_0)\cdot \left(1+\alpha_T(T_0) \cdot (T-T_0)\right) 
\end{align} 
with $\alpha_T(T_0)$ being the (linearized) material temperature coefficient at room temperature $T_0=300\;$K.
Subsequently, it makes sense to use two coils, a temperature stable one made of constantan with many turns an one with few turns and low resistivity (copper).
Figure \ref{coilmeasurement} shows the measured impedance compared to analytical results for beam impedance (Rewall \cite{Mounet2009}). 
The error-bars indicate systematic errors, dominated by $\Delta$, and statistical errors represented by the standard deviation of subsequent DUT and REF measurements.

The coil measurements are not in accordance with an ultrarelativistic beam, but rather with the radial model. The equivalence of the analytical beam impedance results with the radial model for low frequencies is shown in \cite{Niedermayer2012}. One does not have any longitudinal propagation, except the image current in the DUT, which is induced by the magnetic field. Note that for DUTs which consist of two side parts (e.g. collimator jaws) isolated from each other one gets two independently closed eddy current loops. After connecting both sides at their ends one gets a current loop over the whole device, changing the measured impedance significantly. This means that the measurement setup should be chosen exactly as it is seen by the beam in the accelerator.

\section{Material Data Uncertainties}
\label{A4}
Usually the manufacturer of ferrite materials gives material curves only for a particular temperature and without remanence magnetization. Still the permeability and magnetization loss ($\underline{\mu}=\mu'-i\mu''$) curves are mostly specified with an error bar of $\pm 20\%$. There is some physical motivation of the smoothness of such a material curve. Therefore it is sufficient for a worst case estimate, to look at all frequency points for min and max perturbation at once.
\begin{figure}[htb]
	\centering
		\includegraphics[angle=-90, width=.47\textwidth]{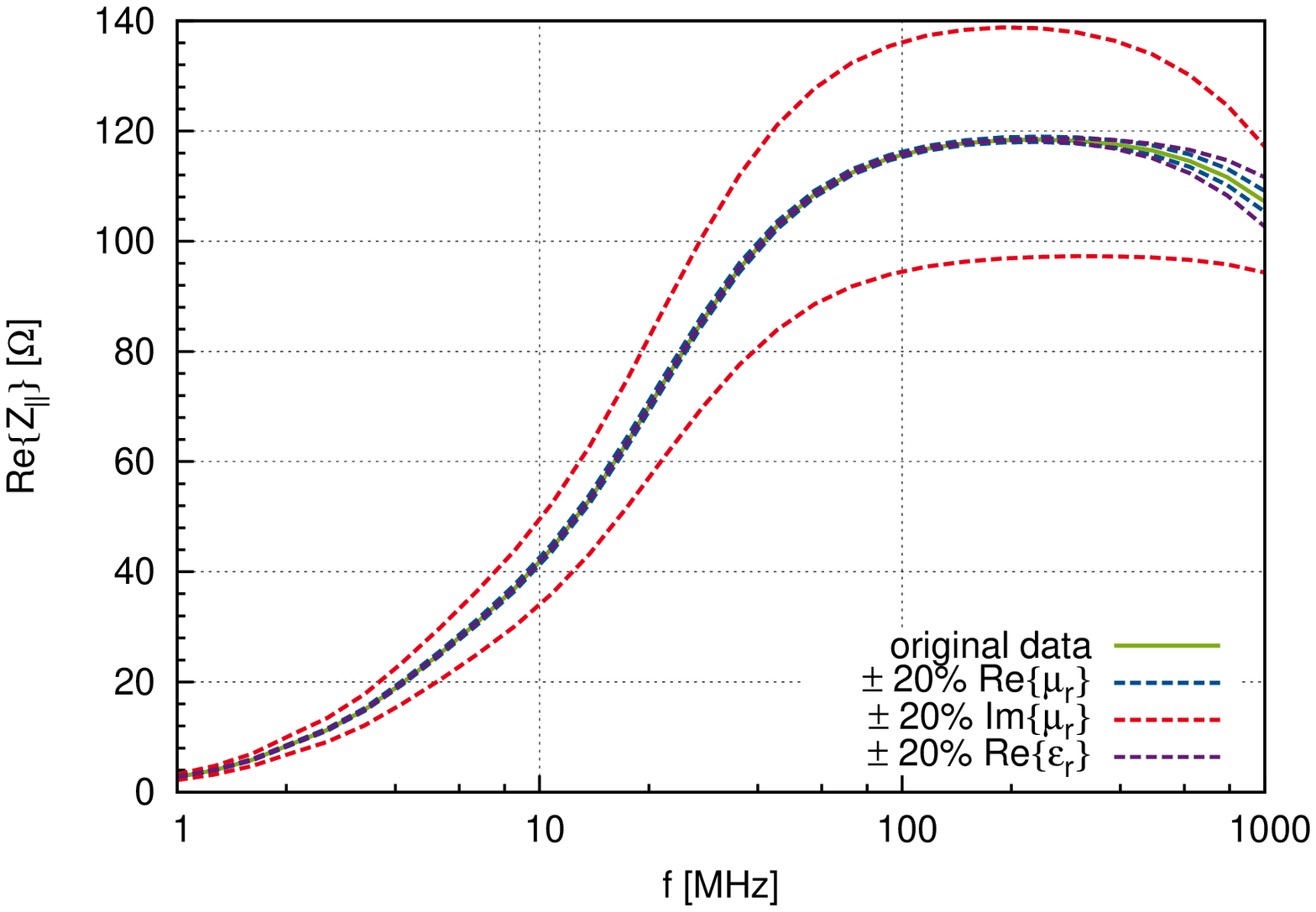}
		\includegraphics[angle=-90, width=.47\textwidth]{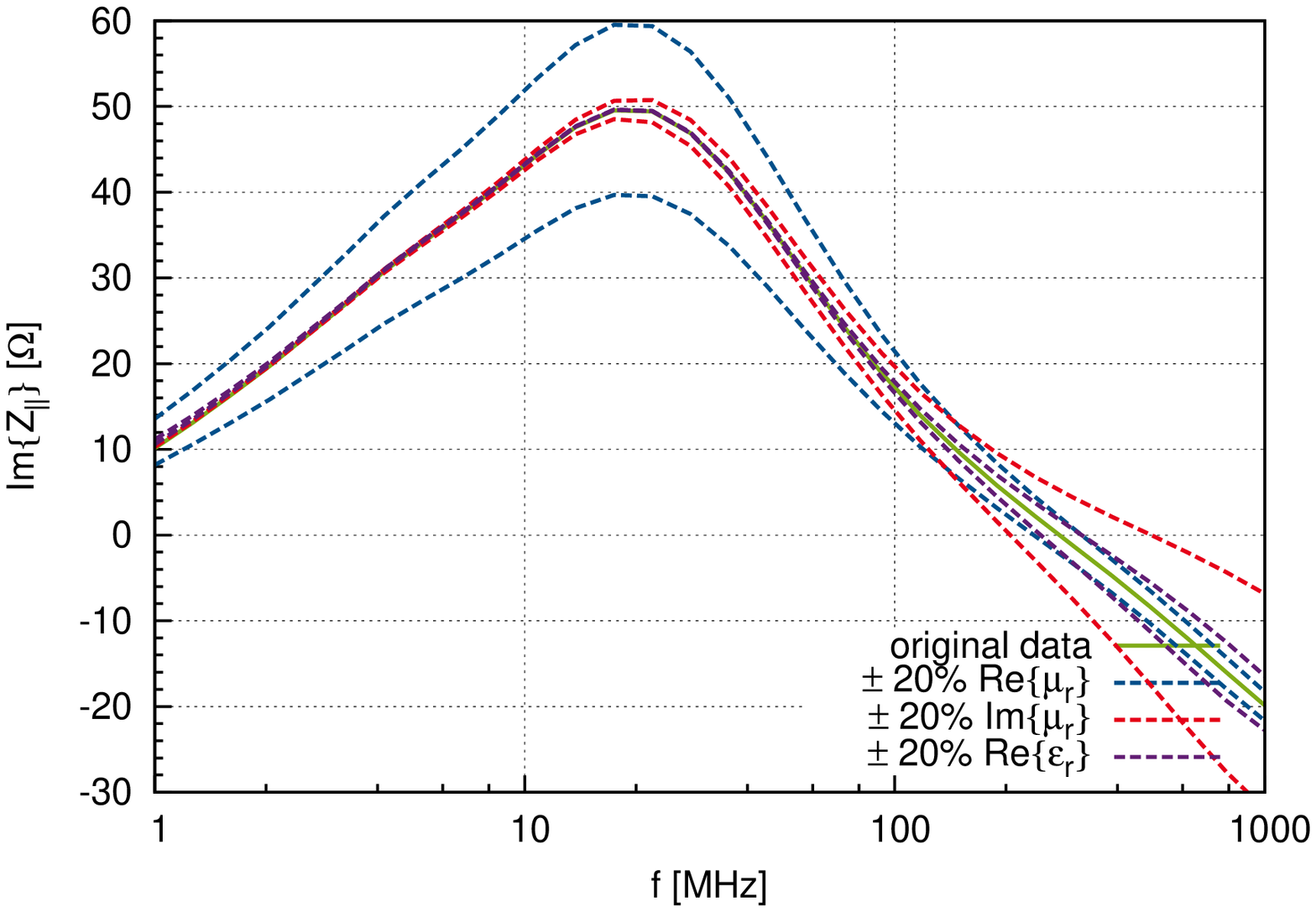}
		\includegraphics[angle=-90, width=.47\textwidth]{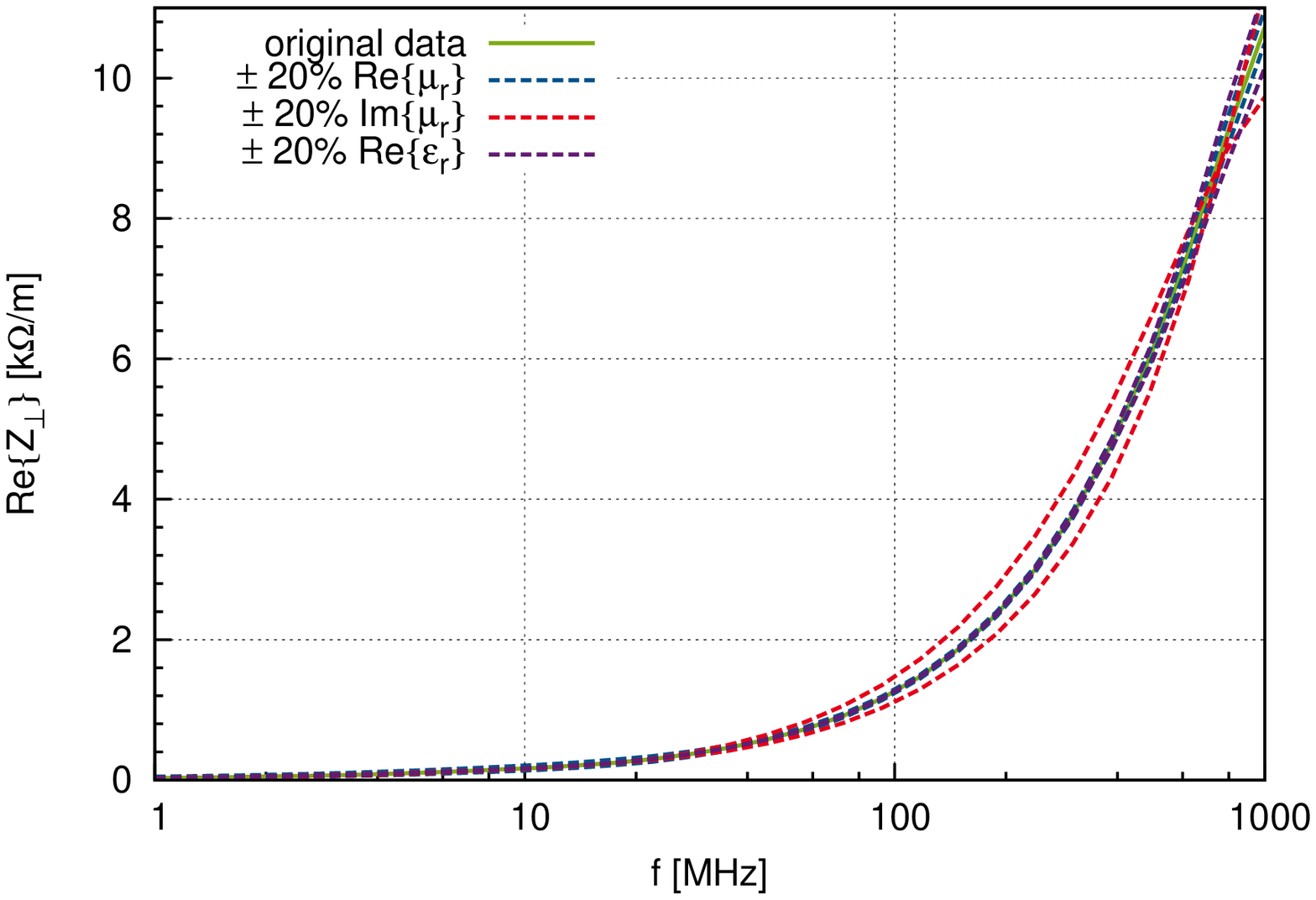}
		\includegraphics[angle=-90, width=.47\textwidth]{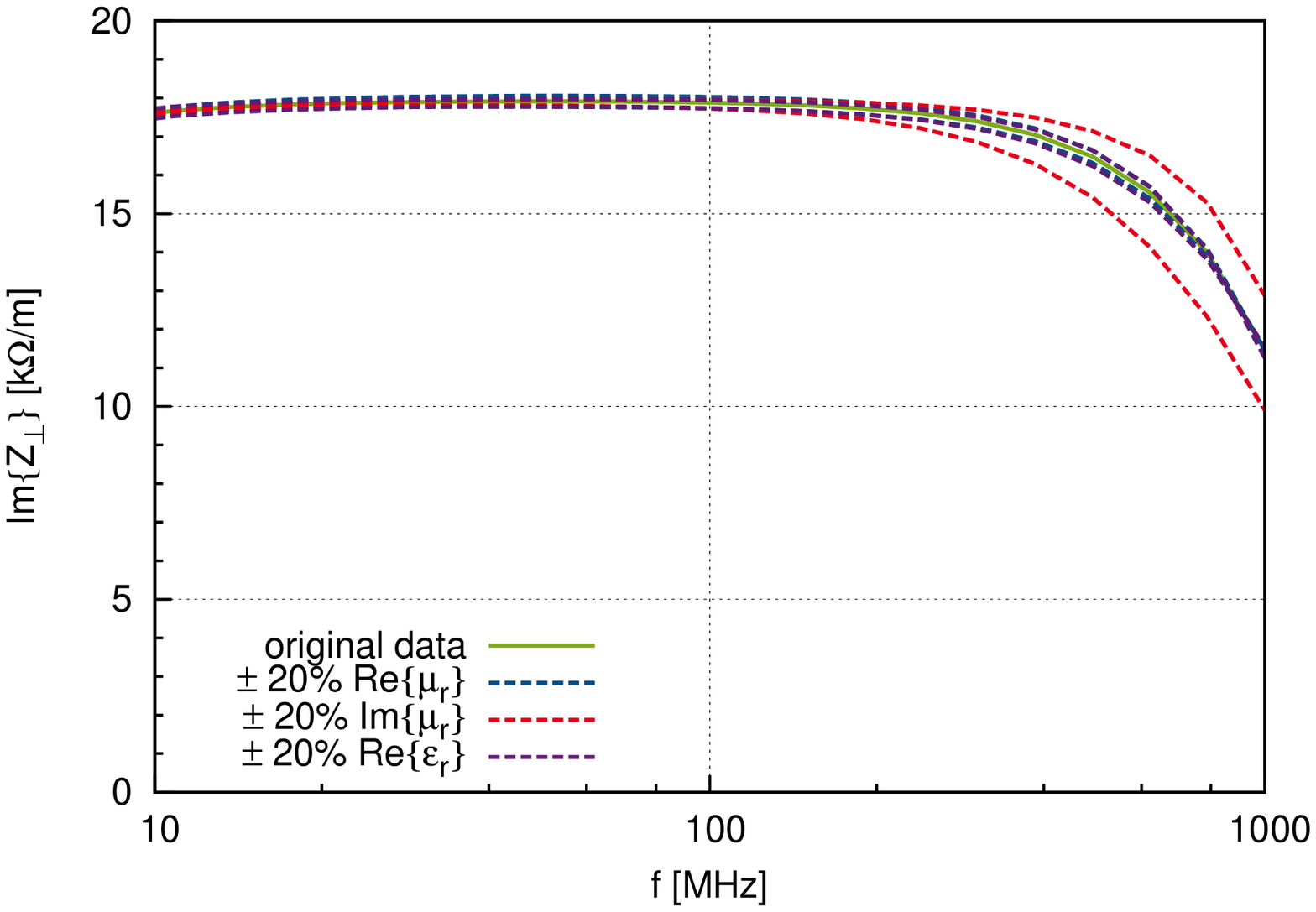}
	\caption{Longitudinal impedance errors from material data deviation}
	\label{material_long}
\end{figure}
Figure \ref{material_long} shows the error propagation in the MWS simulation of the wire measurement. As expected, deviations in $\mu''$ influence mostly the real part of the impedance. The uncertainties in the imaginary part of the impedance is dominated by $\mu'$ for below 100 MHz and above the influences of the (strong) losses prevail. For $Z_\perp$ the error propagation is smaller and dominated by the image current losses due to $\mu''$.

\bibliography{MeasurementsPaper}

\end{document}